\documentclass[10pt,letterpaper]{article}
\usepackage[margin=2.0cm]{geometry}
\usepackage{mathrsfs,amsmath,amssymb,amsthm,mathtools,bm,slashed,booktabs,longtable,hyperref,microtype}
\hypersetup{colorlinks=true,linkcolor=blue,citecolor=blue,urlcolor=blue}
\numberwithin{equation}{section}

\title{\textbf{Toward Dirac Quantization of Teleparallel Gravity with a Coframe--Spin-Connection Pair}}
\author{A. Landry}
\date{\today}
\begin{document}
\maketitle
\begin{abstract}
We develop a branchwise canonical framework for teleparallel $F(T)$ gravity with an independent coframe, Lorentz connection, flatness multiplier and Levi--Civita-coupled Dirac field. Following the algebraic BF and fermionic second-class reductions, the coframe Hessian is classified into ranks $\{0,1,5,6,7\}$. On the regular auxiliary-scalar branch, stabilization of the rank-six primary matrix yields an explicit eight-constraint second-class inverse. Subject to functional independence, invertibility and reduced Euler--Hamilton equivalence, five local gravitational configuration degrees of freedom remain. An inhomogeneous constrained initial-data example exhibits a nonvanishing stabilized bracket between reference-relative spatial torsion observables. We formulate canonical operator prescriptions, a classical BFV charge and a reducible BV master action, using TEGR--Einstein--Dirac theory as a benchmark. The formal all-order quantum obstruction equations distinguish ghost-number-two BFV from ghost-number-one BV cohomology. Anomaly cancellation, a physical representation and regulated measure equivalence remain open.
\end{abstract}

\section{Introduction}
\label{sec:introduction}

The canonical quantization of gravitation requires more than a choice of fundamental geometric variables. It requires a differentiable Hamiltonian, a consistent classification of constraints, and a symplectic structure on which the quantum algebra can be based. The ADM formulation of general relativity provides the standard metric realization of this programme, while the hypersurface-deformation algebra relates its canonical generators to spacetime covariance \cite{ADM1962,Teitelboim1973}. The Dirac--Bergmann theory of constrained systems and its subsequent extensions provide the general machinery for distinguishing gauge generators from second-class constraints and constructing the corresponding reduced brackets \cite{Dirac1964,HenneauxTeitelboim1992}. These requirements are particularly important when alternative geometric formulations of gravity are considered: classical equivalence of field equations does not, by itself, establish equivalence of canonical variables, boundary symplectic structures or quantum measures.

Teleparallel gravity offers a complementary geometric description in which gravitation is encoded in the torsion of a flat, metric-compatible connection rather than in the curvature of the Levi--Civita connection. The teleparallel equivalent of general relativity (TEGR), its nonlinear extensions and their covariant formulation have been extensively investigated \cite{Maluf2013,Bahamonde2023,Krssak2019,Jimenez2019}. The coframe determines the metric and the local orthonormal frame, whereas the Lorentz connection enters the covariant definition of torsion. The conceptual and geometric motivations for treating this pair as the starting point of a teleparallel approach to quantum gravity were developed in Ref.~\cite{Landry2026Quantum}. The present work pursues a concrete canonical realization of that programme. In particular, the coframe and connection are initially retained as independent variational fields, with flatness imposed by a multiplier, rather than imposed by a premature choice of frame.

The distinction between TEGR and nonlinear $F(T)$ gravity is essential. In TEGR, the torsion and Einstein--Hilbert densities differ by a boundary divergence on the flatness surface. In nonlinear theories this relation does not imply equivalence to metric general relativity, and the dependence of the Legendre map on the torsion configuration becomes nontrivial. The covariant formulation of $f(T)$ gravity by Kr\v{s}\v{s}\'ak and Saridakis, the earlier discussion of local Lorentz invariance by Li, Sotiriou and Barrow, and subsequent Hamiltonian investigations clarify the importance of distinguishing paired Lorentz covariance from the constraint structure of the relative frame variables \cite{KrssakSaridakis2016,LiSotiriouBarrow2011,BlixtReview2021,GolovnevGuzman2021}. Restoring manifest covariance does not alone determine the number or class of the independent canonical constraints.

A further ingredient is the implementation of flatness through a Lorentz-algebra-valued two-form multiplier. Four-dimensional BF theory provides the natural framework for this sector, including its Gauss constraint, curvature constraints and Bianchi identities \cite{BlauThompson1991,Birmingham1991}. The multiplier shift is first-stage reducible on shell, so its covariant quantization requires the corresponding ghost-for-ghost structure \cite{BVReducible1983}. Although the isolated BF sector has no local propagating degrees of freedom under suitable topological and boundary assumptions, its coupling to the coframe cannot be discarded without examining the total Gauss moment map. We therefore perform the coupled symplectic quotient explicitly. We also retain a Dirac field, coupled through the Levi--Civita spin connection. This choice is motivated by the established discussions of spinor coupling in torsion-based formulations \cite{Hehl1976,ObukhovPereira2003,Maluf2003} and supplies a definite Einstein--Dirac benchmark. Its velocity-linear contribution shifts the coframe momenta and must be included in the reduced symplectic structure.

The Hamiltonian literature provides several complementary starting points. Maluf and da Rocha-Neto developed a Hamiltonian formulation of TEGR, while Oko\l{}\'ow constructed an ADM-like formulation and derived its constraint algebra \cite{MalufRocha2001,Oko2013,Oko2014}. Ferraro and Guzm\'an studied the teleparallel Hamiltonian structure, and Blixt, Hohmann and Pfeifer classified primary constraints in new general relativity \cite{FerraroGuzman2016,BlixtHohmannPfeifer2019}. For nonlinear $F(T)$ gravity, the analyses of Li, Miao and Miao and of Ferraro and Guzm\'an have led to different gravitational degree-of-freedom counts \cite{LiMiaoMiao2011,FerraroGuzman2018}. These differences motivate a careful separation of the pointwise kinetic Hessian rank, the rank of the primary Poisson matrix and the invertibility of the stabilized functional constraint operator. None of these quantities can generally be substituted for another. In particular, a null primary direction may generate a secondary constraint, and its subsequent stabilization may depend on spatial derivatives and the admissible space of smearings.

The present analysis connects these distinctions to the construction of a reduced quantum algebra. The nonlinear Legendre map is classified through a rank-one modification of the TEGR kinetic matrix, while the auxiliary-scalar formulation provides a separate classification of the mixed primary Poisson matrix. On its regular rank-six stratum, we identify the null direction, derive a lapse-independent secondary representative and invert the stabilized eight-constraint system. The resulting tangent Poisson operator and regulated second-class determinant organize the reduction without quotienting the remaining hypersurface gauge transformations. The analysis also retains the curvature of the fermionic momentum shift, which is relevant to the canonical operator prescriptions.

Spatial torsion is a particularly useful probe of this construction. Although configuration observables commute in the product-reduced chart, the subsequent second-class reduction can induce nontrivial brackets between them. We exhibit this effect explicitly for two smeared reference-relative torsion components on inhomogeneous constrained initial data. The example also tests the regularity coefficients and functional independence of the hypersurface constraints on a specific slice. Its torsion bracket is branch dependent and is not, by itself, an algebra of hypersurface-invariant physical observables.

The quantization problem is addressed from complementary Hamiltonian and covariant perspectives. The reduced symplectic structure supplies the canonical brackets and the starting point for a classical BFV charge, whereas the independent-field BV formulation incorporates the reducibility of the multiplier shift through antifield-dependent terms. At the quantum level, the respective consistency conditions yield formal recursive obstruction equations in ghost numbers two and one. Their cohomological closure follows under compatible regularization assumptions, but their exactness does not. This distinction separates the classical construction from the additional requirements of anomaly cancellation, operator domains, a physical inner product and matching regulated measures.

Throughout, the reduction is local, with compatible boundary conditions and exclusion of relevant zero modes. The gravitational degree-of-freedom count is conditional on functional independence, invertibility and reduced Euler--Hamilton equivalence. The initial-data construction is not an evolution theorem; perturbation spectra, characteristic propagation and strong coupling are outside the scope of this work.

Sections 2--3 establish the geometry and preliminary reductions; Sections 4--5 develop the Legendre structure and constraint stabilization. Sections 6--8 address canonical, BFV--BV and quantum consistency prescriptions, and Section 9 concludes. Appendices A--E provide the kinetic, distributional, constraint-algebra, BV and initial-data calculations.

\section{Covariant teleparallel geometry and action}\label{sec:rev-covariant}
On an oriented four-manifold $M$, take an invertible coframe $h^a=h^a{}_{\mu}dx^\mu$, an independent metric-compatible Lorentz connection $\omega^a{}_b$, and a Lorentz-algebra-valued two-form $\lambda_{ab}$. We take $\eta_{ab}=\mathrm{diag}(-1,1,1,1)$, $h=\det(h^a{}_{\mu})>0$, and $\kappa=8\pi G$. We assume a regular foliation and variations or boundary conditions eliminating surface terms \cite{Maluf2013,Bahamonde2023,Krssak2019,Landry2026Quantum}.

\subsection{Coframe, connection, torsion, and contortion}\label{subsec:rev-geometry}
The metric, curvature and torsion two-forms are
\begin{align}
	g_{\mu\nu}&=\eta_{ab}h^a{}_{\mu}h^b{}_{\nu},\\
	R^a{}_b(\omega)&=d\omega^a{}_b+\omega^a{}_c\wedge\omega^c{}_b,\qquad
	T^a=D_\omega h^a=dh^a+\omega^a{}_b\wedge h^b.\label{eq:rev-RT}
\end{align}
With the derivative index last,
\begin{equation}
	\Gamma^\rho{}_{\nu\mu}=h_a{}^\rho\bigl(\partial_\mu h^a{}_{\nu}+\omega^a{}_{b\mu}h^b{}_{\nu}\bigr),\qquad
	T^\rho{}_{\mu\nu}=\Gamma^\rho{}_{\nu\mu}-\Gamma^\rho{}_{\mu\nu}.
	\label{eq:rev-affine}
\end{equation}
The tetrad postulate implies $R^\rho{}_{\sigma\mu\nu}(\Gamma)=h_a{}^\rho h^b{}_{\sigma}R^a{}_{b\mu\nu}(\omega)$. For the Levi--Civita connection $\mathring\Gamma$ of $g$, define $K^\rho{}_{\nu\mu}=\Gamma^\rho{}_{\nu\mu}-\mathring\Gamma^\rho{}_{\nu\mu}$. Metric compatibility and Eq.~\eqref{eq:rev-affine} give
\begin{equation}
	K_{\rho\nu\mu}=-K_{\nu\rho\mu},\qquad
	T_{\rho\mu\nu}=K_{\rho\nu\mu}-K_{\rho\mu\nu},\label{eq:rev-Kdef}
\end{equation}
Hence
\begin{equation}
	K_{\rho\nu\mu}=\frac12\bigl(T_{\nu\rho\mu}+T_{\mu\rho\nu}-T_{\rho\nu\mu}\bigr),\qquad
	\mathring\omega^a{}_{b\mu}[h]=\omega^a{}_{b\mu}-K^a{}_{b\mu}[h,\omega].\label{eq:rev-K}
\end{equation}
At fixed $h$, $\delta_\omega\mathring\omega[h]=0$; the Dirac derivative uses $\mathring\omega[h]$, not $\omega$ \cite{Hehl1976,ObukhovPereira2003,Maluf2003}.

With $T_\mu=T^\nu{}_{\nu\mu}$, define
\begin{equation}
	T=\frac14T^{\rho\mu\nu}T_{\rho\mu\nu}
	+\frac12T^{\rho\mu\nu}T_{\nu\mu\rho}-T_\rho T^\rho.
	\label{eq:rev-Tscalar}
\end{equation}
Off shell, the curvature difference identity gives
\begin{align}
	R^\rho{}_{\sigma\mu\nu}(\Gamma)&=\mathring R^\rho{}_{\sigma\mu\nu}+2\mathring\nabla_{[\mu}K^\rho{}_{|\sigma|\nu]}+2K^\rho{}_{\ell[\mu}K^\ell{}_{|\sigma|\nu]},\\
	\mathring R&=R(\Gamma)-T+2h^{-1}\partial_\mu(hT^\mu).
	\label{eq:rev-offshell-R}
\end{align}
Our curvature contractions are
$R_{\sigma\nu}=R^\rho{}_{\sigma\rho\nu}$
and $R=g^{\sigma\nu}R_{\sigma\nu}$.
Together with the derivative-last connection convention,
these fix the relative signs of the curvature and
torsion scalars throughout the manuscript
\cite{Maluf2013,Bahamonde2023,Krssak2019}.

On the flatness surface $R(\omega)=0$, the curvature decomposition yields
\begin{equation}
	\mathring R=-T+\frac{2}{h}\partial_\mu(hT^\mu),\qquad
	-\frac{1}{2\kappa}\int_MhT\,d^4x
	=\frac{1}{2\kappa}\int_Mh\mathring R\,d^4x
	-\frac1\kappa\int_{\partial M}hT^\mu\,d\Sigma_\mu.
	\label{eq:rev-TEGRidentity}
\end{equation}
For the canonical role of this boundary term, see \cite{Maluf2013,KrssakPereira2015,MalufRocha2001,Oko2013}.

\subsection{Independent action, multiplier shift, and field equations}\label{subsec:rev-action}
Using the invariant Lorentz pairing $\langle X,Y\rangle=\tfrac12 X_{ab}Y^{ab}$ (ordered antisymmetric pairs), take
\begin{equation}
	S_0=-\frac{1}{2\kappa}\int_MhF(T)\,d^4x
	+\int_M\langle\lambda\wedge R(\omega)\rangle+S_D[h,\Psi,\bar\Psi].
	\label{eq:rev-action}
\end{equation}
With $\Sigma_{ab}=\epsilon_{abcd}h^c\wedge h^d/2$ and oriented volume $\epsilon_{0123}=+1$, one has $\langle\Sigma\wedge R\rangle=\tfrac12\Sigma_{ab}\wedge R^{ab}=\tfrac12 hR(\Gamma)d^4x$. Thus the invertible off-shell TEGR redefinition is
\begin{align}
	\widetilde\lambda_{ab}&=\lambda_{ab}-\Sigma_{ab}/\kappa, \label{eq211}\\
	S_0|_{F=T}&=S_{\rm EH}+S_D+\int_M\langle\widetilde\lambda\wedge R\rangle-\kappa^{-1}\int_{\partial M}hT^\mu d\Sigma_\mu.
	\label{eq:rev-BF-redefinition}
\end{align}
This is an invertible Lagrangian field redefinition; its canonical implementation requires the corresponding momentum transformation, as discussed in Secs.~\ref{subsec:rev-Lorentz} and \ref{subsec:relative_lorentz_reduction}. The BF variation in left-ordered Euler form is
\begin{align}
	\delta S_{\rm BF}&=\int_M\bigl\langle\delta\lambda\wedge R+\delta\omega\wedge D_\omega\lambda\bigr\rangle
	+\int_{\partial M}\langle\lambda\wedge\delta\omega\rangle.
	\label{eq:rev-BFvariation}
\end{align}
The multiplier equation is $R=0$.

Besides paired local Lorentz covariance, the action has the multiplier shift
\begin{equation}
	\delta_\zeta\lambda=D_\omega\zeta,\qquad
	\delta_\zeta(h,\omega,\Psi,\bar\Psi)=0,
	\qquad
	\delta_\zeta S_0=\int_{\partial M}\langle\zeta\wedge R\rangle.
	\label{eq:rev-shift}
\end{equation}
The Bianchi identity $D_\omega R=0$ ensures this symmetry; the replacement $\zeta\mapsto\zeta+D_\omega\chi$ changes $\delta\lambda$ by $D_\omega^2\chi=[R,\chi]$: first-stage \emph{on-shell} reducibility, resolved in Sec.~7 \cite{BlauThompson1991,Birmingham1991,BV1981,BVReducible1983,HenneauxTeitelboim1992}.

\subsection{Levi--Civita Dirac coupling and Lorentz covariance}\label{subsec:rev-dirac}
Adopt $\{\Gamma_a,\Gamma_b\}=-2\eta_{ab}\mathbf1$, $\Gamma_{ab}=\tfrac12[\Gamma_a,\Gamma_b]$, and $\bar\Psi=\Psi^\dagger\Gamma^{\hat0}$ with $(\Gamma^{\hat0})^\dagger=\Gamma^{\hat0}$ and $(\Gamma^I)^\dagger=-\Gamma^I$. The matrices $\alpha^I=\Gamma^{\hat0}\Gamma^I$ are Hermitian. Define
\begin{equation}
	\mathring D_\mu\Psi=\partial_\mu\Psi-\frac14\mathring\omega_{ab\mu}\Gamma^{ab}\Psi,\qquad
	\mathring D_\mu\bar\Psi=\partial_\mu\bar\Psi+\frac14\bar\Psi\mathring\omega_{ab\mu}\Gamma^{ab}.
	\label{eq:rev-spinD}
\end{equation}
Set $\Sigma(\epsilon)=-\epsilon_{ab}\Gamma^{ab}/4$ and $\Omega_\mu=-\mathring\omega_{ab\mu}\Gamma^{ab}/4$. Paired Lorentz covariance is equivalent to
\begin{equation}
	\delta_\epsilon\Psi=\Sigma(\epsilon)\Psi,\quad
	\delta_\epsilon\Omega_\mu=[\Sigma(\epsilon),\Omega_\mu]-\partial_\mu\Sigma(\epsilon),\quad
	\delta_\epsilon(\partial_\mu+\Omega_\mu)\Psi=\Sigma(\epsilon)(\partial_\mu+\Omega_\mu)\Psi.
	\label{eq:rev-spinproof}
\end{equation}
For real $\epsilon$, $\Sigma(\epsilon)^\dagger\Gamma^{\hat0}+\Gamma^{\hat0}\Sigma(\epsilon)=0$. The notation $\Sigma(\epsilon)$ denotes the
spin representation of the Lorentz generator,
and should not be confused with the two-form
$\Sigma_{ab}$ introduced in Eq.~\eqref{eq211}.

Direct Clifford multiplication gives
\begin{align}
	[\Gamma^{ab},\Gamma^c]&=-2(\eta^{bc}\Gamma^a-\eta^{ac}\Gamma^b),\\
	[\Gamma^{ab},\Gamma^{cd}]&=-2(\eta^{bc}\Gamma^{ad}-\eta^{ac}\Gamma^{bd}-\eta^{bd}\Gamma^{ac}+\eta^{ad}\Gamma^{bc}),\\
	[\Sigma(\epsilon),\Gamma^c]&=-\epsilon^c{}_a\Gamma^a,\quad [\Sigma(\epsilon),\Sigma(\eta)]=\Sigma([\epsilon,\eta]).
\end{align}
With $\delta\Gamma^\mu=[\Sigma(\epsilon),\Gamma^\mu]$, the invariant action is
\begin{equation}
	S_D=\int_Mh\left[\frac i2\left(\bar\Psi\Gamma^\mu\mathring D_\mu\Psi
	-(\mathring D_\mu\bar\Psi)\Gamma^\mu\Psi\right)-m\bar\Psi\Psi\right]d^4x,
	\quad\Gamma^\mu=h_a{}^\mu\Gamma^a.
	\label{eq:rev-diracaction}
\end{equation}
The connection-dependent density and its momentum derivative are
\begin{align}
	\mathcal L_{D,\mathring\omega}&=-\frac{ih}{8}\bar\Psi\{\Gamma^\mu,\Gamma^{ab}\}\Psi\,\mathring\omega_{ab\mu}
	=-\frac{ih}{4}\bar\Psi\Gamma^{\mu ab}\Psi\,\mathring\omega_{ab\mu},\label{eq:rev-Lspin}\\
	A_a{}^i&:=\frac{\partial\mathcal L_{D,\mathring\omega}}{\partial\dot h^a{}_i}
	=-\frac{ih}{8}\bar\Psi\{\Gamma^\mu,\Gamma^{bc}\}\Psi\,
	\frac{\partial\mathring\omega_{bc\mu}[h]}{\partial\dot h^a{}_i}.
	\label{eq:rev-Achain}
\end{align}
Define $C_{abc}=\eta_{ad}h_b{}^\mu h_c{}^\nu(\partial_\mu h^d{}_\nu-\partial_\nu h^d{}_\mu)$ and $Z_{abc|di}=\eta_{ad}(h_b{}^0h_c{}^i-h_b{}^ih_c{}^0)$. The Cartan equation gives $\mathring\omega_{abc}=(C_{abc}+C_{bca}-C_{cab})/2$. For $J^{abc}=\bar\Psi\Gamma^{[a}\Gamma^b\Gamma^{c]}\Psi$, cyclic permutations leave the contraction unchanged. Hence the complete raw shift is
\begin{equation}
	\mathcal A^{\rm raw}_d{}^i=-\frac{ih}{8}J^{abc}Z_{abc|di}=-\frac{ih}{4}\eta_{ad}J^{abc}h_b{}^0h_c{}^i.
	\label{eq:rev-Acompact}
\end{equation}
The spin-coordinate correction is derived in Sec.~\ref{sec:rev-fermions} \cite{BirrellDavies1982,ParkerToms2009,MalufRocha2001,Oko2014}.

\paragraph{Off-shell Lorentz Noether identity.}\label{subsec:rev-noether}
Write the bulk variation with left-ordered bosonic variations as
\begin{equation}
	\delta S_0=\int_M\left[\delta h^a\wedge E_a+\langle\delta\omega\wedge E_\omega\rangle
	+\langle\delta\lambda\wedge R\rangle+E_\Psi\delta\Psi+\delta\bar\Psi E_{\bar\Psi}\right]+\int_{\partial M}\Theta.
	\label{eq:rev-Euler}
\end{equation}
For each Lorentz generator $X_I$, invariance gives \cite{HenneauxTeitelboim1992,GolovnevGuzman2021}
\begin{multline}
	0\equiv\langle X_I,D_\omega E_\omega\rangle+(X_I)^a{}_bh^b\wedge E_a+\langle[X_I,\lambda],R\rangle\\+E_\Psi\Sigma(X_I)\Psi-\bar\Psi\Sigma(X_I)E_{\bar\Psi}.
\end{multline}

\section{Canonical formulation and constraint structure}\label{sec:rev-canonical}
Assume $M\simeq\mathbb R\times\Sigma$ and decompose $h^a{}_0=Nn^a+N^ih^a{}_i$, with $n_an^a=-1$, $n_ah^a{}_i=0$, $q_{ij}=\eta_{ab}h^a{}_ih^b{}_j$. We retain temporal primary momenta and odd canonical pairs; the coframe Legendre map is derived in Sec.~\ref{sec:TEGR-Hamiltonian-benchmark} \cite{Dirac1964,MalufRocha2001,BlixtHohmannPfeifer2019,BlixtReview2021}.

\subsection{Exact spatial BF Legendre reduction}\label{subsec:rev-BF}
For $\epsilon^{0123}=+1$ and $\lambda_{ab}=\tfrac12\lambda_{ab\mu\nu}dx^\mu\wedge dx^\nu$, the BF density (six independent Lorentz pairs) is
\begin{equation}
	\mathcal L_{\rm BF}=\frac12\epsilon^{ijk}\lambda_{abjk}
	(\dot\omega^{ab}{}_i-D_i\omega^{ab}{}_0)
	+\frac12\epsilon^{ijk}\lambda_{ab0i}R^{ab}{}_{jk}.
	\label{eq:rev-BFsplit}
\end{equation}
Define $B_{ab}{}^i=\tfrac12\epsilon^{ijk}\lambda_{abjk}$ and $\mathcal F_{ab}{}^i=\tfrac12\epsilon^{ijk}R_{abjk}$. Then
\begin{equation}
	\mathcal L_{\rm BF}=B_{ab}{}^i\dot\omega^{ab}{}_i
	-B_{ab}{}^i(D_i\omega_0)^{ab}+\lambda_{ab0i}\mathcal F^{ab i},\qquad
	\Pi_{ab}{}^i=B_{ab}{}^i.
	\label{eq:rev-BFmom}
\end{equation}
For independent $B_{ab}{}^i$ with momenta $p_B^{ab}{}_i$, the constraints $\Phi_{ab}{}^i=\Pi_{ab}{}^i-B_{ab}{}^i$ and $\varphi^{ab}{}_i=p_B^{ab}{}_i$ have the explicit bosonic bracket matrix
\begin{equation}
	C_{AB}(x,y)=\begin{pmatrix}0&-I_{18}\\ I_{18}&0\end{pmatrix}\delta^3(x-y),\qquad
	C^{-1}_{AB}(x,y)=\begin{pmatrix}0&I_{18}\\-I_{18}&0\end{pmatrix}\delta^3(x-y).
	\label{eq:rev-BFmatrix}
\end{equation}
The exact Dirac reduction sets $B=\Pi$, $p_B=0$ while preserving
\begin{equation}
	\{\omega^{ab}{}_i(x),\Pi_{cd}{}^j(y)\}_{D,\mathrm{BF}}
	=\delta^{ab}{}_{cd}\delta_i{}^j\delta^3(x-y).
	\label{eq:rev-BFbracket}
\end{equation}
Here $\delta^{ab}{}_{cd}$ is the independent-pair identity; $\Pi$ remains dynamical at this stage.

\subsection{Flatness, Bianchi identities, and reducibility}\label{subsec:rev-flat}
The vanishing momenta of $\lambda_{ab0i}$ produce the spatial curvature constraints
\begin{equation}
	\mathcal F_{ab}{}^i=\frac12\epsilon^{ijk}R_{abjk}\approx0,
	\qquad D_i\mathcal F_{ab}{}^i\equiv0.
	\label{eq:rev-flat}
\end{equation}
On the regular local patch, six Bianchi relations leave twelve independent flatness constraints. Together with six Gauss constraints, these are first class in isolated BF theory:
\begin{equation}
	N_{\mathrm{BF,local}}=\frac12\bigl[36-2(6+18-6)\bigr]=0.
	\label{eq:rev-BFcount}
\end{equation}
The coupled Gauss condition and quotient are treated in Sec.~\ref{sec:dirac_bergmann}; global and boundary modes are excluded \cite{BlauThompson1991,Birmingham1991,DonnellyFreidel2016}.

\subsection{Lorentz cotangent lift and temporal constraint chains}\label{subsec:rev-Lorentz}
The extended Lorentz cotangent lift is
\begin{multline}
	G_L[\epsilon]=\int_\Sigma d^3x\bigl[
	\pi_a{}^i\epsilon^a{}_bh^b{}_i
	-\Pi_{ab}{}^i(D_i\epsilon)^{ab}
	+p_\lambda\cdot[\epsilon,\lambda]
	+\pi_\Psi\delta_\epsilon\Psi
	+\delta_\epsilon\bar\Psi\,\pi_{\bar\Psi}\bigr]\\
	+G_{\rm temporal}[\epsilon]+Q_{\partial\Sigma}[\epsilon].
	\label{eq:rev-GL}
\end{multline}
Its brackets generate $\delta_\epsilon h^a{}_i=\epsilon^a{}_bh^b{}_i$, $\delta_\epsilon\omega_i=-D_i\epsilon$, and the spinor transformation of Eq.~\eqref{eq:rev-spinproof}. Kinematical closure alone does not establish stabilization \cite{Dirac1964,HenneauxTeitelboim1992,GolovnevGuzman2021}.

After BF reduction define $G[\epsilon]=\int\langle\epsilon,D_i\Pi^i\rangle$ and $\mathscr F[\zeta]=\int\langle\zeta_i,\mathcal F^i\rangle$. Then
\begin{equation}
	H_{\rm BF}=-G[\omega_0]-\mathscr F[\lambda_0]+\int_{\partial\Sigma}\langle\Pi^i,\omega_0\rangle dS_i.
\end{equation}
For $H_T=H_{gD}+H_{\rm BF}+\int u^\alpha C_\alpha+\int u^r p_r$, temporal stabilization gives
\begin{align}
	\dot p_{\lambda_0}^i&=\mathcal F^i,\\
	\dot p_{\omega_0}&=D_i\Pi^i-\delta_{\omega_0}(H_{gD}+\int u^\alpha C_\alpha),\\
	\dot p_{h_0^a}&=-\delta_{h_0^a}(H_{gD}+\int u^\alpha C_\alpha).
\end{align}
The derivatives hold canonical coordinates and multipliers fixed. Spatial flatness is preserved because $H_{gD}$ and $C_\alpha$ are independent of $\Pi$ in this chart. 

\paragraph{Boundary-related momentum translation.}
The multiplier shift induces $Q^i_{ab}=-(2\kappa)^{-1}\epsilon^{ijk}\Sigma_{abjk}$. A canonical translation on the reduced spatial chart is generated by
\begin{equation}
	\mathcal G=\int\langle Q^i,\omega_i\rangle,\qquad \widetilde\Pi^i=\Pi^i+Q^i,\quad\widetilde\pi_A=\pi_A+\delta\mathcal G/\delta h^A.
\end{equation}
The one-form changes by $\delta\mathcal G$; the distinct relative boundary translation is given in Sec.~\ref{sec:dirac_bergmann} \cite{Dirac1964,BlixtReview2021,GolovnevGuzman2021}.

\subsection{Fermionic symplectic structure and Dirac reduction}\label{sec:rev-fermions}
For the isolated Grassmann block, work in a local time-gauge frame, $h^{\hat0}{}_i=0$, $h^I{}_i=e^I{}_i$, and use $h=N\sqrt q$ \cite{HenneauxTeitelboim1992,BirrellDavies1982,ParkerToms2009}.

\subsubsection{Half-density spinors and the reduced bracket}\label{subsec:rev-halfdensity}
The explicit spinor-velocity contribution to the symmetrized action is
\begin{equation}
	\mathcal L_{D,\mathrm{exp}}^{\mathrm{vel}}=\frac{i\sqrt q}{2}
	(\psi^\dagger\dot\psi-\dot\psi^\dagger\psi).
	\label{eq:rev-fermvel}
\end{equation}
For $\xi=q^{1/4}\psi$, the $\partial_t\ln q$ terms cancel, giving
\begin{equation}
	\mathcal L_{D,\mathrm{exp}}^{\mathrm{vel}}=\frac i2
	(\xi^\dagger\dot\xi-\dot\xi^\dagger\xi),\qquad
	\Theta_F=\frac i2\int_\Sigma d^3x
	(\xi^\dagger\delta\xi-\delta\xi^\dagger\xi).
	\label{eq:rev-ThetaF}
\end{equation}
With a fixed symmetric graded bracket for elementary odd canonical partners, the primary constraints are
\begin{equation}
	\chi_A=p_{\xi_A}-\frac{i}{2}\xi_A^\dagger
	\approx 0,\qquad
	\bar{\chi}_A=p_{\xi_A^\dagger}
	-\frac{i}{2}\xi_A\approx 0.
\end{equation}
The nonzero elementary coordinate--momentum brackets are
\begin{equation}
	\{\xi_A(x),p_{\xi_B}(y)\}_{\rm gr}=\delta_{AB}\delta^3(x-y),\qquad
	\{\xi_A^\dagger(x),p_{\xi_B^\dagger}(y)\}_{\rm gr}=\delta_{AB}\delta^3(x-y).
	\label{eq:rev-elementary}
\end{equation}
Ordering $(\chi_A,\bar\chi_A)$, their second-class matrix and inverse are
\begin{equation}
	C=-i\begin{pmatrix}0&I_4\\I_4&0\end{pmatrix}\delta^3,
	\qquad C^{-1}=i\begin{pmatrix}0&I_4\\I_4&0\end{pmatrix}\delta^3.
	\label{eq:rev-fermC}
\end{equation}
The Dirac subtraction then gives
\begin{equation}
	\{\xi_A(x),\xi_B^\dagger(y)\}_{D,F}
	=-i\delta_{AB}\delta^3(x-y),\qquad
	\{\xi_A,\xi_B\}_{D,F}
	=\{\xi_A^\dagger,\xi_B^\dagger\}_{D,F}=0.
\end{equation}
See Sec.~\ref{sec:canonical_quantization} for quantization \cite{Thiemann2007,Rovelli2004}.

\subsubsection{Spin-frame momentum shift and mixed symplectic brackets}\label{subsec:rev-rot}
Choose a local spin lift $S[h_i]$ mapping the reference normal to $n^a$, without eliminating boost coordinates. Set $\Psi=S q^{-1/4}\xi$, $\bar\Psi=q^{-1/4}\bar\xi S^{-1}$ and $\bar\xi=\xi^\dagger\Gamma^{\hat0}$. Since $S^{-1}\Gamma^0 S=N^{-1}\Gamma^{\hat0}$, the explicit spinor one-form becomes
\begin{equation}
	\Theta_{F,\rm exp}=\Theta_F+\int\mathcal B_A\delta h^A,\qquad \mathcal B_A=\frac i2\bar\xi\{\Gamma^{\hat0},S^{-1}\partial_A S\}\xi.
\end{equation}
The boost section can have rotational Maurer--Cartan components; the total shift is
\begin{equation}
	\mathcal A_A=\mathcal A_A^{\rm raw}[h,S q^{-1/4}\xi]+\mathcal B_A,\qquad \pi_A=P_A+\mathcal A_A.
	\label{eq:rev-total-A}
\end{equation}
The spin lift induces no temporal or BF momentum.
\paragraph{Lorentz projections.}
At a normal-adapted evaluation point, Eq.~\eqref{eq:rev-Acompact} gives
\begin{equation}
	\mathcal A^{\rm raw}_{\hat0}{}^i=0,\qquad \mathcal A^{\rm raw}_I{}^i=\frac i4\xi^\dagger\Gamma^{IJ}\xi e_J{}^i.
\end{equation}
The independent rotation coefficients are $i\xi^\dagger\Gamma^{IJ}\xi/2$ for $I<J$. The residual rotation charge is $G_F[\theta]=-i\int\xi^\dagger\theta_{IJ}\Gamma^{IJ}\xi/4$, generating $\{\xi,G_F\}_*=-\theta_{IJ}\Gamma^{IJ}\xi/4$. The adapted-spinor transformation also includes $\delta S[h_i]$.
\paragraph{Mixed brackets.}\label{subsec:rev-mixed}
The reduced local bulk one-form is $\Theta=\int(\pi_A\delta h^A+\langle\Pi^i,\delta\omega_i\rangle)+\Theta_F$. Only $\pi_A$ is Darboux. Write $\mathcal A_A=\xi^\dagger m_A[h]\xi$. Reality of the velocity-linear fermionic term implies $m_A^\dagger=m_A$ in the adapted spinor chart. Direct expansion gives
\begin{align}
	\{\xi(x),P_A(y)\}_*&=im_A\xi\,\delta^3(x-y),\\
	\{\xi^\dagger(x),P_A(y)\}_*&=-i\xi^\dagger m_A\,\delta^3(x-y),\\
	\{P_A(x),P_B(y)\}_*&=\frac{\delta\mathcal A_B(y)}{\delta h^A(x)}-\frac{\delta\mathcal A_A(x)}{\delta h^B(y)}\notag\\&\quad-i\xi^\dagger[m_A,m_B]\xi\,\delta^3(x-y).
\end{align}
For ultralocal $m_A$, define the matrix-valued curvature
\begin{equation}
 \mathcal F_{AB}=\partial_A m_B-\partial_B m_A-i[m_A,m_B],\qquad
 \{P_A(x),P_B(y)\}_*=\xi^\dagger\mathcal F_{AB}\xi\,\delta^3(x-y).
 \label{eq:rev-shift-curvature}
\end{equation}
The identity $\nabla^{(m)}_{[A}\mathcal F_{BC]}=0$, with $\nabla^{(m)}_A=\partial_A-i[m_A,\cdot]$, follows from the matrix Jacobi identity. This is a configuration-space curvature associated
with the fermionic momentum shift, rather than
the spacetime curvature $R(\omega)$ constrained
by the BF multiplier. Here $\partial_A$ differentiates with respect to $h^A$ at fixed adapted spinor. For derivative-dependent coefficients the functional kernel above must be retained. The velocity-linear shift leaves both the coframe Hessian and the BF block unchanged. Sections~6--7 and Appendix~\ref{app:torsion-brackets} use this partial chart \cite{Dirac1964,HenneauxTeitelboim1992,Thiemann2007}.

\paragraph{Bracket conventions.}
The subscripts ${\rm gr}$, $D,\mathrm{BF}$ and $D,F$ denote the elementary graded bracket and the algebraic BF and fermionic Dirac reductions, respectively. The star denotes the product-reduced bracket; the subscript $4$ denotes the subsequent paired-primary reduction (six constraints), and ${\rm st}$ the complete second-class reduction of Sec.~\ref{sec:dirac_bergmann}.

\section{TEGR benchmark and nonlinear Legendre structure}\label{sec:TEGR-Hamiltonian-benchmark}
\subsection{TEGR benchmark and ADM kinetic structure}\label{subsec:TEGR-boundary}
The multiplier redefinition of Sec.~2 separates the TEGR bulk action into Einstein--Dirac and BF sectors. Adding $d\mu$ to the Lagrangian changes the potential by $\delta\mu$ modulo a spatial divergence, leaving the bulk presymplectic current unchanged. The boundary representative and canonical translations must be chosen consistently; original and ADM momenta need not coincide \cite{ADM1962,MalufRocha2001,Oko2013,DonnellyFreidel2016}.

\paragraph{Local Lorentz coordinates in the separated representative.}\label{subsec:TEGR-Hessian}
Choose a reference coframe $\bar h(q,N,N^i)$ and write $h=\Lambda\bar h$, $\Psi=S(\Lambda)\psi$. Paired covariance removes the six Lorentz coordinates $\ell^I$ from the Einstein--Dirac action at fixed separated BF variables. Thus

\begin{equation}
	p_{\ell I}\approx0,\qquad \{p_{\ell I}(x),p_{\ell J}(y)\}=0,\qquad \{p_{\ell I},H_{\rm sep}\}=0.
\end{equation}
The primary matrix vanishes in these coordinates (weakly in a body basis). These are independent coframe--spinor transformations at fixed separated BF variables, not the nonlinear mixed constraints of Sec.~\ref{sec:dirac_bergmann} \cite{Dirac1964,HenneauxTeitelboim1992,GolovnevGuzman2021}.

\paragraph{ADM momenta and kinetic Hessian.}\label{subsec:adm-momenta-revised}
Let $V_{ij}=\dot q_{ij}-\mathcal L_{\vec N}q_{ij}$ and $K_{ij}=V_{ij}/(2N)$. The gravitational kinetic density and momenta are
\begin{align}
	\mathcal L_{\rm kin}&=\frac{\sqrt q}{8\kappa N}\mathcal G^{ij,kl}V_{ij}V_{kl},\\
	\mathcal G^{ij,kl}&=q^{i(k}q^{l)j}-q^{ij}q^{kl},\\
	P_g^{ij}&=\frac{\sqrt q}{2\kappa}(K^{ij}-q^{ij}K),\quad K_{ij}=\frac{2\kappa}{\sqrt q}(P_{gij}-\tfrac12q_{ij}P_g).
\end{align}
The coframe velocity map $Lv=2h_{a(i}v^a{}_{j)}$ is surjective, since $v^a{}_i=h^a{}_k q^{kj}X_{ji}/2$ maps to any symmetric $X$. Its six-dimensional kernel consists of $v^a{}_i=\epsilon^a{}_bh^b{}_i$. The DeWitt form has eigenvalues $1$ on traceless tensors and $-2$ on the trace. Thus $W_{\rm ADM}=\sqrt q L^T\mathcal G L/(4\kappa N)$ has rank six. At unit lapse and orthonormal coframe with zero shift, $V=2\operatorname{sym}v$ for the spatial velocity matrix $v$, so
\begin{equation}
	\mathcal L_{\rm kin}=\frac{1}{2\kappa}\bigl[\operatorname{tr}((\operatorname{sym}v)^2)-(\operatorname{tr}v)^2\bigr]=-\frac{T_{\rm kin}}{2\kappa}.
\end{equation}
This proves equality of the quadratic Hessians in the adapted chart of Sec.~\ref{subsec:coframe_hessian_rank}, with normalized spectrum $(-2,1,1,1,1,1,0,0,0,0,0,0)$. The trace sign is not a physical ghost count. The velocity-linear shift \eqref{eq:rev-total-A} leaves the Hessian unchanged \cite{ADM1962,MalufRocha2001,BlixtHohmannPfeifer2019}.

\subsection{Einstein--Dirac Hamiltonian and hypersurface constraints}\label{subsec:TEGR-algebra}
The pure metric densities are
\begin{align}
	\mathcal H_g(P_g)&=\frac{2\kappa}{\sqrt q}(P_g^{ij}P_{gij}-\tfrac12P_g^2)-\frac{\sqrt q}{2\kappa}{}^{(3)}R,\\
	\mathcal H_{gi}(P_g)&=-2q_{ij}{}^{(3)}D_kP_g^{jk}.
\end{align}
For an explicit local reference-triad chart, use the half-density $\xi=q^{1/4}\psi$ and let $a^{ij}=\mathcal A^{\rm raw}_I{}^k\partial e^I{}_k/\partial q_{ij}$ evaluated in that reference triad. Then the total metric momentum is $p^{ij}=P_g^{ij}+a^{ij}$. Define $\mathcal L_{D,0}$ by evaluating the full Dirac density in the reference coframe with $\dot q=\dot\xi=\dot\xi^\dagger=0$. This is explicitly
\begin{equation}
	\mathcal L_{D,0}=h\left[\frac i2\bar\psi\Gamma^i\overleftrightarrow{\partial_i}\psi-m\bar\psi\psi-\frac i8\bar\psi\{\Gamma^\mu,\Gamma^{ab}\}\psi\,\mathring\omega_{ab\mu}^{(0)}\right],
\end{equation}
where $\mathring\omega^{(0)}$ is obtained by the Cartan formula of Sec.~2 with zero metric velocities, and $\psi=q^{-1/4}\xi$. There are no lapse or shift velocities. The exact Legendre transform in this chart is
\begin{equation}
	\mathcal H_{ED}=N\mathcal H_g(p-a)+N^i\mathcal H_{gi}(p-a)-\mathcal L_{D,0},
\end{equation}
modulo the usual spatial boundary terms. This expression retains the spin-frame transport contribution, including the momentum shift in the gravitational quadratic form. Lapse and shift constraint densities are the functional derivatives of the integrated Hamiltonian with respect to $N,N^i$, holding $p,q,\xi$ fixed. In particular one must not use $p=P_g$ unless the chosen spin-frame shift vanishes. This gives a local Hamiltonian prescription without assuming a flat-space matter Hamiltonian.

With geometrical generators in the internal-gauge-treated Einstein--Dirac representative, the standard hypersurface-deformation algebra is \cite{Teitelboim1973,HenneauxTeitelboim1992,Oko2014}
\begin{align}
	\{D[V],D[W]\}_D&=D[[V,W]],\\
	\{D[V],H[N]\}_D&=H[\mathcal L_V N],\\
	\{H[N],H[M]\}_D&=D[q^{ij}(N\partial_j M-M\partial_j N)].
\end{align}
Here $H,D$ denote the total generators, including matter. The algebra is the geometrical benchmark; a component-by-component verification of its mixed matter brackets in the reference-triad chart is not claimed. With internal variables retained, compensating frame generators may be needed. The metric sector has twelve canonical dimensions and four first-class hypersurface constraints, hence two local gravitational configuration degrees of freedom. The separated BF sector has zero local propagating modes on a regular contractible patch. This count concerns TEGR with compatible boundary conditions, not arbitrary nonlinear branches.

\subsection{Coframe Hessian and exhaustive rank stratification}
\label{subsec:coframe_hessian_rank}

Let $\mathscr B(X,Y)$ denote the polarization of the
quadratic torsion scalar,
\begin{equation}
	\mathscr B(X,Y)
	=\frac12\left[T(X+Y)-T(X)-T(Y)\right].
\end{equation}
At fixed coframe and connection, write the torsion tensor as
\begin{equation}
	X=X_0+E_Av^A,
	\qquad
	v^A=\dot h^a{}_i,
\end{equation}
where $A=(a,i)$ labels the twelve spatial coframe
velocities. Then
\begin{equation}
	T=T_0+2s^Tv+v^TMv,
	\qquad
	M_{AB}=\mathscr B(E_A,E_B),
	\qquad
	s_A=\mathscr B(X_0,E_A).
\end{equation}
Introduce
\begin{equation}
	q_A=s_A+M_{AB}v^B,
	\qquad
	c=\frac{h}{\kappa}.
\end{equation}
The symbol $q_A$ denotes the torsion-gradient covector
and must not be confused with the spatial metric.

Subtracting the complete fermionic momentum shift gives
\begin{equation}
	P_A=-cF_Tq_A.
\end{equation}
Consequently, the exact spatial coframe Hessian is
\begin{equation}
	\boxed{
		W_{AB}
		=-c\left(F_TM_{AB}+2F_{TT}q_Aq_B\right).
	}
	\label{eq:sec6_exact_hessian}
\end{equation}

The fermionic action is velocity-linear in the coframe
sector and therefore does not modify this Hessian.

In an adapted orthonormal frame, with unit lapse and
vanishing shift at the evaluation point, the TEGR
quadratic matrix takes the form
\begin{equation}
	M=\operatorname{diag}(0_3,D,S,S,S),
\end{equation}
where
\begin{equation}
	D=
	\begin{pmatrix}
		0&1&1\\
		1&0&1\\
		1&1&0
	\end{pmatrix},
	\qquad
	S=-\frac12
	\begin{pmatrix}
		1&1\\
		1&1
	\end{pmatrix}.
\end{equation}
Its spectrum is
\begin{equation}
	\operatorname{spec}(M)
	=
	\{2,-1,-1,-1,-1,-1,0,0,0,0,0,0\}.
\end{equation}

Thus $\operatorname{rank}M=6$, and its kernel consists
of the six infinitesimal Lorentz directions of the
spatial coframe
\cite{Maluf2013,MalufRocha2001,
	BlixtHohmannPfeifer2019,FerraroGuzman2016}.

\paragraph{Proposition 1: exhaustive pointwise rank
	classification.}

Let
\begin{equation}
	\mathcal K=\ker M,\qquad
	\mathcal R=\mathcal K^\perp,
\end{equation}
and let $M^+$ denote the inverse of $M$ on $\mathcal R$,
extended by zero on $\mathcal K$. Define
\begin{equation}
	\mathscr D
	=1+\frac{2F_{TT}}{F_T}q^TM^+q
\end{equation}
whenever $F_T\neq0$.

The possible pointwise ranks of the spatial coframe
Hessian are
\begin{equation}
	\boxed{
		\operatorname{rank}W=
		\begin{cases}
			7,&F_TF_{TT}\neq0,\ q_{\mathcal K}\neq0,\\
			6,&F_T\neq0,\ F_{TT}=0,\\
			6,&F_T\neq0,\ q_{\mathcal K}=0,\
			\mathscr D\neq0,\\
			5,&F_T\neq0,\ q_{\mathcal K}=0,\
			\mathscr D=0,\\
			1,&F_T=0,\ F_{TT}\neq0,\ q\neq0,\\
			0,&F_T=0,\ F_{TT}qq^T=0.
		\end{cases}
	}
	\label{eq:sec6_exhaustive_rank}
\end{equation}

\paragraph{Proof.}

For $F_TF_{TT}\neq0$ and $q_{\mathcal K}\neq0$,
projection of $Wx=0$ onto $\mathcal K$ gives
$q^Tx=0$. Projection onto $\mathcal R$ subsequently
gives $x_{\mathcal R}=0$. Hence
\begin{equation}
	\ker W=\mathcal K\cap q_{\mathcal K}^{\perp},
\end{equation}
which has dimension five.

For $q_{\mathcal K}=0$, the matrix determinant lemma
applied to the invertible restriction $M_{\mathcal R}$
gives
\begin{equation}
	\det\left(
	M_{\mathcal R}
	+\frac{2F_{TT}}{F_T}q_{\mathcal R}q_{\mathcal R}^T
	\right)
	=
	\det(M_{\mathcal R})\mathscr D.
\end{equation}
A rank-one update removes at most one nonzero direction.
At $\mathscr D=0$, the additional null direction is
proportional to $M^+q$. Finally, when $F_T=0$,
the Hessian is proportional to $qq^T$, and therefore
has rank one or zero. This exhausts the possibilities. The classification is exhaustive at the pointwise
algebraic level. A Hamiltonian branch additionally
requires a constant-rank neighborhood and tangency
of the stabilized evolution to the corresponding
constraint stratum
\cite{Dirac1964,HenneauxTeitelboim1992,BlixtHohmannPfeifer2019,BlixtReview2021}.

\hfill$\square$

On the rank-seven branch put $b=\|q_K\|$ and $\gamma=2F_{TT}/F_T$. In an orthonormal basis of $\mathcal R\oplus\operatorname{span}(q_K)$, block elimination gives
\begin{equation}
\det W_7=(-cF_T)^7\gamma b^2\det M_R=4c^7F_T^6F_{TT}b^2,
\end{equation}
where $\det M_R=-2$ in the adapted frame. The inverse is
\begin{equation}
W_7^{-1}=-\frac1{cF_T}\begin{pmatrix}M_R^{-1}&-M_R^{-1}q_R/b\\-q_R^TM_R^{-1}/b&\mathscr D/(\gamma b^2)\end{pmatrix}.
\end{equation}
Thus $\mathscr D=0$ is not singular on this branch.

\subsection{Branchwise Legendre inversion and Hamiltonian
	representatives}
\label{subsec:branchwise_legendre_inversion}

Let $z_I$ span the six Lorentz directions of $\ker M$ and
define
\begin{equation}
	b_I=z_I^Ts,\qquad f=F_T.
\end{equation}

On the TEGR branch, the primary Lorentz relations are
\begin{equation}
	C_I=z_I^TP+cb_I\approx0.
\end{equation}
On the regular rank-seven branch, the kernel momentum
satisfies
\begin{equation}
	P_{\mathcal K}=-cf\,s_{\mathcal K}.
\end{equation}
The five independent primary relations are therefore
\begin{equation}
	C_\alpha=w_\alpha^TP\approx0,
	\qquad
	w_\alpha\in\mathcal K\cap s_{\mathcal K}^\perp.
\end{equation}

If $b_6\neq0$, one may choose
\begin{equation}
	w_\alpha=z_\alpha-\frac{b_\alpha}{b_6}z_6,
	\qquad \alpha=1,\ldots,5.
\end{equation}
The local Legendre inversion is
\begin{align}
	f&=-\frac{s_{\mathcal K}^TP_{\mathcal K}}
	{c\,s_{\mathcal K}^Ts_{\mathcal K}},\\
	T_*&=(F_T)^{-1}(f),\\
	v_{\mathcal R}
	&=M^+\left(-\frac{P_{\mathcal R}}{cf}
	-s_{\mathcal R}\right),\\
	s_{\mathcal K}^Tv_{\mathcal K}
	&=\frac12\left(
	T_*-T_0-q_{\mathcal R}^TM^+q_{\mathcal R}
	+s_{\mathcal R}^TM^+s_{\mathcal R}
	\right),
	\label{eq:sec6_rank7_inversion}
\end{align}
where
\begin{equation}
	q_{\mathcal R}=-\frac{P_{\mathcal R}}{cf}.
\end{equation}
The remaining five null velocities are arbitrary primary
multipliers.

For $s_{\mathcal K}=0$ and $F_T\neq0$, the six primary
relations are $P_{\mathcal K}\approx0$, and $f$ is
determined implicitly by
\begin{equation}
	f=
	F_T\left(
	T_0+
	\frac{P_{\mathcal R}^TM^+P_{\mathcal R}}{c^2f^2}
	-s_{\mathcal R}^TM^+s_{\mathcal R}
	\right).
	\label{eq:sec6_rank6_inversion}
\end{equation}
The derivative of the corresponding implicit equation
with respect to $f$, evaluated on its solution, is
$\mathscr D$. The local inversion is therefore regular
when $\mathscr D\neq0$.

On either regular branch, the Legendre transform can be evaluated without retaining the velocities. Define the on-branch Legendre expression
$U(f)=fT_\ast-F(T_\ast)$.
For a linear function $F(T)=aT+b$, this reduces
to the constant $U=-b$ at $f=a$; no invertible
auxiliary-scalar Legendre map is implied in
that case. Substitution of the regular velocity solution gives
\begin{equation}
 P^Tv_*=-\frac{cf}{2}\left[T_*-T_0+(q_{\mathcal R}-s_{\mathcal R})^TM^+(q_{\mathcal R}-s_{\mathcal R})\right].
\end{equation}
Consequently,
\begin{equation}
\boxed{\begin{aligned}
\mathcal H_{gD}={}&-\frac{P_{\mathcal R}^TM^+P_{\mathcal R}}{2cf}-P_{\mathcal R}^TM^+s_{\mathcal R}\\
&+\frac{cf}{2}(T_0-s_{\mathcal R}^TM^+s_{\mathcal R})-\frac c2 U(f)-\mathcal L_{D,0}.
\end{aligned}}
\label{eq:sec6_branch_hamiltonian}
\end{equation}
Here $f$ is reconstructed by the appropriate branch inversion, and $\mathcal L_{D,0}$ is the velocity-independent Dirac density. Different choices of null velocities change the Hamiltonian only by primary terms.

For momentum-independent smearings and coefficients,
write
\begin{equation}
	C[u]=\int_\Sigma d^3x\,(u^A\pi_A+J_u),
	\qquad
	X_u=u^A\frac{\delta}{\delta h^A}.
\end{equation}
The product-chart bracket is
\begin{align}
	\{C[u],C[v]\}_*
	={}&-\int_\Sigma d^3x\,[u,v]^A\pi_A
	-X_uJ_v+X_vJ_u
	\nonumber\\
	&+\{J_u,J_v\}_{D,F},
	\label{eq:sec6_general_primary_bracket}
\end{align}
where the functional derivatives include all spatial
derivative contributions.

In the rank-seven chart, let
\begin{equation}
	L_I=z_I^TP,\qquad
	r_\alpha=\frac{b_\alpha}{b_6},
	\qquad
	C_\alpha=L_\alpha-r_\alpha L_6.
\end{equation}
The brackets of these constraints must include the
derivatives of $r_\alpha$. In particular, the terms
proportional to $L_6$ cannot be discarded, because
$L_6$ is not itself a primary constraint.

This observation motivates the auxiliary representation
below, in which the complete mixed primary matrix
admits a substantially simpler evaluation
\cite{Dirac1964,HenneauxTeitelboim1992,
	FerraroGuzman2018,GolovnevGuzman2021}.

\subsection{Auxiliary-scalar formulation and regular
	constraint pairs}
\label{subsec:auxiliary_scalar_reduction}\label{subsec:rev5-aux}

Assume $F_{TT}\neq0$ on a locally invertible branch and
introduce
\begin{equation}
	\varphi=F_T(T),
	\qquad
	U(\varphi)=\varphi T(\varphi)-F(T(\varphi)).
\end{equation}
Then
\begin{equation}
	U'(\varphi)=T(\varphi),
	\qquad
	U''(\varphi)=\frac{1}{F_{TT}}.
\end{equation}
The gravitational density becomes
\begin{equation}
	\mathcal L_{g,\mathrm{aux}}
	=-\frac{h}{2\kappa}
	\left(\varphi T-U(\varphi)\right).
\end{equation}

In the original auxiliary representative, the primary
relations are
\begin{equation}
	p_\varphi\approx0,\qquad
	C_I=z_I^TP+c\varphi b_I\approx0,
\end{equation}
with
\begin{equation}
	\{p_\varphi(x),C_I(y)\}_*
	=-c(y)b_I(y)\delta^3(x-y).
	\label{eq:sec6_scalar_lorentz_pair}
\end{equation}

For $b^2=b_Ib_I\neq0$, define
\begin{equation}
	C_\parallel=\frac{b_IC_I}{b^2}.
\end{equation}
The pair $(p_\varphi,C_\parallel)$ is second class,
with weak bracket
\begin{equation}
	\{p_\varphi(x),C_\parallel(y)\}_*
	\approx-c(y)\delta^3(x-y).
\end{equation}
Its elimination reconstructs
\begin{equation}
	\varphi_*=-\frac{b_IL_I}{cb^2},
	\qquad L_I=z_I^TP.
\end{equation}
For observables independent of $\varphi$ and $p_\varphi$,
the corresponding scalar-pair Dirac correction vanishes.

On an admissible $b_I=0$ stratum, the scalar consistency
condition instead gives the candidate secondary
\begin{equation}
	S_\varphi=T_*(\varphi)-U'(\varphi)\approx0,
\end{equation}
where
\begin{equation}
	T_*(\varphi)
	=T_0-s_{\mathcal R}^TM^+s_{\mathcal R}
	+\frac{P_{\mathcal R}^TM^+P_{\mathcal R}}
	{c^2\varphi^2}.
\end{equation}
At fixed original momenta, $-\partial_\varphi S_\varphi=U''+2P_{\mathcal R}^TM^+P_{\mathcal R}/(c^2\varphi^3)$. Substituting $P_{\mathcal R}=-c\varphi q_{\mathcal R}$ gives, on the scalar equation,
\begin{equation}
	\{p_\varphi(x),S_\varphi(y)\}_*
	=
	\frac{\mathscr D}{F_{TT}}\delta^3(x-y).
	\label{eq:sec6_scalar_secondary_bracket}
\end{equation}

Thus the scalar-secondary pair is regular when
$\mathscr D\neq0$. Its elimination preserves the
brackets between scalar-independent original variables,
although the reconstructed scalar generally has
nontrivial brackets with those variables.

For $\mathscr D=0$, the scalar consistency equation
cannot be solved by the same inverse. Further
compatibility and tangency conditions must be examined.
Likewise, the locus $\varphi=0$ requires a separate
Legendre analysis.

The off-shell curvature identity gives, modulo a
boundary divergence,
\begin{align}
	\mathcal L_{g,\mathrm{aux}}+\mathcal L_{\mathrm{BF}}
	={}&
	\frac{h}{2\kappa}
	\left(\varphi\mathring R+U\right)
	+\frac{h}{\kappa}T^\mu\partial_\mu\varphi
	\nonumber\\
	&+\mathcal L_{\mathrm{BF}}
	\left[\omega,\lambda-\frac{\varphi\Sigma}{\kappa}\right].
	\label{eq:sec6_auxiliary_curvature_identity}
\end{align}
The derivative coupling prevents the direct extension
of the TEGR bulk decoupling argument to nonlinear
$F(T)$ gravity.

For the quadratic model,
\begin{equation}
	F(T)=T+\alpha T^2,\qquad
	U(\varphi)=\frac{(\varphi-1)^2}{4\alpha}.
\end{equation}
The limit $\alpha\rightarrow0$ is singular in this
auxiliary chart and must be treated through the
separate TEGR representative
\cite{FerraroGuzman2016,FerraroGuzman2018,
	BlixtReview2021,GolovnevGuzman2021}.

\section{Dirac--Bergmann analysis and regular nonlinear
	constraint reduction}
\label{sec:dirac_bergmann}

We apply the preliminary reductions of Sec.~3 and the auxiliary Legendre map of Sec.~4 on a regular local patch, with differentiable generators and the boundary assumptions of Sec.~2. The product-reduced bracket is denoted by $\{\cdot,\cdot\}_*$ \cite{Dirac1964,HenneauxTeitelboim1992,BlixtHohmannPfeifer2019,GolovnevGuzman2021}.

\subsection{Coupled BF reduction and canonical variables}
\label{subsec:relative_lorentz_reduction}

Choose a smooth local reference coframe
$\bar h(q,N,N^i)$ and a corresponding spin trivialization.
Write
\begin{equation}
	h=\Lambda\bar h,\qquad
	\Psi=S(\Lambda)\psi,
\end{equation}
and introduce the relative connection
\begin{equation}
	A_\mu=\Lambda^{-1}(\partial_\mu+\omega_\mu)\Lambda.
\end{equation}
The Levi--Civita Dirac action is independent of
$\Lambda$ in these spin coordinates.

The torsion-vector difference is
\begin{equation}
	T^\mu-\bar T^\mu
	=\bar h_a{}^\nu\bar h^{b\mu}A^a{}_{b\nu},
	\qquad
	J^\mu=h(T^\mu-\bar T^\mu).
\end{equation}
Using the curvature identity, one obtains the first-order
representative
\begin{align}
	\mathcal L'={}&
	-\frac{h}{2\kappa}(\varphi\bar T-U)
	+\mathcal L_D[\bar h,\psi]
	\nonumber\\
	&+\mathcal L_{\mathrm{BF}}
	\left[\omega,\lambda-\frac{\varphi\Sigma}{\kappa}\right]
	+\frac1\kappa J^\mu\partial_\mu\varphi.
	\label{eq:sec6_relative_action}
\end{align}
After identifying the BF variables through the algebraic multiplier redefinition, the difference between the two first-order representatives is
the divergence $\kappa^{-1}\partial_\mu(\varphi J^\mu)$.

Since $J^0$ contains no velocities, the associated
boundary translation is generated locally by
\begin{equation}
	\mathcal G_{\partial}
	=\frac1\kappa\int_\Sigma d^3x\,\varphi J^0.
\end{equation}
At fixed reference section, $J^0=-\kappa B^i(A_i)$. With $\eta=\Lambda^{-1}\delta\Lambda$ and $\delta A_i=D_i^A\eta+\Lambda^{-1}\delta\omega_i\Lambda$, integration by parts gives
\begin{align}
\delta\mathcal G_\partial=\int_\Sigma\big[&-B^i(A_i)\delta\varphi-\varphi B^i{}_{,m}(A_i)\delta q^m\notag\\
&+a(\eta)-\varphi\beta(\eta)-\varphi B^i(\Lambda^{-1}\delta\omega_i\Lambda)\big].
\end{align}
This is the momentum translation for addition of $d\mathcal G_\partial/dt$; its inverse has the opposite sign. The algebraic multiplier redefinition in the preceding action is performed first, with the BF momentum defined in that representative.

On a contractible regular flat patch, the connection
can be represented locally as
\begin{equation}
	\omega=g^{-1}dg.
\end{equation}
Put $\theta=g^{-1}\delta g$ and dress the coframe and spinor by $g$ and its local spin lift. Then $\delta\omega_i=D_i^\omega\theta$. In the consistently translated canonical chart, invariance of the matter cotangent potential gives
\begin{align}
\Theta_{hD}&=\widetilde\Theta_{hD}-\int_\Sigma\langle\mathcal J_{hD},\theta\rangle,\\
\Theta_{\rm BF}&=-\int_\Sigma\langle D_i^\omega\Pi^i,\theta\rangle
+\int_{\partial\Sigma}\langle\Pi^i,\theta\rangle dS_i.
\end{align}
The total Gauss moment map is $\mathcal G=D_i^\omega\Pi^i+\mathcal J_{hD}$. On $\mathcal G=0$ the bulk pullback is $\widetilde\Theta_{hD}$. To exhibit the quotient explicitly, define $\widehat\Pi^i=g\Pi^ig^{-1}$ and $\widehat{\mathcal J}=g\mathcal J_{hD}g^{-1}$. Then
\begin{equation}
\partial_i\widehat\Pi^i=-\widehat{\mathcal J},\qquad
\widehat\Pi^i=-\partial^i\Delta^{-1}\widehat{\mathcal J}+\epsilon^{ijk}\partial_jY_k.
\end{equation}
Here $\Delta$ is an auxiliary coordinate Laplacian with an admissible inverse and compatible source (in particular zero integrated source on a closed compact slice); the transverse decomposition assumes exactness of the spatial complex with the selected boundary conditions. The flatness generator shifts $\widehat\Pi^i$ by a curl, removing $Y$ modulo $Y\mapsto Y+d\chi$.  Thus the coupled quotient retains the dressed coframe and spinor but no independent local BF canonical pair. This argument excludes harmonic modes, global holonomies and boundary charges. The auxiliary Laplacian is used only to construct
a local representative of the Gauss solution.
Changing this representative by an admissible
divergence-free contribution amounts locally to
a multiplier-shift gauge transformation.
No auxiliary metric is introduced into the
physical reduced symplectic structure
\cite{BlauThompson1991,Birmingham1991,
	DonnellyFreidel2016}.

In a local flat representative,
\begin{equation}
	A_i=\Lambda^{-1}\partial_i\Lambda.
\end{equation}
Here $\Lambda$ denotes the surviving relative Lorentz
variable, rather than an additional independent
paired-gauge degree of freedom.

Define
\begin{equation}
	H^{\nu\mu}(X)
	=\frac{h}{\kappa}
	\bar h_a{}^\nu\bar h^{b\mu}X^a{}_b,
	\qquad
	B^i(X)=H^{0i}(X),
\end{equation}
and
\begin{equation}
	a(X)=B^i(X)\partial_i\varphi.
\end{equation}
In the adapted reference frame,
\begin{equation}
	B^i(K_I)=\frac{\sqrt q}{\kappa}e_I{}^i,
	\qquad
	B^i(J_I)=0.
\end{equation}

Let $\Omega=\Lambda^{-1}\dot\Lambda$. The Lorentz and
scalar velocity terms are
\begin{equation}
	a(\Omega)-B^i(A_i)\dot\varphi.
\end{equation}
The corresponding body-momentum primary constraints are
therefore
\begin{equation}
	\boxed{
		C_X=L_X-a(X)\approx0,
		\qquad
		P_\varphi=p'_\varphi+B^i(A_i)\approx0.
	}
	\label{eq:sec6_relative_primaries}
\end{equation}
The reference metric Legendre map is regular for
$\varphi\neq0$.

The reduced bulk symplectic potential takes the form
\begin{equation}
	\Theta_*=\int_\Sigma d^3x\,
	\left[
	p^{ij}\delta q_{ij}
	+\langle L,\Lambda^{-1}\delta\Lambda\rangle
	+p'_\varphi\delta\varphi
	\right]+\Theta_F.
	\label{eq:sec6_relative_symplectic_potential}
\end{equation}

This construction is local. Nontrivial flat holonomies,
boundary symplectic structures, and global spin
trivializations require a separate treatment
\cite{Krssak2019,KrssakSaridakis2016,
	GolovnevGuzman2021,DonnellyFreidel2016}.

\subsection{Mixed primary constraint algebra and rank
	classification}
\label{subsec:mixed_primary_classification}

Choose Lorentz generators satisfying
\begin{align}
	[J_I,J_J]&=\epsilon_{IJK}J_K,\\
	[J_I,K_J]&=\epsilon_{IJK}K_K,\\
	[K_I,K_J]&=-\epsilon_{IJK}J_K.
\end{align}
The body momenta satisfy
\begin{equation}
	\{L_X(x),L_Y(y)\}_*
	=-L_{[X,Y]}(x)\delta^3(x-y).
\end{equation}
Consequently,
\begin{equation}
	\{C_X(x),C_Y(y)\}_*
	\approx-a([X,Y])\delta^3(x-y).
\end{equation}

In the ordered basis $(K_1,K_2,K_3,J_1,J_2,J_3)$,
introduce
\begin{equation}
	a_I=B^i(K_I)\partial_i\varphi,
	\qquad
	E(a)v=a\times v.
\end{equation}
The Lorentz primary matrix is
\begin{equation}
	D_6=
	\begin{pmatrix}
		0&E(a)\\
		E(a)&0
	\end{pmatrix},
	\qquad
	E(a)=
	\begin{pmatrix}
		0&-a_3&a_2\\
		a_3&0&-a_1\\
		-a_2&a_1&0
	\end{pmatrix}.
	\label{eq:sec6_D6_matrix}
\end{equation}
Since
\begin{equation}
	E(a)^2=aa^T-a^2I_3,
	\qquad a^2=a^Ta,
\end{equation}
one obtains
\begin{equation}
	\det(\lambda I_6-D_6)
	=\lambda^2(\lambda^2+a^2)^2.
\end{equation}
For $a\neq0$, the rank is four and
\begin{equation}
	\ker D_6=\operatorname{span}\{(a,0),(0,a)\}.
\end{equation}
For $a=0$, the matrix vanishes.

The mixed scalar--Lorentz bracket follows from
\begin{equation}
	\delta_XA_i=\partial_iX+[A_i,X].
\end{equation}
For compactly supported smearings,
\begin{align}
	\{P_\varphi[f],C[X]\}_*
	&=
	\int_\Sigma d^3x\,
	\left[
	B^i(X)\partial_if
	+fB^i(\partial_iX+[A_i,X])
	\right]
	\nonumber\\
	&=\int_\Sigma d^3x\,f\beta(X),
\end{align}
where
\begin{equation}
	\boxed{
		\beta(X)=-(\partial_iB^i)(X)+B^i([A_i,X]).
	}
	\label{eq:sec6_beta_definition}
\end{equation}
The derivative-of-delta contributions cancel exactly.

The complete seven-primary matrix is therefore
\begin{equation}
	\boxed{
		C_7=
		\begin{pmatrix}
			D_6&-\beta\\
			\beta^T&0
		\end{pmatrix}.
	}
	\label{eq:sec6_C7_matrix}
\end{equation}
All entries multiply $\delta^3(x-y)$ on the primary
surface.

Comparison with the original auxiliary chart gives
\begin{equation}
	\beta(X)=-cb_X.
\end{equation}
Equivalently, the coefficient of the relative Lorentz
velocity $\Omega$ in the original torsion scalar is
$-2\kappa\beta(\Omega)/h$ on the flatness surface.
Thus $\beta\neq0$ corresponds to the regular rank-seven
coframe branch under the assumptions above.

\paragraph{Proposition 2: primary rank classification.}

Write
\begin{equation}
	r_K=a^T\beta_K,\qquad
	r_J=a^T\beta_J.
\end{equation}
Then
\begin{equation}
	\boxed{
		\operatorname{rank}C_7=
		\begin{cases}
			6,&a\neq0,\ r_K^2+r_J^2>0,\\
			4,&a\neq0,\ r_K=r_J=0,\\
			2,&a=0,\ \beta\neq0,\\
			0,&a=0,\ \beta=0.
		\end{cases}
	}
	\label{eq:sec6_primary_rank_classification}
\end{equation}

\paragraph{Proof.}
For $A=a^2>0$, let $P_0$ be the orthogonal projector onto $\ker D_6$, $B=\beta^T\beta$ and $\sigma=\beta^TP_0\beta$. The resolvent is
\begin{equation}
(\lambda I-D_6)^{-1}=\frac{P_0}{\lambda}+\frac{\lambda(I-P_0)+D_6}{\lambda^2+A}.
\end{equation}
The scalar Schur complement, using $\beta^TD_6\beta=0$, gives the polynomial identity
\begin{equation}
\boxed{\det(\lambda I_7-C_7)=\lambda(\lambda^2+A)[\lambda^4+(A+B)\lambda^2+A\sigma].}
\end{equation}
Thus the regular singular values (each repeated twice) are $\sqrt A$ and
\begin{equation}
s_\pm=\sqrt{\frac{A+B\pm\sqrt{(A-B)^2+4A(B-\sigma)}}{2}}.
\end{equation}
The singular values and the pseudodeterminant
refer to the chosen Euclidean normalization
of the primary constraint basis.
The rank classification is invariant under
admissible invertible changes of constraint basis. Since $A\sigma=r_K^2+r_J^2$, the rank is six for $\sigma>0$ and four otherwise. At $a=0$ the scalar augmentation alone has rank two or zero. On the rank-six stratum the nonzero singular values give $\operatorname{pdet}C_7=A^2\sigma$. On lower-rank strata the product of nonzero eigenvalues must be evaluated separately. \hfill$\square$

For $\beta\neq0$, the scalar pair may first be eliminated
using
\begin{equation}
	C_\parallel=\frac{\beta^TC}{\beta^T\beta}.
\end{equation}
If the columns of $Q$ span $\beta^\perp$, the five
remaining primary constraints have weak matrix
\begin{equation}
	D_5=Q^TD_6Q.
\end{equation}
Its possible ranks are four, two, and zero.

On the generic rank-six primary stratum, a null vector
of $C_7$ is
\begin{equation}
	\boxed{
		n=(r_Ja,-r_Ka,0)^T.
	}
	\label{eq:sec6_primary_null_vector}
\end{equation}

The kinetic rank of the original coframe Hessian and
the rank of the auxiliary primary matrix are distinct
classifications. In particular,
\begin{equation}
	a_I=\frac{\sqrt q}{\kappa}e_I{}^i
	F_{TT}\partial_iT
\end{equation}
on the scalar equation, so homogeneous torsion can
produce $a=0$ without determining the full nonlinear
physical spectrum.

\subsection{Secondary stabilization and exceptional
	degeneracy surfaces}
\label{subsec:secondary_stabilization}

The Lorentz Euler equation supplies an explicit source
for primary stabilization. Define
\begin{equation}
	k^\nu(X)=H^{\nu\mu}(X)\partial_\mu\varphi.
\end{equation}
Variation of the relative Lorentz field gives
\begin{equation}
	-\partial_\nu k^\nu(X)
	+k^\nu([A_\nu,X])=0.
	\label{eq:sec6_lorentz_euler_equation}
\end{equation}
Separating the Lorentz and scalar velocities yields
\begin{equation}
	0=f_X+D_{6,XY}\Omega^Y-\beta(X)\dot\varphi,
\end{equation}
where
\begin{align}
	f_X={}&-\dot B^i(X)\partial_i\varphi
	-\partial_i\left[H^{ij}(X)\partial_j\varphi\right]
	\nonumber\\
	&+H^{ij}([A_i,X])\partial_j\varphi
	+a([\Lambda^{-1}\omega_0\Lambda,X]).
	\label{eq:sec6_primary_sources}
\end{align}
Here $\dot B^i$ is evaluated through the regular
reference metric Legendre inversion, including the
fermionic momentum shift.

The scalar Euler equation is
\begin{equation}
	-\frac{h}{2\kappa}(\bar T-U')
	-\frac1\kappa\partial_\mu J^\mu=0.
\end{equation}
Its Lorentz-velocity coefficient is $+\beta(\Omega)$.
Together, these equations reproduce the mixed primary
matrix modulo the temporal and flatness equations.

On the rank-six primary stratum, contraction with
the null vector gives the candidate secondary
\begin{equation}
	S=n^Xf_X\approx0.
	\label{eq:sec6_null_secondary}
\end{equation}

After the scalar-pair reduction, denote the remaining
five primaries by $\Phi_\alpha$ and their Hamiltonian
sources by
\begin{equation}
	g_\alpha=\{\Phi_\alpha,H_c\}_D.
\end{equation}
Their consistency equation is
\begin{equation}
	D_5u+g\approx0.
\end{equation}
On the compatible rank-four stratum, its general
solution is
\begin{equation}
	u=-D_5^+g+\mu\widehat n,
\end{equation}
where $Q\widehat n=n$ in the Lorentz components.
Preservation of the secondary gives
\begin{equation}
	0\approx Y+Z\mu,
	\label{eq:sec6_final_stabilization}
\end{equation}
with
\begin{align}
	Y&=\{S,H_c\}_D
	-\{S,\Phi\}_DD_5^+g,\\
	Z\mu&=\{S,\Phi\}_D[\mu\widehat n].
\end{align}

The operator $Z$ must be considered together with the
temporal consistency equations. A suitable inverse
fixes the final multiplier; otherwise, its left null
smearings yield further compatibility conditions.

\paragraph{Lapse-independent secondary representative.}

In the local BF-reduced chart, define
\begin{equation}
	g_i=\partial_i\varphi,
	\qquad
	R^{ij}(X)=\frac{\sqrt q}{\kappa}
	e_I{}^ie_J{}^jX^{IJ}.
\end{equation}
Let $P^{ij}$ denote the metric momentum after subtraction
of the reference fermionic shift. The regular metric
inversion is
\begin{equation}
	K_{ij}=
	\frac{2\kappa}{\varphi\sqrt q}
	\left(P_{ij}-\frac12q_{ij}P\right).
\end{equation}

The normal Lorentz source for unit lapse and zero
shift is
\begin{align}
	f_1(X)={}&
	-2B^i{}_{,q_{kl}}(X)K_{kl}g_i
	-\partial_i\left[R^{ij}(X)g_j\right]
	\nonumber\\
	&+R^{ij}([A_i,X])g_j.
	\label{eq:sec6_unit_normal_source}
\end{align}
The derivatives hold the Lorentz generator fixed.

The contraction of the lapse-gradient contribution
with the primary null direction vanishes because
the rotational component of $n$ is parallel to $a$,
whereas the lapse-gradient current is perpendicular
to $a$. Consequently, the secondary admits the
lapse-independent representative
\begin{equation}
	\boxed{
		S=n^Xf_1(X)\approx0.
	}
	\label{eq:sec6_projectable_secondary}
\end{equation}

This representative is appropriate for the temporal
projectability and stabilized deformation analysis.

\paragraph{Exceptional strata.}

The generic stabilization must not be extended by
continuity to the loci
\begin{equation}
	a=0,\qquad
	r_K=r_J=0,\qquad
	\beta=0,\qquad
	\varphi=0,\qquad F_{TT}=0.
\end{equation}
The locus $\mathscr D=0$ is exceptional for the scalar inversion on the $\beta=0$ branch, not by itself for the rank-seven Hessian with nonzero kernel projection. When the rank of $D_5$ is two or zero, the primary
null space has dimension three or five, respectively.
Every independent null direction supplies a
compatibility equation, and tangency to the selected
rank stratum must also be imposed.

Similarly, a zero of the residual stabilization
operator does not by itself establish an additional
first-class constraint. It may instead produce a
tertiary condition, a differential multiplier
equation, or a failure of the chosen local chart.

These exceptional cases require independent
Dirac--Bergmann analyses
\cite{Dirac1964,HenneauxTeitelboim1992,
	FerraroGuzman2018,BlixtReview2021}.

\subsection{Complete second-class reduction on the regular
	branch}
\label{subsec:complete_regular_reduction}

Consider the primary stratum
\begin{equation}
	a^2>0,\qquad r_K^2+r_J^2>0.
\end{equation}
Let $P_0$ be the orthogonal projector onto $\ker D_6$,
\begin{equation}
	P_0=
	\begin{pmatrix}
		P_a&0\\
		0&P_a
	\end{pmatrix},
	\qquad
	P_a=\frac{aa^T}{a^2}.
\end{equation}
Define
\begin{equation}
	\beta_0=P_0\beta,\qquad
	\sigma=\beta_0^T\beta_0
	=\frac{r_K^2+r_J^2}{a^2}>0,
\end{equation}
and
\begin{equation}
	L_0=I_6-\frac{\beta\beta_0^T}{\sigma}.
\end{equation}

\paragraph{Proposition 3: intrinsic primary inverse.}

The Moore--Penrose inverse of the regular primary
matrix is
\begin{equation}
	\boxed{
		Q_7=
		\begin{pmatrix}
			L_0^TD_6^+L_0&\beta_0/\sigma\\
			-\beta_0^T/\sigma&0
		\end{pmatrix},
		\qquad
		D_6^+=-\frac{D_6}{a^2}.
	}
	\label{eq:sec6_intrinsic_primary_inverse}
\end{equation}

\begin{proof}
The identities $D_6^2=-a^2(I-P_0)$, $D_6\beta_0=0$ and $\beta_0^T\beta=\sigma$ imply
\begin{align*}
D_6D_6^+&=I-P_0, & L_0\beta&=0, & \beta_0^TL_0&=0,\\
D_6L_0^T&=D_6, & L_0D_6&=D_6, & P_0-\beta_0\beta_0^T/\sigma&=\nu_L\nu_L^T.
\end{align*}
Direct multiplication yields $C_7Q_7=Q_7C_7=I_7-\nu\nu^T$ and $Q_7\nu=C_7\nu=0$. These identities give all four Moore--Penrose conditions.
\end{proof}

The auxiliary Euclidean inner product only selects
a complement of the null line. For bracket numerators
annihilating that line, the resulting contraction
with $Q_7$ is independent of the selected complement.

Choose six independent paired primary constraints
$\chi_P$ spanning $\nu^\perp$, and denote their
invertible bracket matrix by $D_P$. Their elimination
defines
\begin{equation}
	\{A,B\}_4
	=
	\{A,B\}_*
	-\{A,\chi_P\}_*D_P^{-1}\{\chi_P,B\}_*.
	\label{eq:sec6_six_primary_reduction}
\end{equation}
The subscript $4$ retains the notation of the
rank-four Lorentz reduction together with the
previous scalar-pair elimination.

Use the unit residual pair $\chi_R=(C_\nu,S_u)$, with $w=\|n\|^{-1}=(a^2\sqrt\sigma)^{-1}$, $C_\nu=\nu^T\Phi$, $S_u=wS$ and $\Phi=(C_I,P_\varphi)$. Since
\[
\|n\|^2
=a^2(r_K^2+r_J^2)
=(a^2)^2\sigma,
\]
the normalization is smooth on the regular
rank-six primary stratum. On the stabilized surface,
\begin{equation}
\mathcal B=\begin{pmatrix}0&-Z_\nu^\dagger\\Z_\nu&J_u\end{pmatrix},\qquad Z_\nu=\{S_u,C_\nu\}_4,\quad J_u=\{S_u,S_u\}_4.
\label{eq:sec6_residual_matrix}
\end{equation}
Appendices B--C compute the polynomial-secondary kernels first. Their normalized versions satisfy
\begin{equation}
 Z_\nu(x,y)\approx z_\nu(x)\delta^3(x-y),\qquad z_\nu=w^2z,\qquad J_u\approx wJ_*w.
\label{eq:sec6_residual_ultralocal_bracket}
\end{equation}
For $J_*=2j_*^i\partial_i+\partial_i j_*^i$, this gives $J_u=2w^2j_*^i\partial_i+\partial_i(w^2j_*^i)$ weakly. Terms from brackets of $w$ multiply constraints. The adjoint is taken on the chosen smearing domain.

If $Z_\nu$ admits a suitable inverse, then
\begin{equation}
	\boxed{
		\mathcal B^{-1}=
		\begin{pmatrix}
			Z_\nu^{-1}J_u(Z_\nu^\dagger)^{-1}
			&Z_\nu^{-1}\\
			-(Z_\nu^\dagger)^{-1}&0
		\end{pmatrix}.
	}
	\label{eq:sec6_residual_inverse}
\end{equation}
The inverse is understood on compatible smearing spaces. On a bounded patch, a sufficient condition for the multiplication inverse is that $z_\nu$ and $z_\nu^{-1}$ have the regularity required by the differential kernels, with $|z_\nu|$ bounded away from zero. For example, smooth bounded coefficients with bounded first derivatives allow the displayed inverse to map $H^1\oplus L^2$ to $L^2\oplus H^1$ (with compatible boundary domains). Pointwise nonvanishing alone does not imply bounded invertibility on arbitrary global domains. The stated Sobolev mapping concerns the residual
constraint operator and its inverse. The full stabilized Poisson operator additionally requires differentiability of the observables and compatibility of all functional derivatives with the selected smearing domains \cite{Dirac1964,HenneauxTeitelboim1992}.

The complete stabilized Dirac bracket is
\begin{align}
	\{A,B\}_{\mathrm{st}}
	={}&\{A,B\}_4
	\nonumber\\
	&-\{A,\chi_R\}_4
	\mathcal B^{-1}
	\{\chi_R,B\}_4,
	\qquad
	\chi_R=(C_\nu,S_u).
	\label{eq:sec6_stabilized_dirac_bracket}
\end{align}

Equivalently, collecting all eight constraints into
$\chi_A$, one has
\begin{equation}
	\{A,B\}_{\mathrm{st}}
	=
	\{A,B\}_*
	-\{A,\chi_A\}_*
	(C_8^{-1})^{AB}
	\{\chi_B,B\}_*.
\end{equation}

The equivalence of the sequential and direct
reductions follows from the block Schur inverse.
On a regular functional domain, the induced bracket
satisfies the Jacobi identity and annihilates the
eight eliminated constraints.

This reduction concerns the gravitational
second-class sector. The temporal first-class
constraints and their gauge conditions must still
be treated separately
\cite{Dirac1964,HenneauxTeitelboim1992,HT1990}.

\subsection{Hypersurface generators and reference covariance}
Write $\Gamma_{\rm SC}=\{\chi=0\}$ and $\Gamma_{\rm phys}=\{\chi=0,\mathcal H_\perp=\mathcal H_i=0\}$. In the deformation calculation below, weak equality denotes restriction to $\Gamma_{\rm SC}$, without imposing the four hypersurface constraints. The latter are imposed only in the physical reduction. This distinction is required for the BFV construction.
\label{subsec:stabilized_deformation_algebra}

The local BF-reduced normal Hamiltonian density is
\begin{align}
	\mathcal H_\perp={}&
	\frac{2\kappa}{\varphi\sqrt q}
	\left(P^{ij}P_{ij}-\frac12P^2\right)
	+\frac{\sqrt q}{2\kappa}
	\left(\varphi\,{}^{(3)}T-U\right)
	\nonumber\\
	&-\frac1\kappa
	\partial_i(\sqrt q\,\varphi t^i)
	-R^{ij}(A_i)g_j+\mathcal H_F,
	\label{eq:sec6_normal_density}
\end{align}
where $t^i$ is the reference spatial torsion vector,
$g_i=\partial_i\varphi$, and $\mathcal H_F$ is the
reference Dirac Hamiltonian density.

Using the spatial curvature identity, this becomes
\begin{align}
	\mathcal H_\perp={}&
	\mathcal K
	-\frac{\sqrt q}{2\kappa}
	\left(\varphi\,{}^{(3)}R+U\right)
	\nonumber\\
	&-\left(c_qt^i+R^{ji}(A_j)\right)g_i
	+\mathcal H_F,
	\label{eq:sec6_normal_curvature_density}
\end{align}
where
\begin{equation}
	c_q=\frac{\sqrt q}{\kappa},
	\qquad
	\mathcal K=
	\frac{2\kappa}{\varphi\sqrt q}
	\left(P^{ij}P_{ij}-\frac12P^2\right).
\end{equation}
This is a spatial identity, not an additional temporal
boundary transformation.

For a spatial vector field $V$, define the reference
triad compensator
\begin{equation}
	r_V=
	\left(
	\mathcal L_Ve-\delta_qe[\mathcal L_Vq]
	\right)e^{-1}.
\end{equation}
It is a spatial Lorentz rotation. The compensated
spatial generator is
\begin{align}
	D[V]={}&
	\int_\Sigma d^3x\,
	\big[
	p^{ij}\mathcal L_Vq_{ij}
	+\langle L,V^iA_i+r_V\rangle
	\nonumber\\
	&\hspace{30mm}
	+p'_\varphi V^ig_i
	\big]
	+\Theta_F(\delta_V).
	\label{eq:sec6_compensated_spatial_generator}
\end{align}

Under the regular Legendre-projectability and boundary hypotheses of Appendix~\ref{app:spatial_ideal}, the compensated cotangent lift preserves the second-class ideal, so its Dirac subtraction vanishes on $\Gamma_{\rm SC}$.

Let
\begin{equation}
	H[N]=\int_\Sigma d^3x\,N\mathcal H_\perp.
\end{equation}
On the primary surface, its primary bracket numerator
has the form
\begin{equation}
	b[N]=\{H[N],\Phi\}_*
	=Nu+\partial_iN\,v^i.
\end{equation}
The Lorentz components are
\begin{equation}
	u_X=-f_1(X),\qquad
	v_X^i=R^{ij}(X)g_j.
\end{equation}
In the boost--rotation basis,
\begin{equation}
	v_K^i=0,\qquad
	v_J^i=a\times e^i,
\end{equation}
and the scalar component satisfies the triad identity
\begin{equation}
	v_\varphi^i=e_I{}^i\beta_{K_I}.
\end{equation}

The null secondary ensures that the normal bracket
numerators annihilate the primary null line.
Furthermore,
\begin{equation}
	(v^i)^TQ_7v^j=0.
\end{equation}
Writing
\begin{equation}
	s_i=N\partial_iM-M\partial_iN,
\end{equation}
the primary Dirac correction reduces to
\begin{equation}
	\Delta_P[N,M]
	\approx\int_\Sigma d^3x\,s_iJ_P^i,
\end{equation}
where
\begin{equation}
	\boxed{
		J_P^i=u^TQ_7v^i
		\approx-e_I{}^if_1(K_I).
	}
	\label{eq:sec6_primary_deformation_current}
\end{equation}
The final null-primary--secondary correction vanishes
weakly on the stabilized surface for these normal
generators.

\paragraph{Metric and relative Lorentz contributions.}

Choose a reference section with symmetric metric
variation at the configuration under consideration.
The curvature, spatial torsion, and primary metric
currents are, respectively,
\begin{align}
	J_{\mathrm{curv}}^i
	&=-2D_jP^{ij}+\frac{4}{\varphi}P^{ij}g_j,\\
	J_{\mathrm{tor}}^i
	&=-\frac{2}{\varphi}P^{ij}g_j,\\
	J_{\mathrm{prim,met}}^i
	&=-\frac{2}{\varphi}P^{ij}g_j.
\end{align}
Their sum is therefore
\begin{equation}
	J_{\mathrm{met}}^i=-2D_jP^{ij}.
\end{equation}
All explicit scalar-gradient corrections to this
metric current cancel.

The remaining relative Lorentz--scalar contribution
is
\begin{equation}
	\boxed{
		J_\Lambda^i
		=
		q^{ik}\left[
		B^j(A_k)g_j-B^j(A_j)g_k
		\right].
	}
	\label{eq:sec6_relative_deformation_current}
\end{equation}
It agrees with the relative Lorentz--scalar part of
the spatial generator on the primary surface.

\paragraph{Fermionic contribution.}

At any fixed smooth metric configuration, one may
choose a local reference section with symmetric
metric variation. In this section, the reference
fermionic momentum shift vanishes at that
configuration, although its metric derivative need
not vanish.

Appendix~\ref{app:fermionic_deformation} gives the cancellation of shift-curvature and mixed metric terms. The remaining matter bracket is

\begin{align}
	\{H_F[N],H_F[M]\}_{D,F}
	={}&\frac{i}{2}\int_\Sigma d^3x\,
	\left[
	\xi^\dagger V^i\partial_i\xi
	-(V^i\partial_i\xi^\dagger)\xi
	\right]
	\nonumber\\
	&-i\int_\Sigma d^3x\,
	\xi^\dagger\Sigma(r_V)\xi,
	\label{eq:sec6_fermionic_deformation_current}
\end{align}
where
\begin{equation}
	V^i=q^{ij}s_j,\qquad
	\Sigma(X)=-\frac14X_{IJ}\Gamma^{IJ}.
\end{equation}
The mass contribution cancels by the Clifford
anticommutator. The last term is precisely the
reference spin-frame compensation.

\paragraph{Proposition 4: stabilized deformation algebra.}

Combining the metric, relative Lorentz, and fermionic
currents gives
\begin{align}
	\{D[V],D[W]\}_{\mathrm{st}}
	&\approx D[[V,W]],\\
	\{D[V],H[N]\}_{\mathrm{st}}
	&\approx H[\mathcal L_VN],\\
	\boxed{
		\{H[N],H[M]\}_{\mathrm{st}}
		\approx
		D[q^{ij}(N\partial_jM-M\partial_jN)].
	}
	\label{eq:sec6_stabilized_HDA}
\end{align}
These identities are understood on the regular
second-class manifold $\Gamma_{\rm SC}$, before
imposing the hypersurface constraints themselves.
They therefore supply the reduced first-class
algebra required by the BFV construction of Sec.~7
\cite{Teitelboim1973,HenneauxTeitelboim1992,HT1990}. The weak equalities refer to $\Gamma_{\rm SC}$ in the local BF-reduced chart, before imposing $H=D=0$. They are identities of the induced bracket on that reduced surface under the stated hypotheses.

This is the classical geometrical algebra, including the spin-frame compensation \cite{ADM1962,Teitelboim1973,Oko2013,Oko2014,HenneauxTeitelboim1992}.

\paragraph{Stabilized Dirac brackets and reference-section
	independence}
\label{subsec:reference_section_independence}

For two reference sections related by a smooth local rotation with no explicit time dependence, write

\begin{equation}
	\widetilde e=O(q,x)e,
	\qquad O\in SO(3),
\end{equation}
with a smooth local spin lift $U=S(O)$ and no explicit
time dependence.

The same physical coframe is represented by
\begin{equation}
	\widetilde\Lambda=\Lambda O^{-1},
	\qquad
	\widetilde\xi=U\xi.
\end{equation}
Introduce
\begin{equation}
	\alpha_{ij}=O^{-1}\frac{\partial O}{\partial q_{ij}}.
\end{equation}
The finite canonical lift is
\begin{align}
	\widetilde L_{\operatorname{Ad}_OX}&=L_X,\\
	\widetilde p^{ij}
	&=p^{ij}+L_{\alpha_{ij}}
	-i\xi^\dagger\Sigma(\alpha_{ij})\xi.
	\label{eq:sec6_reference_canonical_lift}
\end{align}

\paragraph{Proposition 5: reference-section covariance.}

The transformation
\eqref{eq:sec6_reference_canonical_lift} preserves the
complete partially reduced symplectic potential.
Furthermore, the shifted metric momenta satisfy
\begin{equation}
	\widetilde P^{ij}=P^{ij}+C_{\alpha_{ij}}.
	\label{eq:sec6_reference_shifted_momentum}
\end{equation}
The normal potential is invariant, and the difference
between the kinetic representatives belongs to the
primary constraint ideal.

Consequently, the primary and stabilized constraint
surfaces are mapped into one another, and the
compensated spatial generator is invariant.

The canonical pullback and covariance of the primary ideal are proved in Appendix~\ref{app:reference_section_covariance}.

For a canonical reference transformation $\mathcal T$ and invertible constraint change $\widetilde\chi=M\chi$, the weak congruence $\widetilde C_8\approx MC_8M^\dagger$ gives
\begin{equation}
\{\mathcal T_*A,\mathcal T_*B\}_{\widetilde{\mathrm{st}}}\approx\mathcal T_*\{A,B\}_{\mathrm{st}}.
\label{eq:sec6_reference_covariant_dirac_bracket}
\end{equation}
The deformation algebra may therefore be evaluated in a symmetric reference section and transported to any related regular section.

For torsion observables the general eight-constraint subtraction reads
\begin{equation}
\{T_A,T_B\}_{\rm st}=\{T_A,T_B\}_*-\{T_A,\chi_C\}_*(C_8^{-1})^{CD}\{\chi_D,T_B\}_*.
\label{eq:sec6_stabilized_torsion_bracket}
\end{equation}
Kernel integrations are implicit; Sec.~6 evaluates the spatial torsion numerator.
The induced bracket is Poisson on the regular second-class surface; physical observables require the remaining first-class reduction
\cite{Dirac1964,HenneauxTeitelboim1992,
	Rovelli2004,Dittrich2007,DonnellyFreidel2016}.

\subsection{Regularity and physical degrees of freedom}
\label{subsec:regular_physical_count}

The preceding analysis identifies a regular local
nonlinear branch characterized by
\begin{equation}
	F_{TT}\neq0,\qquad
	\varphi\neq0,\qquad
	a^2>0,\qquad
	\sigma>0,\qquad
	z_\nu\neq0.
	\label{eq:sec6_regular_branch_conditions}
\end{equation}
The nonvanishing of $\sigma$ implies $\beta\neq0$,
and hence corresponds to the regular rank-seven
coframe Legendre branch.

Assume the local BF quotient, functional independence of the eight second-class constraints, invertibility on the chosen domain, reduced Euler--Hamilton equivalence and four independent first-class temporal chains. The spatial variables $(q_{ij},\Lambda,\varphi)$ have $6+6+1=13$ configuration components; lapse and shift add four canonical pairs. The seven primaries and their single secondary constitute the eight second-class constraints.

\paragraph{Theorem: conditional regular-branch physical
	count.}

Under all the preceding quotient, regularity, functional independence, Euler--Hamilton equivalence and first-class closure assumptions, the local nonlinear
gravitational configuration degree-of-freedom count is
\begin{equation}
	\boxed{
		N_g=
		\frac12\left(34-8-2\times8\right)=5.
	}
	\label{eq:sec6_five_physical_dof}
\end{equation}

Equivalently, treating lapse and shift as external
multipliers gives
\begin{equation}
	N_g=\frac12(26-8-2\times4)=5.
\end{equation}
The original rank-seven formulation equivalently gives $(32-6-2\times8)/2=5$. This count excludes odd fermionic dimensions and the locally eliminated BF sector. Appendix~\ref{app:regular_background_evolution} tests regularity and independence on explicit data, not on all strata \cite{FerraroGuzman2018,IzumiOng2013,GolovnevGuzman2021,BlixtReview2021}.

\paragraph{Regular constraint measure}
\label{subsec:regular_constraint_measure}
The primary pseudodeterminant established above gives
\begin{equation}
\boxed{\operatorname{pdet}C_7=\det D_P=(a^2)^2\sigma=\|n\|^2.}
\label{eq:sec6_primary_pseudodeterminant}
\end{equation}
Here $D_P$ is the restriction to an orthonormal basis of $\nu^\perp$. The determinant refers to this Euclidean normalization, not to an invariant density by itself.

For the residual matrix \eqref{eq:sec6_residual_matrix}, a block-column exchange gives $\det\mathcal B=(\det Z_\nu)^2$ at finite regulator, independently of $J_u$.

The complete stabilized second-class measure therefore
factorizes as
\begin{equation}
	\boxed{
		\sqrt{|\operatorname{Det}C_8|}
		=
		\sqrt{|\operatorname{Det}D_P|}
		|\operatorname{Det}Z_\nu|.
	}
	\label{eq:sec6_complete_measure_factorization}
\end{equation}
For the weak ultralocal residual bracket
$Z_\nu(x,y)\approx z_\nu(x)\delta^3(x-y)$, this becomes
formally
\begin{equation}
	\boxed{
		\sqrt{|\operatorname{Det}C_8|}
		\sim\prod_x a^2(x)\sqrt{\sigma(x)}\,|z_\nu(x)|.
	}
	\label{eq:sec6_regular_local_measure}
\end{equation}
The unit-pair factor equals $|z|/\|n\|$; the secondary delta distribution compensates this rescaling. These are regulated products, not continuum determinants defined independently of a regulator.

The insertion is
\begin{equation}
\mathcal D\mu_{\mathrm{SC}}=\mathcal Dz\,\delta[\chi]\sqrt{|\operatorname{Det}C_8|}.
\label{eq:sec6_second_class_insertion}
\end{equation}
Its invariance under invertible constraint-basis changes follows from the opposite delta-function and determinant Jacobians (Appendix~\ref{app:constraint_measure_variations}). The constraint-basis invariance concerns the
complete insertion
$\delta[\chi]\sqrt{|\operatorname{Det}C_8|}$.
Neither the determinant factor nor the delta
distribution is separately invariant under
a general invertible constraint redefinition.
The cancellation is exact at a compatible
finite regulator
\cite{HenneauxTeitelboim1992,HT1990}.

The determinant variation, derived in Appendix~\ref{app:constraint_measure_variations}, is
\begin{equation}
\delta\ln\sqrt{|\operatorname{Det}C_8|}=\operatorname{Tr}_\Sigma\left[\frac{\delta z}{z}-\frac{\delta(a^2)}{a^2}-\frac{\delta\sigma}{2\sigma}\right].\label{eq:sec6_measure_variation}
\end{equation}
For the scalar pairs of
Sec.~\ref{subsec:auxiliary_scalar_reduction}, the
corresponding determinant factors cancel the
constraint Jacobians when the auxiliary scalar is
eliminated in a compatible finite regularization.

The zeros of $a^2$, $\sigma$, or $z_\nu$ indicate a
breakdown of the present regular chart. They do not
establish that exceptional configurations carry zero
physical weight. Such configurations require their
own constraint reductions and functional measures
\cite{Dirac1964,HenneauxTeitelboim1992,HT1990,
	BlixtReview2021}.

\paragraph{Regular initial data.}\label{subsec:regular_inhomogeneous_background}
Appendix~\ref{app:regular_background_evolution} constructs nonstationary constrained data with $a^2\sigma z_0\ne0$ and tests functional independence. Persistence requires a sufficiently smooth evolution; no existence theorem is inferred.

\section{Formal canonical quantization and quantum torsion}
\label{sec:canonical_quantization}
We quantize only after specifying the regular symplectic reduction of Sec.~\ref{sec:dirac_bergmann}. The following local prescriptions distinguish the partial product chart from the stabilized algebra \cite{Dirac1964,HenneauxTeitelboim1992,Thiemann2007}.

\subsection{Reduced canonical brackets and representations}
\label{subsec:quantum_kinematics}
Use the partial product Darboux chart of Sec.~\ref{subsec:rev-mixed}, with BF and fermionic brackets given in Sec.~\ref{sec:rev-canonical}. The shifted momentum $P_A=\pi_A-\mathcal A_A$ is not Darboux. For a specified quantization map
$\mathcal Q_\hbar$ on the reduced phase space,
the formal semiclassical correspondence reads
\[
[\mathcal Q_\hbar(A),\mathcal Q_\hbar(B)]_{\rm gr}
=
i\hbar\,\mathcal Q_\hbar
\bigl(\{A,B\}_{\rm red}\bigr)
+O(\hbar^2).
\]
The existence of such a map with suitable domains
and ordering properties is an additional
quantization requirement.

For any regular Darboux coordinates $(Q^I,P_I)$ of a specified reduced symplectic manifold, the formal correspondence is $[\widehat A,\widehat B]_{\rm gr}=i\hbar\widehat{\{A,B\}_{\rm red}}+O(\hbar^2)$, with the graded commutator for odd variables. In a local canonical chart one may seek
\begin{equation}
	[\widehat Q^I(x),\widehat P_J(y)]=i\hbar\delta^I_J\delta^3(x-y),\qquad
	[\widehat Q^I,\widehat Q^J]=[\widehat P_I,\widehat P_J]=0.
	\label{eq:q_canonical_CCR}
\end{equation}
A candidate functional representation is
\begin{equation}
	(\widehat Q^I\Phi)[Q]=Q^I\Phi[Q],\qquad
	(\widehat P_I\Phi)[Q]=-i\hbar\frac{\delta\Phi}{\delta Q^I}+\mathscr A_I[Q]\Phi[Q].
	\label{eq:q_functional_rep}
\end{equation}
For a scalar wavefunctional and a finite-dimensional positive density $\mu$, formal symmetry on compactly supported states requires $\mathscr A_I=a_I-i\hbar\partial_I\ln\mu/2$ with real $a_I$. Indeed integration by parts gives $(-i\hbar\partial_I)^\dagger=-i\hbar(\partial_I+\partial_I\ln\mu)$. Moreover $[\widehat P_I,\widehat P_J]=-i\hbar(\partial_I a_J-\partial_J a_I)$, so commuting momenta require a closed $a$. In the partial coframe chart, $\widehat P_A=\widehat\pi_A-\widehat{\mathcal A}_A$; this is not a representation of the fully stabilized algebra. The positive configuration density is not generally the momentum-dependent second-class determinant. The partially reduced spinor algebra gives
\begin{equation}
	\{\widehat\xi_A(x),\widehat\xi_B^\dagger(y)\}_+=\hbar\delta_{AB}\delta^3(x-y),
	\label{eq:q_CAR}
\end{equation}
with vanishing same-type anticommutators. At a finite regulator with configuration-independent CAR, put $B(m)=\widehat\xi^\dagger m\widehat\xi$. The CAR imply $[B(m),B(n)]=\hbar B([m,n])$ and $[\widehat\xi,B(m)]=\hbar m\widehat\xi$. Thus
\begin{align}
[\widehat h^A,\widehat P_B]&=i\hbar\delta^A_B,&
[\widehat\xi,\widehat P_A]&=-\hbar m_A\widehat\xi,\\
[\widehat\xi^\dagger,\widehat P_A]&=\hbar\widehat\xi^\dagger m_A,&
[\widehat P_A,\widehat P_B]&=i\hbar B(\mathcal F_{AB})-i\hbar(\partial_Aa_B-\partial_Ba_A).
\end{align}
Here $\mathcal F_{AB}=\partial_Am_B-\partial_Bm_A-i[m_A,m_B]$. The last identity follows by expanding $P=\pi-B(m)$; the density connection is exact. The finite-regulator representation above realizes
the product-reduced algebra.
A representation of the stabilized algebra must
instead implement the Dirac bracket associated
with the eight second-class constraints, or use
local Darboux coordinates on the corresponding
reduced symplectic manifold
\cite{Dirac1964,HenneauxTeitelboim1992,
	Thiemann2007}.

Moreover $[\widehat P_A,B(X)]=-i\hbar B(\partial_AX-i[m_A,X])$, so the momentum Jacobi identity reduces to the matrix Bianchi identity (and $d^2a=0$). These identities are exact in this regulated representation, not assertions about continuum contact terms. We retain $\hbar$ explicitly in this section. A Fock representation requires a reference state or a choice of complex structure, generally background dependent \cite{BirrellDavies1982,ParkerToms2009}. None of these formulas specifies a physical inner product.

\paragraph{Branchwise and stabilized Dirac brackets}
\label{subsec:q-branchwise}
For an independent even second-class set $\chi$, the bracket is
\begin{equation}\label{eq:q-dirac-general}
\{A,B\}_{\chi}=\{A,B\}_*-\{A,\chi\}_*\mathbb C^{-1}\{\chi,B\}_*,\qquad \mathbb C=\{\chi,\chi\}_*.
\end{equation}
All products are kernel compositions on the specified domain. Both regular scalar pairs of Sec.~\ref{subsec:auxiliary_scalar_reduction} have the form $(p_\phi,X)$, with upper-right block $d=-c$ or $d=\mathscr D/F_{TT}$ respectively. Their inverse is
\begin{equation}
\begin{pmatrix}0&d\\-d^\dagger&J\end{pmatrix}^{-1}=
\begin{pmatrix}(d^\dagger)^{-1}Jd^{-1}&-(d^\dagger)^{-1}\\d^{-1}&0\end{pmatrix},\qquad J=\{X,X\}_*.
\end{equation}
Direct multiplication gives the identity even for differential $J$. In particular the correction is
\begin{align}\label{eq:q-pair-correction}
\Delta\{A,B\}={}&-\{A,p_\phi\}(d^\dagger)^{-1}Jd^{-1}\{p_\phi,B\}\notag\\
&+\{A,p_\phi\}(d^\dagger)^{-1}\{X,B\}-\{A,X\}d^{-1}\{p_\phi,B\}.
\end{align}
Scalar-independent original variables retain their brackets under this pair reduction, but the reconstructed scalar is not an independent oscillator. On the regular rank-six mixed-primary branch, use the six-primary reduction and residual inverse of Eqs.~\eqref{eq:sec6_six_primary_reduction}--\eqref{eq:sec6_stabilized_dirac_bracket}. Use the unit pair $(C_\nu,S_u)$ and its kernels $Z_\nu,J_u$; polynomial-secondary numerators are rescaled by $w=\|n\|^{-1}$. For the explicit differential upper-left inverse entry, see Appendix~\ref{app:torsion-brackets}.
Equivalently the stabilized Poisson operator is $\mathbb J_{\rm st}=\mathbb P\mathbb J_{\rm prod}\mathbb P^\dagger$, with the tangent projector defined in Appendix~\ref{app:poisson-projector}. The stabilized bracket, not the original CCR, governs the semiclassical correspondence. Exceptional strata require separate stabilization; a singular inverse alone implies neither a quantum transition nor a superselection rule \cite{Dirac1964,HenneauxTeitelboim1992,HT1990}.

\subsection{Stabilized torsion operators}
\label{subsec:quantum_torsion}
The spatial torsion and its candidate configuration operator are
\begin{align}
	T^a{}_{ij}&=2\partial_{[i}h^a{}_{j]}+2\omega^a{}_{b[i}h^b{}_{j]},\label{eq:q_torsion_classical}\\
	\widehat T^a{}_{ij}&=2\partial_{[i}\widehat h^a{}_{j]}+2(\widehat\omega^a{}_{b[i}\widehat h^b{}_{j]})_{\rm reg}.\label{eq:q_torsion_operator}
\end{align}
For compactly supported antisymmetric test densities define
\begin{equation}
	\widehat T[f]=\tfrac12\int_\Sigma d^3x\, f_a{}^{ij}\widehat T^a{}_{ij}.
	\label{eq:q_smeared_torsion}
\end{equation}
In the partial product chart, functional differentiation gives the formal commutators (Appendix~\ref{app:torsion-brackets})
\begin{align}
	[\widehat T^a{}_{ij}(x),\widehat\pi_b{}^k(y)]&=2i\hbar\delta^a_b\delta^k_{[j}\partial^x_{i]}\delta^3(x-y)+2i\hbar\widehat\omega^a{}_{b[i}\delta^k_{j]}\delta^3(x-y),\label{eq:q_torsion_pi}\\
	[\widehat T^a{}_{ij}(x),\widehat\Pi^k_{cd}(y)]&=2i\hbar\delta^k_{[i}\delta^a_{[c}\widehat h_{d]j]}\delta^3(x-y),\label{eq:q_torsion_Pi}
\end{align}
with the independent-pair identity and index normalization of Sec.~3. Equivalently, for an arbitrary Lie-algebra-valued connection variation $\eta_i$, the second bracket smeared against its canonical momentum generator is $2i\hbar\eta^a{}_{b[i}h^b{}_{j]}$; this fixes the normalization independently of pair notation.

The scalar pair reductions above leave these classical brackets unchanged because $T^a{}_{ij}$, $\pi$ and $\Pi$ are independent of $\phi,p_\phi$. This conclusion applies only to those pairs. For any further independent even second-class constraints $\chi$, the torsion--momentum bracket becomes
\begin{equation}
	\{T[f],P[g]\}_{\chi}=\{T[f],P[g]\}_*-\{T[f],\chi_m\}_*(\mathbb C^{-1})^{mn}\{\chi_n,P[g]\}_*.
\end{equation}
Here $P[g]$ denotes any specified smeared momentum, with its original bracket used in the first term. The same formula gives a potentially nonzero torsion--torsion bracket even when the original configuration brackets vanish. Thus the kinematical torsion commutators do not determine a physical torsion spectrum. Temporal torsion requires the branchwise inverse Legendre map and may retain undetermined primary multipliers; it is not an independent configuration operator.

On the regular initial data of Appendix~\ref{app:regular_background_evolution}, two relative torsion components have the explicit nonzero bracket
\begin{equation}\label{eq:explicit-torsion-example}
\left\{\int f\mathscr T^0{}_{12},\int h\mathscr T^1{}_{13}\right\}_{\rm st}
=-\frac{\kappa}{g}\int (\partial_2f)(\partial_1h)\,d^3x.
\end{equation}
The nonvanishing bracket is a statement about
the regular second-class reduced algebra.
It does not establish noncommutativity of
fully gauge-invariant Dirac observables.
Constructing relational torsion observables
would additionally require the treatment of
the remaining hypersurface gauge transformations
\cite{Dittrich2007,Rovelli2004}. This equality is evaluated on the benchmark, not a central relation throughout phase space; its leading quantum symbol is $i\hbar$ times the displayed bracket. Internal Lorentz covariance does not imply invariance under hypersurface deformations; relational localization is discussed below \cite{Dittrich2007,DonnellyFreidel2016,Rovelli2004}.

\paragraph{Fully stabilized spatial torsion brackets}
\label{subsec:fully-stabilized-torsion}
Work in the local BF-reduced patch of Sec.~\ref{sec:dirac_bergmann}, with compactly supported smearings. The relative spatial torsion is
\begin{equation}
	\mathscr T^a{}_{ij}=(\Lambda^{-1})^a{}_bT^b{}_{ij}
	=2\partial_{[i}\bar h^a{}_{j]}+2A^a{}_{b[i}\bar h^b{}_{j]}.
\end{equation}
It is invariant under the paired internal Lorentz gauge but transforms under changes of reference section. In the product canonical chart it depends only on configurations. For a fixed body-algebra basis $X_I$, the primary variation $\delta_X A_i=D_i^AX$, $\delta_X\bar h_i=0$ gives
\begin{align}
	U^a{}_{ij,I}(x,y)&=\{\mathscr T^a{}_{ij}(x),C_I(y)\}_*\nonumber\\
	&=2(X_I)^a{}_b\bar h^b{}_{[j}(x)\partial^x_{i]}\delta^3(x-y)
	+2[A_{[i},X_I]^a{}_b\bar h^b{}_{j]}(x)\delta^3(x-y),\\
	U^a{}_{ij,\phi}(x,y)&=\{\mathscr T^a{}_{ij}(x),P_\phi(y)\}_*=0.
\end{align}
For an antisymmetric compactly supported test density $f_a^{ij}$, write $T_f=\int f_a^{ij}\mathscr T^a{}_{ij}$. Unlike the earlier convention with a factor $1/2$, this definition sums both antisymmetric index orders without that factor. All numerators below use the normalization of $T_f$; a single independent component is recovered by assigning half the desired weight to each ordered pair with opposite signs. Integration by parts yields the primary numerator
\begin{equation}
	U_{f,I}=-2\partial_i(f_a^{ij}(X_I)^a{}_b\bar h^b{}_j)
	+2f_a^{ij}[A_i,X_I]^a{}_b\bar h^b{}_j.
	\label{eq:torsion-primary-smeared}
\end{equation}
The scalar component vanishes. If $E$ is the $7\times6$ orthonormal complement frame used in Sec.~\ref{subsec:complete_regular_reduction}, the paired numerator is $U_fE$ weakly, including for a field-dependent $E$. Since $\{T_f,T_g\}_*=0$, the paired-primary correction is
\begin{equation}
	\{T_f,T_g\}_4\approx U_f E D_P^{-1}E^TU_g^T=U_fQ_7U_g^T.
\end{equation}
Products denote integration and contraction of kernels. The numerator $U_f$ need not annihilate the null line; the displayed contraction uses the chosen orthogonal complement. In general the Lorentz--Lorentz block $L_0^TD_6^+L_0$ is nonzero, so no universal vanishing follows; the contraction must be retained.

Define $c_f=\{T_f,C_\nu\}_4$ and $s_f=\{T_f,S_u\}_4$. Using the residual inverse of Sec.~\ref{subsec:complete_regular_reduction},
the weak stabilized bracket is
\begin{align}
	\{T_f,T_g\}_{\rm st}\approx{}&U_fQ_7U_g^T
	+c_fZ_\nu^{-1}J_u(Z_\nu^\dagger)^{-1}c_g^T\nonumber\\
	&+c_fZ_\nu^{-1}s_g^T-s_f(Z_\nu^\dagger)^{-1}c_g^T.
	\label{eq:fully-stabilized-torsion}
\end{align}
In particular $c_f\approx U_f\nu$. The secondary numerator can be evaluated without solving the multiplier equations. Let $m$ label the independent metric components, $B^i_{,m}$ denote their derivatives, and $G^{mn}=\partial(2K_n)/\partial p_m$ at fixed configurations and spinor. The secondary of Sec.~\ref{subsec:secondary_stabilization} has momentum dependence $-2K_m B^i_{,m}(n)g_i$. Thus
\begin{align}
	s_f^{(0)}&=\{T_f,S\}_*
	=-\int_\Sigma \frac{\delta T_f}{\delta q_m}G^{m\ell}B^i_{,\ell}(n)g_i,\\
	s_f&\approx w\left(s_f^{(0)}-U_fQ_7\{\Phi,S\}_*\right).
\end{align}
The secondary is independent of body momenta in this representative. The metric derivative in the first line is explicitly defined by
\begin{equation}
	\delta_q T_f=-2\int_\Sigma (D_i^Af^{ij})_a\,\delta_q\bar h^a{}_j,
\end{equation}
where the derivative on $f$ is the dual connection. These expressions retain the derivatives of the smearing functions and the field dependence of the null vector in the secondary bracket. Appendix~\ref{app:torsion-brackets} gives the explicit coefficient formulas; on the initial slice of Appendix~\ref{app:regular_background_evolution}, $J_u=0$ and the inverse residual block is ultralocal. The factors $Z_\nu^{-1}$ can become singular near exceptional strata, but numerator cancellations must be examined before attributing a physical divergence to any observable \cite{Dirac1964,HenneauxTeitelboim1992,GolovnevGuzman2021}.

\paragraph{Reduced representations.}
\label{subsec:reduced-torsion-quantum}
For any functional $F$, define a local extension $F^\sharp=F-\{F,\chi_a\}_*(\mathbb C_8^{-1})^{ab}\chi_b$, with kernel contractions understood. Direct differentiation gives $\{F^\sharp,\chi_a\}_*\approx0$ and $\{F^\sharp,G^\sharp\}_*\approx\{F,G\}_{\rm st}$. The Jacobi identity follows from the induced symplectic structure on the regular second-class surface, rather than from treating the field-dependent inverse matrix as a constant. In a finite-dimensional regular symplectic regulator one may choose local Darboux coordinates and apply canonical quantization there. A candidate semiclassical correspondence for smeared even torsion observables is
\begin{equation}
	[\widehat T_f,\widehat T_g]=i\hbar\,\widehat{\{T_f,T_g\}_{\rm st}}+O(\hbar^2).
\end{equation}
This is a semiclassical condition, not an exact operator identity. The product-chart representation need not represent the stabilized algebra. Ordering and composite-operator regularization are required; local Darboux existence does not establish a continuum representation \cite{Thiemann2007,Rovelli2004,Dittrich2007,DonnellyFreidel2016}.

\paragraph{Reference-independent and relational observables.}
\label{sec:new-relational}
In the regular chart of Sec.~\ref{subsec:reference_section_independence}, $\widetilde{\mathscr T}=O\mathscr T$ and $\widetilde e=Oe$. Consequently the spatial and normal projections
\begin{equation}
	t^k{}_{ij}=e_I{}^k\mathscr T^I{}_{ij},\qquad t^\perp{}_{ij}=\mathscr T^{\hat0}{}_{ij}
\end{equation}
are independent of the reference section. They are invariant under paired internal Lorentz transformations, but not under general hypersurface deformations. Their primary numerators follow from Sec.~\ref{subsec:fully-stabilized-torsion} by replacing the test density by the corresponding projected test density: the extra metric dependence has zero bracket with every primary. In the secondary numerator, however, the metric derivative must also act on the inverse triad, $\delta e^{-1}=-e^{-1}(\delta e)e^{-1}$. Appendix~\ref{app:torsion-brackets} gives the resulting operator prescription.

For admissible clocks $X^\mu$, the matrix $M^\mu{}_A=\{X^\mu,G_A\}_{\rm st}$ must be invertible on the first-class surface. A relational observable is the value of a functional at the local gauge-orbit intersection $X=x$; its bracket is the relational completion of the gauge-fixed bracket, not generally the ungauged stabilized bracket \cite{Dittrich2007,Rovelli2004,HenneauxTeitelboim1992}.
\subsection{Quantum constraints and physical states}
\label{subsec:quantum_constraints}
The curvature operator in the partial connection representation is formally
\begin{equation}
	\widehat R^{ab}{}_{ij}=2\partial_{[i}\widehat\omega^{ab}{}_{j]}+2(\widehat\omega^a{}_{c[i}\widehat\omega^{cb}{}_{j]})_{\rm reg}.
	\label{eq:q_curvature_operator}
\end{equation}
The flatness and paired Gauss conditions must be implemented with the reducibility and coupled quotient of Secs.~\ref{subsec:rev-flat} and \ref{sec:dirac_bergmann}; the BF Gauss term alone is not the coupled generator. The regular reduced branch retains the four hypersurface constraints.

If an independent stabilized first-class set $\mathcal C_\alpha$ exists on the chosen regular branch, Dirac quantization seeks states satisfying
\begin{equation}
	\widehat{\mathcal C}_\alpha[f^\alpha]\,|\Psi_{\rm phys}\rangle=0
	\label{eq:q_physical_states}
\end{equation}
for admissible smearings on a common domain. On the regular reduced branch these are the four hypersurface generators. Kinetic null directions are not automatically first class, and multiplier-fixing equations are not physical-state conditions \cite{Dirac1964,HenneauxTeitelboim1992,GolovnevGuzman2021}.

\paragraph{Quantum closure and physical inner product.}
\label{subsec:quantum_closure}
For a stabilized first-class algebra with structure functions $f_{\alpha\beta}{}^\gamma$, a regulated representation should satisfy
\begin{equation}
	[\widehat{\mathcal C}_\alpha[f],\widehat{\mathcal C}_\beta[g]]=i\hbar\widehat{\mathcal C_\gamma[f_{\alpha\beta}{}^\gamma(f,g;Q,P)]}+\widehat{\mathscr A}_{\alpha\beta}[f,g].
	\label{eq:q_anomaly}
\end{equation}
The first term requires an explicit ordering prescription, and the anomaly must vanish or be consistently removable on the physical domain. The Jacobi identities of the classical reduced bracket are necessary but not sufficient for this quantum condition. Boundary charges, regulator-dependent terms and structure functions must be treated consistently \cite{Teitelboim1973,HenneauxTeitelboim1992,Thiemann2007}.

The classical BFV charge is constructed in Sec.~\ref{subsec:regular-bfv}; quantum nilpotency and the physical inner product are addressed in Sec.~\ref{sec:effective_field_theory}.

\section{Covariant quantization and reducible BV formulation}\label{sec:internalBVsection}
\label{sec:covariant-functional}
We compare the independent-field BV description with the regular canonical BFV reduction. All functional determinants are understood with a compatible regulator, boundary conditions and zero-mode prescription \cite{HenneauxTeitelboim1992,HT1990,BV1981,BVReducible1983}.

\subsection{Independent-field measure and flatness constraint}
For $\Phi=(h,\omega,\lambda,\Psi,\bar\Psi)$ the multiplier representation is
\begin{equation}
	Z[J]=\frac{1}{\mathcal V_{\rm gauge}}\int\mathcal D\Phi\,\mu[\Phi]
	\exp\{i(S_0[\Phi]+J\cdot\Phi)/\hbar\}.
\end{equation}
In a finite regulator on a regular canonical stratum, let $\chi$ be independent second-class constraints, $G$ independent first-class constraints, and $g$ admissible gauge conditions. The corresponding phase-space expression is formally
\begin{align}
	Z_{\rm can}={}&\int\mathcal Dz\,\delta[\chi]\,
	\sqrt{\operatorname{sdet}\{\chi,\chi\}}\,
	\delta[G]\delta[g]\,\left|\det\{g,G\}_{D}\right|\nonumber\\
	&\times\exp\left(\frac{i}{\hbar}\int dt\,[\Theta(z)\cdot\dot z-H_c]\right).
	\label{eq:rev8-canonical-measure}
\end{align}
Here the displayed gauge determinant refers to even independent gauge generators; graded gauge sectors require the corresponding Berezinian. The square root is understood with a consistent orientation or phase prescription. Reducible generators must first be resolved,
rather than inserted redundantly in this
expression \cite{HenneauxTeitelboim1992,HT1990}.

The algebraic BF determinant is constant; the fermionic reduction retains the shifted symplectic potential of Sec.~\ref{sec:rev-fermions}.

The formal oscillatory integral over the real multiplier gives
\begin{equation}
	\int\mathcal D\lambda\,e^{i\int_M\langle\lambda,R\rangle/\hbar}
	\propto\delta[R(\omega)].
\end{equation}
The Bianchi identity makes this delta functional redundant; the proportionality includes gauge volumes and zero-mode factors. Flat connections may have global holonomy. A local pure-gauge parametrization is therefore not a replacement for the global sum over flat sectors \cite{BlauThompson1991,Birmingham1991,Krssak2019}.

\subsection{Reducible BF symmetry and the internal BV master action}\label{sec:internalBVrevised}
The internal minimal BV complex resolves the on-shell multiplier reducibility of Sec.~\ref{subsec:rev-action} \cite{BV1981,BVReducible1983,HenneauxTeitelboim1992,Birmingham1991}. The Lorentz ghost $c\in\Omega^0(M,\mathfrak g)$ and shift ghost $\eta\in\Omega^1(M,\mathfrak g)$ are odd of ghost number one; $\rho\in\Omega^0(M,\mathfrak g)$ is even of ghost number two. Here $\mathfrak g=\mathfrak{so}(1,3)$. The antifield degrees and signs are specified in Appendix~\ref{app:BV-proof}.

In the total-degree algebra of Appendix~\ref{app:BV-proof}, introduce the superfields

\begin{equation}\label{eq:BVrevisedAB}
	\mathcal A=c+\omega+\lambda^*+\eta^*-\rho^*,\qquad \mathcal B=-\rho+\eta+\lambda+\omega^*-c^*.
\end{equation}
The minimal functional is
\begin{equation}\label{eq:BVrevisedmasterBF}
	S_{\mathrm{BF}}^{\mathrm{BV}}=
	\int_M\left\langle\mathcal B,
	d\mathcal A+\tfrac12[\mathcal A,\mathcal A]\right\rangle_{(4,0)}.
\end{equation}
It contains the classical BF term, the ghost and reducibility couplings,
and the essential quadratic-antifield vertex
\begin{equation}\label{eq:BVrevisedquadratic}
	S_2=-\tfrac12\int_M
	\big\langle\rho,[\lambda^*,\lambda^*]\big\rangle_{(4,0)}.
\end{equation}
The canonical component cancellation in Appendix~\ref{app:BV-proof} shows that
\begin{equation}\label{eq:BVrevisedCME}
	(S_{\mathrm{BF}}^{\mathrm{BV}},S_{\mathrm{BF}}^{\mathrm{BV}})=0
\end{equation}
for compactly supported variations or boundary conditions eliminating the
surface term.  Appendix~\ref{app:BV-proof} matches all ten canonical transformations to the signed superfield projections, including the quadratic-antifield correction. The super-Bianchi identities therefore imply canonical BF nilpotency; coupled antifield equations also contain matter and representation derivatives.

For the coframe and odd Dirac fields, use $s=(S,\cdot)$ and the canonical component ordering
\begin{equation}\label{eq:BVrevisedrep}
 S_{\rm rep}=\int_M\sum_{Y=h,\Psi,\bar\Psi}(-1)^{\epsilon_Y}\langle Y^*,sY\rangle_{\rm can},
 \quad sh=ch,\quad s\Psi=\Sigma(c)\Psi,\quad s\bar\Psi=\bar\Psi\Sigma(c).
\end{equation}
Here $\epsilon_Y$ is the Grassmann parity and the ghost in the last expression is ordered on the right. Lorentz invariance of
$S_{\mathrm{grav}}+S_D$ and closure of these representation operators
imply, in the same canonical convention,
\begin{equation}\label{eq:BVrevisedfullCME}
	S_{\mathrm{int}}^{\mathrm{BV}}=
	S_{\mathrm{grav}}+S_D+S_{\mathrm{BF}}^{\mathrm{BV}}+S_{\mathrm{rep}},
	\qquad (S_{\mathrm{int}}^{\mathrm{BV}},S_{\mathrm{int}}^{\mathrm{BV}})=0.
\end{equation}
This minimal internal master equation is classical; the diffeomorphism extension is treated in Appendix~\ref{app:BV-proof} \cite{BarnichBrandtHenneaux2000,DonnellyFreidel2016}. The classical master equation establishes the nilpotency of the corresponding BV
differential. Properness additionally requires the appropriate
rank condition on the stationary surface and
a complete resolution of the independent gauge
generators and their reducibility relations.
These conditions are not inferred solely from
the master equation
\cite{BV1981,BVReducible1983,HenneauxTeitelboim1992}.

\subsection{Hamiltonian BFV charge and classical nilpotency}
\label{subsec:regular-bfv}
After restriction to the regular second-class surface, Eq.~\eqref{eq:sec6_stabilized_HDA} supplies the first-class algebra with field-dependent structure functions. Let $G_A$ denote the four reduced hypersurface generators (excluding the lapse and shift primary momenta), with condensed spatial indices, and write $\{G_A,G_B\}_{\rm st}=U_{AB}{}^C G_C$ on this surface. Introduce odd ghosts $c^A$ of ghost number one and odd conjugate momenta $\mathcal P_A$ of ghost number minus one, with the even graded bracket $\{c^A,\mathcal P_B\}=\delta^A_B$. The minimal BFV charge begins with
\begin{equation}
	\Omega= c^AG_A-\frac12 c^Ac^BU_{AB}{}^C\mathcal P_C+\Omega_{\geq2},
	\label{eq:regular-bfv-charge}
\end{equation}
where $\Omega_{\geq2}$ has ghost-momentum degree at least two. With $sF=\{\Omega,F\}$, the ghost-momentum-independent part of the ghost transformations is
\begin{align}
	sc^\perp&=-c^i\partial_i c^\perp,\\
	sc^i&=-c^j\partial_jc^i-q^{ij}c^\perp\partial_jc^\perp.
\end{align}
These follow by evaluating the bilinear deformation
bracket on the odd smearing functions; the factor
$1/2$ cancels the two antisymmetric contributions.

A nonminimal sector and gauge fermion $\Psi_{\rm BFV}$
give
\[
H_{\rm BFV}
=H_{\rm inv}+\{\Omega,\Psi_{\rm BFV}\}_{\rm BFV}
\]
once the charge is nilpotent.
Here $H_{\rm inv}$ denotes a BRST-invariant extension
of the reduced Hamiltonian, satisfying
\[
\{\Omega,H_{\rm inv}\}_{\rm BFV}=0.
\]
Its restriction to the original ghost-free sector
reproduces the corresponding reduced Hamiltonian,
with boundary terms treated consistently.
Temporal primary chains and boundary Hamiltonians
require their own extensions
\cite{HenneauxTeitelboim1992,HT1990}.

\paragraph{Classical nilpotency test.}
\label{sec:new-bfv}
Restrict throughout to the regular second-class manifold, not the first-class zero set. Assume the deformation identities of Sec.~\ref{subsec:stabilized_deformation_algebra} hold on this manifold. Write a descriptor as $u=(N,V)$ and define
\begin{equation}
	[u,v]_q=(V^i\partial_i M-W^i\partial_iN,\ [V,W]^i+q^{ij}(N\partial_jM-M\partial_jN)),
	\quad v=(M,W).
\end{equation}
Appendix~\ref{app:metric_lorentz_currents} establishes $\{q^{ij},q^{kl}\}_{\rm st}=0$, $\{q^{ij},H[N]\}_{\rm st}=-2NK^{ij}$ and $\{q^{ij},D[V]\}_{\rm st}=\mathcal L_Vq^{ij}$ on this manifold. For field-independent descriptors the coefficient Jacobi identity is
\begin{equation}
	\sum_{\rm cyc}\left([u,[v,w]_q]_q+\{G[u],[v,w]_q\}_{\rm st}\right)=0.
\end{equation}
The second term acts only on the field-dependent coefficients. Three spatial descriptors give the vector-field Jacobi identity; two spatial and one normal give the representation identity for Lie derivatives on scalars. For one spatial and two normal descriptors, put $s=N\,dM-M\,dN$. The ordinary nested brackets sum to $[V,q^{-1}s]-q^{-1}\mathcal L_Vs=(\mathcal L_Vq^{-1})s$, canceled by $\{D[V],q^{-1}\}s$. For three normals the scalar coefficient is a cyclic sum of $-q^{ij}(M\partial_jL-L\partial_jM)\partial_iN$, which vanishes by symmetry. The vector coefficient is $2K^{ij}\sum_{\rm cyc}N(M\partial_jL-L\partial_jM)=0$. These smeared identities include distributional derivatives by duality with compactly supported tests.

The ghost-momentum-degree-zero term in the BFV self-bracket cancels by closure, the degree-one term by the coefficient Jacobi identity, and the degree-two term by the vanishing brackets of the metric structure functions. There are no higher terms in the self-bracket of the displayed quadratic-ghost charge. Hence
\begin{align}
	\Omega_{\rm min}=\int_\Sigma\big[&c^\perp\mathcal H_\perp+c^i\mathcal H_i-c^i\partial_i c^\perp\mathcal P_\perp\nonumber\\
	&-(c^j\partial_jc^i+q^{ij}c^\perp\partial_jc^\perp)\mathcal P_i\big],\qquad
	\{\Omega_{\rm min},\Omega_{\rm min}\}_{\rm st}=0.
\end{align}
The truncation in ghost-momentum degree is specific to this reduced hypersurface algebra, its coefficient Jacobi identity and its commuting metric structure functions; it does not follow from first-class closure alone. This is a classical local result for the four reduced hypersurface generators, conditional on the regularity and closure hypotheses of Sec.~\ref{subsec:stabilized_deformation_algebra}. It neither restores the temporal primary chains nor proves regulator-preserved quantum nilpotency \cite{HenneauxTeitelboim1992,HT1990,BarnichBrandtHenneaux2000}.
\subsection{Gauge fixing, reduced measure and quantum consistency}
On a contractible flat patch the paired Lorentz gauge can be chosen by setting the local pure-gauge group variable to the identity. Imposing all components of $\omega=0$ as independent gauge conditions before resolving flatness would overcount. The metric harmonic gauge fixes linearized diffeomorphisms, with operator $\Box\delta_\mu{}^\nu$ around Minkowski. The multiplier shift can be treated with a background co-closed gauge $\bar D^\dagger\lambda=0$ and the ghost-for-ghost complex. In a Euclidean auxiliary regulator with flat $\bar D$, the operators on forms are $\Delta_p=\bar D^\dagger\bar D+\bar D\bar D^\dagger$; on coexact one-forms the shift variation is $\bar D^\dagger\bar D\zeta$. Exact one-forms are reducibility directions. Harmonic modes require separate treatment. This identifies the operators, not the complete Lorentzian determinant ratio \cite{FP1967,BRST1975,Birmingham1991,BVReducible1983}.

For a proper quantum BV action $W$ satisfying the regulated master equation of Sec.~\ref{subsec:quantum_consistency_UV}, a gauge fermion $\mathcal X$ defines the formal integral
\begin{equation}
 Z_{\mathcal X}=\int\mathcal D\Phi\,\mu_{\rm red}\exp\left(\frac{i}{\hbar}W\big|_{\Phi^*=\delta\mathcal X/\delta\Phi}\right).
\end{equation}
The measure denotes a compatible induced density, not an additional determinant in an already reduced integral. At tree level $W=S_{\rm BV}$. Gauge independence requires the quantum conditions below.

\paragraph{Complete second-class measure on the regular branch}
\label{subsec:complete-regular-measure}
The local BF and fermionic reductions are understood to have been performed, and temporal multiplier pairs are suppressed. In the product chart the formal Liouville--Berezin measure is
\begin{equation}
	d\mu_* =\mathcal Dq\,\mathcal Dp\,\mathcal D\Lambda\,\mathcal DL\,
	\mathcal D\phi\,\mathcal Dp'_\phi\,\mathcal D\xi\,\mathcal D\xi^\dagger.
\end{equation}
The group and body-momentum factors denote the canonical cotangent-bundle measure, not a flat measure in arbitrary group coordinates. The eight even constraints of Sec.~\ref{sec:dirac_bergmann} define the induced measure
\begin{equation}
	d\mu_{\rm st}=d\mu_*\,\delta[\chi]\sqrt{|\operatorname{Det}\mathbb C_8|},
\end{equation}
with the determinant of Eq.~\eqref{eq:sec6_complete_measure_factorization}. The fermionic second-class factor is already incorporated in the reduced Berezin measure; the eliminated BF constraints and their gauge volume are not counted again.

Write $\mathcal H_\mu=(\mathcal H_\perp,\mathcal H_i)$ for the remaining four generators, and choose admissible local gauges $G^\mu$ with invertible $\mathcal M^\mu{}_{\nu}=\{G^\mu,\mathcal H_\nu\}_{\rm st}$. A formal gauge-fixed canonical integral is
\begin{align}
	\mathcal Z_G={}&\int d\mu_{\rm st}\,\delta[G]\,\delta[\mathcal H]\,
	|\operatorname{Det}\mathcal M|\exp\left(\frac{i}{\hbar}\int dt\,\Theta_*(\dot Z)\right).
\end{align}
The delta functionals pull back $\Theta_*$ and implement lapse and shift integration once. Boundary Hamiltonians must be included when present. This expression assumes admissible gauges and a compatible regulator \cite{HenneauxTeitelboim1992,HT1990,FP1967,BVReducible1983}.

The unit-pair determinant and basis-invariance identities are derived in Sec.~\ref{subsec:regular_constraint_measure} and Appendix~\ref{app:constraint_measure_variations}. At finite regulator the auxiliary odd Gaussian $\int d\eta\exp(\eta^a C_{ab}\eta^b/2)=\operatorname{Pf}C$ represents the signed Pfaffian, not its absolute value; these auxiliary variables are not BFV ghosts. The induced Liouville--Berezin measure is preserved by reduced Hamiltonian flows. This does not establish a BRST-invariant continuum regulator or equality of canonical and BV quantum measures \cite{HenneauxTeitelboim1992,HT1990,BarnichBrandtHenneaux2000}.

\section{Quantum consistency and semiclassical benchmarks}\label{sec:effective_field_theory}
The local TEGR equivalence provides a check of the canonical construction, not a proof of quantum equivalence. In a torsion-free Minkowski patch the linear metric perturbation has the Fierz--Pauli quadratic action \cite{FierzPauli1939,Weinberg1965}. For $F(0)=0$ and $F_T(0)=1$, the nonlinear terms do not alter this quadratic action because $T$ starts at second perturbative order. The independent flatness sector introduces no local graviton pole. The linear matter coupling is $\delta S_D=\frac12\int d^4x\,\delta g_{\mu\nu}T_D^{\mu\nu}$, with $\partial_\mu T_D^{\mu\nu}=0$ on the free Dirac equations. This is the linear gravitational Ward identity.

\subsection{Effective counterterms and composite operators}\label{subsec:teleparallel_EFT}
The TEGR benchmark admits the usual low-energy Einstein--Dirac effective action with curvature-squared counterterms and matter operators \cite{Donoghue1994,Burgess2004}. In the nonlinear paired formulation the counterterm basis must respect diffeomorphism and paired Lorentz symmetry and the flatness implementation; independent torsion contractions, covariant derivatives of torsion and matter operators may occur. No closure of the $F(T)$ ansatz under renormalization is assumed. Products of torsion operators require smearing, regularization and composite-operator renormalization \cite{BirrellDavies1982,ParkerToms2009}. The relational construction in Sec.~\ref{sec:new-relational} supplies a classical localization prescription, not a quantum representation.

\subsection{Order-by-order quantum obstructions}\label{subsec:quantum_consistency_UV}
Classical closure does not establish the regulated quantum master equation or the canonical anomaly condition.

For a compatible regulated BV Laplacian with $\Delta_\mu^2=0$, set $s=(S,\cdot)$ and $W=S+\hbar M_1+\hbar^2M_2+\cdots$. With $M_0=S$, the quantum master equation gives
\begin{equation}
sM_n=i\Delta_\mu M_{n-1}-\tfrac12\sum_{k=1}^{n-1}(M_k,M_{n-k}),\quad n\geq1.
\end{equation}
Write the right-hand side as $\mathfrak A_n$ and let $\mathscr E(W)=\tfrac12(W,W)-i\hbar\Delta_\mu W$. For even $W$, Jacobi and BV compatibility give $(W,(W,W))=0$ and $\Delta_\mu(W,W)=-2(W,\Delta_\mu W)$. Nilpotency of $\Delta_\mu$ then implies
\begin{equation}
 (W,\mathscr E)-i\hbar\Delta_\mu\mathscr E=0.
\end{equation}
If the equation holds through order $n-1$, its defect for $W_{n-1}=S+\sum_{k=1}^{n-1}\hbar^kM_k$ is $-\hbar^n\mathfrak A_n+O(\hbar^{n+1})$. The leading coefficient of the identity gives $s\mathfrak A_n=0$. Thus a next-order solution exists precisely if $[\mathfrak A_n]=0$ in ghost-number-one cohomology of the admissible functionals (locally $H^{1,4}(s\mid d)$ for local counterterms). No such vanishing has been established here. In the Hamiltonian formulation the independent test is $\widehat\Omega^2=0$ on a common invariant domain, not merely the classical BFV identity. For an associative graded star product at a compatible regulator, write $[A,B]_{\star,\rm gr}/(i\hbar)=\sum_{r\ge0}\hbar^r C_r(A,B)$, with $C_0$ the extended BFV Poisson bracket. For an odd ghost-number-one charge $\Omega_\hbar=\sum_{k\ge0}\hbar^k\Omega_k$, classical nilpotency defines $s_{\rm BFV}=C_0(\Omega_0,\cdot)$. The order-$n$ equation is
\begin{equation}
s_{\rm BFV}\Omega_n=-\mathfrak B_n,\qquad
\mathfrak B_n=\tfrac12\sum_{r+k+l=n,\ k,l<n}C_r(\Omega_k,\Omega_l).
\end{equation}
For example $\mathfrak B_1=C_1(\Omega_0,\Omega_0)/2$ and $\mathfrak B_2=C_0(\Omega_1,\Omega_1)/2+C_1(\Omega_0,\Omega_1)+C_2(\Omega_0,\Omega_0)/2$. The graded Jacobi identity gives $[\Omega_\hbar,[\Omega_\hbar,\Omega_\hbar]_{\star,\rm gr}]_{\star,\rm gr}=0$. If lower orders vanish, its leading defect implies $s_{\rm BFV}\mathfrak B_n=0$. Thus the next correction exists precisely when $[\mathfrak B_n]=0$ in the admissible ghost-number-two BFV cohomology. Neither this vanishing nor an identification with ghost-number-one BV cohomology is assumed. A physical inner product on the resulting cohomology (or an admissible rigging map) must be constructed independently. These conditions and a matching regulated measure are prerequisites for a quantum equivalence claim \cite{BV1981,BarnichBrandtHenneaux2000,HT1990}. The relevant obstruction classes depend on the
chosen space of admissible functionals.
In a local perturbative formulation, one must
distinguish unrestricted BRST cohomology from
local BRST cohomology modulo total derivatives.
Consequently, formal cohomological closure does
not guarantee the existence of admissible local
counterterms
\cite{BarnichBrandtHenneaux2000,HenneauxTeitelboim1992}.

\section{Conclusions and outlook}
The independent coframe--spin-connection formulation provides a branchwise canonical foundation for teleparallel $F(T)$ gravity. The coupled BF quotient and fermionic reduction retain the relevant momentum shifts and recover the TEGR--Einstein--Dirac benchmark \cite{ADM1962,Dirac1964,HenneauxTeitelboim1992}. On the regular nonlinear branch, the explicit stabilized inverse and tangent Poisson operator determine the reduced brackets; the five-gravitational-degree-of-freedom count remains conditional on the stated functional assumptions.

The inhomogeneous initial-data example exhibits a nonvanishing bracket between reference-relative torsion observables that commute before second-class stabilization. This supplies a concrete input for canonical quantization, alongside the regulated determinant. The classical BFV and reducible BV constructions \cite{BV1981,BVReducible1983} lead to distinct all-order quantum obstruction equations, whose cohomological closure does not establish anomaly cancellation.

Further progress requires control of these obstruction classes, BV properness, operator domains, a physical inner product and compatible regulated measures. Exceptional strata, global holonomies and boundary sectors require separate analyses. The present framework establishes classical and formal quantum foundations rather than a completed quantum theory.

\appendix
\section{Lorentz projections and explicit kinetic matrix}
\label{app:rev-calculation-A}

\subsection{Kinetic contractions and pseudoinverse}
\label{app:hessian-rank}
At an adapted unit coframe let $u_i=X^0{}_{0i}$ and $V_{ij}=X^i{}_{0j}$ denote the purely velocity-dependent components. The three contractions in the torsion scalar, evaluated on the velocity insertion alone, are respectively
\begin{equation}
	\tfrac12 u^Tu-\tfrac12\operatorname{tr}(V^TV),\qquad
	\tfrac12 u^Tu-\tfrac12\operatorname{tr}(V^2),\qquad
	(\operatorname{tr}V)^2-u^Tu.
\end{equation}
Their sum is $(\operatorname{tr}V)^2-\operatorname{tr}[(\operatorname{sym}V)^2]$. This yields the matrix of Sec.~\ref{subsec:coframe_hessian_rank} in the ordered basis
\[ (01,02,03,11,22,33,12,21,13,31,23,32). \] Here $\mathbf1=(1,1,1)^T$. The pseudoinverse has diagonal block $D^{-1}=-I_3+\mathbf1\mathbf1^T/2$, off-diagonal blocks $S^+=S$, and vanishing boost block. Rank is invariant under regular velocity-basis congruence, consistently with Sec.~\ref{subsec:adm-momenta-revised} \cite{MalufRocha2001,Oko2013,BlixtHohmannPfeifer2019}.

In the auxiliary chart the Hessian is $-c\operatorname{diag}(\phi M,0)$, of rank six for $\phi\ne0$ and zero otherwise.

\subsection{Homogeneous and inhomogeneous kinetic tests}
For $F=T+\alpha T^2$, homogeneous diagonal data have $T=2\sum_{i<j}H_iH_j$, $q_{\mathcal K}=0$ and $q^TM^+q=T$, hence $\mathscr D=(1+6\alpha T)/(1+2\alpha T)$. For flat FLRW ($T=6H^2$), the eigenvalues of $-W/c$ are $-(1+12\alpha H^2)$ (fivefold), $2(1+36\alpha H^2)$ (simple), and zero (sixfold). For $\alpha<0$ the rank-five and rank-one loci are respectively $H^2=-1/(36\alpha)$ and $H^2=-1/(12\alpha)$. These are kinetic, not physical, degeneracies \cite{IzumiOng2013,FerraroGuzman2018,Bahamonde2023,BlixtReview2021}.

For a contrasting pointwise configuration choose $X^1{}_{12}=s$ and its antisymmetric partner, all other components zero, at the unit coframe. The torsion scalar is zero. Adding the boost insertion $X^0{}_{02}=u$ changes the trace contribution to $-(s+u)^2$ while the first two contractions contribute $s^2+u^2$. Thus $T=-2su$ and $q_{02}=-s$. For $s\ne0$, $F=T+\alpha T^2$ and $\alpha\ne0$, the Hessian has rank seven. This is a local coframe jet, not a field-equation solution.

\section{Distributional torsion brackets and their domain of validity}
\label{app:torsion-brackets}
\subsection{Canonical derivation in the partially reduced chart}
In the product chart of Sec.~3, functional differentiation gives ($\delta_{xy}=\delta^3(x-y)$)
\begin{align}
	\{T^a{}_{ij}(x),\pi_b{}^k(y)\}_*&=
	2\delta^a_b\delta^k_{[j}\partial^x_{i]}\delta_{xy}
	+2\omega^a{}_{b[i}(x)\delta^k_{j]}\delta_{xy},\label{eq:appC-pi}\\
	\{T^a{}_{ij}(x),\Pi^k_{cd}(y)\}_*&=
	2\delta^k_{[i}\delta^a_{[c}h_{d]j]}(x)\delta_{xy}.\label{eq:appC-Pi}
\end{align}
With the weight-one-half antisymmetric projector (rescale for independent pairs),
\begin{equation}
	\{T,T\}_*=\{T,h\}_*=\{T,\omega\}_*=\{T,\xi\}_*=0.
\end{equation}
These are not physical brackets \cite{Dirac1964,HenneauxTeitelboim1992,GolovnevGuzman2021}.

\paragraph{Smeared brackets and boundary terms.}
For compactly supported antisymmetric test fields $f_a{}^{ij}$ define
\begin{equation}
	T[f]=\tfrac12\int_\Sigma f_a{}^{ij}T^a{}_{ij},\quad
	P[g]=\int_\Sigma g^a{}_i\pi_a{}^i,\quad
	P_\omega[u]=\int_\Sigma\langle u_i,\Pi^i\rangle .
\end{equation}
Normalize $P_\omega[u]$ to generate $\delta\omega_i=u_i$. Then
\begin{align}
	\{T[f],P[g]\}_*&=\int_\Sigma f_a{}^{ij}D_i g^a{}_j\\
	&=-\int_\Sigma(D_i f_a{}^{ij})g^a{}_j
	+\int_{\partial\Sigma}dS_i\, f_a{}^{ij}g^a{}_j,\\
	\{T[f],P_\omega[u]\}_*&=\int_\Sigma f_a{}^{ij}u^a{}_{bi}h^b{}_j.
\end{align}
The test density carries the dual Lorentz representation. Also $\delta_uR_{ij}=2D_{[i}u_{j]}$, $\{T[f],\mathscr F[\zeta]\}_*=0$ for fixed smearings, and $D_\omega T^a=R^a{}_b\wedge h^b$ off shell \cite{BlauThompson1991,Birmingham1991,DonnellyFreidel2016}.

\paragraph{Shifted momenta.}
For the shift of Sec.~3, $\{T,\mathcal A_A\}_*=0$, hence $\{T,P_A\}_*=\{T,\pi_A\}_*$. The mixed momentum brackets of Sec.~\ref{subsec:quantum_kinematics} must nevertheless be retained.

\subsection{Exact scalar-pair reduction of the torsion algebra}
On either regular auxiliary branch of Sec.~4.5, let $X=C_\parallel$ or $X=S_\phi$, and set $d=\{p_\phi,X\}_*$, $J=\{X,X\}_*$. The inverse of the second-class block is
\begin{equation}
	\mathcal B^{-1}
	=
	\begin{pmatrix}
		(d^\dagger)^{-1}Jd^{-1}
		&
		-(d^\dagger)^{-1}
		\\[2pt]
		d^{-1}
		&
		0
	\end{pmatrix},
	\qquad
	J=\{X,X\}_* .
\end{equation}
The dagger denotes the formal adjoint on the
selected smearing domain, including the
interchange of kernel arguments.
For ultralocal finite-dimensional blocks,
it reduces to the ordinary transpose. The zero lower-right block preserves brackets of functionals commuting with $p_\phi$, including the original torsion--momentum brackets.

For the reconstructed scalar $\phi_*(z)$, differentiating $X(z,\phi_*(z))=0$ and using $d=-\delta X/\delta\phi$ (with transposed kernel arguments as appropriate) yields
\begin{equation}
	\{T,\phi_*\}_{\rm pair}=-(\delta X/\delta\phi)^{-1}\{T,X\}_*\big|_{\phi_*},
\end{equation}
where the inverse acts on the scalar output index. On the rank-seven branch define $b^2=\sum_I b_I^2$, $c_h=h/\kappa$ and $w^A=b^Iz_I^A/b^2$. Then $X=w^AP_A+c_h\phi$, and explicitly
\begin{equation}
	\{T^a{}_{ij}(x),\phi_*(y)\}_{\rm pair}
	=-c_h(y)^{-1}w^{bk}(y)\{T^a{}_{ij}(x),\pi_b{}^k(y)\}_*.
\end{equation}
On the regular $b=0$ scalar-secondary branch, $X=S_\phi$ and $\delta X/\delta\phi=-\mathscr D/F_{TT}$. Its momentum derivative is $2(M^+P_R)^A/(c_h^2\phi^2)$, giving
\begin{equation}
	\{T(x),\phi_*(y)\}_{\rm pair}=
	\frac{2F_{TT}}{\mathscr D c_h^2\phi_*^2}(M^+P_R)^A(y)\{T(x),\pi_A(y)\}_*.
\end{equation}
Further branch constraints may modify this bracket; the scalar inverse fails at $\mathscr D=0$ or on singular Legendre strata.

\subsection{Stabilized torsion kernels}
For the relative torsion $\tau_{ij}$ of Sec.~\ref{subsec:fully-stabilized-torsion}, the primary flow gives
\begin{align}
	\{\tau_{ij},C[X]\}_*&=2(D^A_{[i}X)\bar h_{j]},\\
	\{T_{ij},C[X]\}_*&=\Lambda(X\tau_{ij}+2(D^A_{[i}X)\bar h_{j]}).
\end{align}
These are relative Lorentz variations, not paired gauge transformations. Although $\{T,P_\varphi\}_*=0$, the translated scalar constraint contains configuration terms, so the scalar-pair correction to arbitrary momentum brackets must be retained. The original-chart scalar-pair result above requires the canonical translation of all observables.

For $A=T[f]$ the unprojected secondary row follows directly from Eqs.~\eqref{eq:appF_unit_normal_source} and \eqref{eq:appF_projectable_secondary}:
\begin{equation}
	\{T[f],S(y)\}_*=-\frac{\delta T[f]}{\delta q^m(y)}n^I B^i{}_{,l}(X_I)g_i(W^{-1})^{lm}(y).
\end{equation}
The functional derivative acts also on the test field. Project this row and multiply by $w$ on the stabilized surface to obtain the unit-secondary numerator $s_T$; momentum rows require the full first variation below. Multiplication by $z_\nu^{-1}$ does not commute with derivatives. Physical observables require further gauge reduction \cite{Dirac1964,HenneauxTeitelboim1992,Dittrich2007}.

\paragraph{Secondary kernel and differential order.}
Write $S=-2K_m u_m+v$, where $u_m=B^i_{,m}(n)g_i$ and $v=-\partial_i(R^{ij}(n)g_j)+R^{ij}(\partial_i n+[A_i,n])g_j$. For arbitrary variations,
\begin{align}
	\delta S&=-2u_m\delta K_m-2K_m\delta u_m+\delta v,\\
	\delta u_m&=(\delta B^i_{,m})(n)g_i+B^i_{,m}(\delta n)g_i+B^i_{,m}(n)\partial_i\delta\phi,\\
	\delta n&=(\delta r_J a+r_J\delta a,-\delta r_K a-r_K\delta a,0),\\
	\delta r_X&=\delta a^T\beta_X+a^T\delta\beta_X,\\
	\delta\beta(X)&=-\partial_i\delta B^i(X)+\delta B^i([A_i,X])+B^i([\delta A_i,X]).
\end{align}
Here $\delta A_i=D_i^A(\Lambda^{-1}\delta\Lambda)$ in the flat gauge. Antisymmetry of $R^{ij}$ cancels the Hessian of $\phi$, and the two derivatives of $n$ cancel identically. Hence $v=-(\partial_iR^{ij})(n)g_j+R^{ij}([A_i,n])g_j$, where the derivative acts on the fixed-basis coefficients. The secondary is a first-jet function with algebraic momentum dependence. In particular $\delta K_{ij}=\delta[2\kappa(P_{ij}-q_{ij}P/2)/(\phi\sqrt q)]$, including $\delta P=\delta(p-a_F)$. Define the row differential operators $L_S=dS$, $L_\Phi=d\Phi$ in the product chart and let $J$ be its graded Poisson tensor. Then the exact kernel compositions are
\begin{equation}
	E=L_\Phi J L_S^\dagger,\quad J_0=L_S J L_S^\dagger,\quad J_*=J_0+E^\dagger Q_7 E.
\end{equation}
The dagger denotes the formal graded adjoint under the smearing pairing; $T$ denotes the pointwise matrix transpose. In particular the metric part of $L_S$ in the momentum direction is $-4\kappa(u_{kl}-q_{kl}q^{ij}u_{ij}/2)/(\phi\sqrt q)$. For the projected torsion, $L_t$ includes the inverse-triad derivative in Sec.~\ref{subsec:quantum_torsion}. Its bracket is $L_t J_{\rm st}L_t^\dagger$, with the sequential inverse of Sec.~\ref{subsec:fully-stabilized-torsion}. The primary contribution has order-two candidate symbol $U_1(k)Q_7 U_1(k)^\dagger$, where $(U_1)^a{}_{ij,I}=2ik_{[i}(X_I)^a{}_b\bar h^b{}_{j]}$. The dagger is the Fourier adjoint. Combine all terms, including derivatives of $z^{-1}$, before assigning a principal symbol \cite{Dirac1964,HenneauxTeitelboim1992}.

\paragraph{Order and coefficients of the secondary self-bracket.}
In a product Darboux chart write $\delta S=A_A\delta Q^A+B_A^i\partial_i\delta Q^A+C^A\delta p_A+\delta_FS$. The spin dependence is algebraic. For compactly supported smearings $f,h$, direct integration by parts gives
\begin{equation}
\{S[f],S[h]\}_*=\int j_0^i(f\partial_i h-h\partial_i f),\qquad j_0^i=C^AB_A^i.
\end{equation}
Thus $J_0=2j_0^i\partial_i+\partial_i j_0^i$. Both $S$ and the primaries depend on first configuration jets and algebraically on momenta, so $E=E_0+E_1^i\partial_i$. More explicitly, writing $L_\Phi=(D+F^i\partial_i,G)$ and $L_S=(A+B^i\partial_i,C)$ in the even canonical blocks, one has
\begin{align}
E_1^i&=F^i C^T+G(B^i)^T,\\
E_0&=DC^T+F^i\partial_i C^T-GA^T+G\partial_i(B^i)^T+E_F,
\end{align}
where $E_F$ is the algebraic graded spin contraction. These expressions are in local Darboux coordinates. Alternatively the body-primary flow gives the coefficients directly. Put $u^{mn}(X)=B^i{}_{,q_{mn}}(X)g_i$, $k=(a,0)$, $j=(0,a)$, and $\mathscr D(U,V)=U^{mn}V_{mn}-\tfrac12\operatorname{tr}U\operatorname{tr}V$. For $\mathcal K=\{S,\Phi\}_*=k^i\partial_i+k^0$ one finds
\begin{align}
k_X^i&=f_1(k)B^i([X,j])-f_1(j)B^i([X,k])+R^{ij}([X,n])g_j,\\
k_X^0&=f_1(k)B^i([[A_i,X],j])-f_1(j)B^i([[A_i,X],k])\notag\\
&\quad+R^{ij}([[A_i,X],n])g_j-\frac{4\kappa}{\varphi\sqrt q}\mathscr D(u(n),u(X)).
\end{align}
Indeed $\delta_X P^{mn}=u^{mn}(X)$ and $\delta_X A_i=D_i^AX$ for external smearings. For the scalar flow set $b^i_I=B^i(K_I)$, $V^{mn}=B^\ell{}_{,q_{mn}}(A_\ell)$ and
\begin{equation}
N^i=((\beta_J^Tb^i)a+r_Jb^i,-(\beta_K^Tb^i)a-r_Kb^i,0)^T.
\end{equation}
Then $\delta_\rho n=N^i\partial_i\rho$, $\delta_\rho P=-\rho V$, and
\begin{align}
k_\varphi^i&=f_1(N^i)-2K_{mn}B^i{}_{,q_{mn}}(n)-(\partial_\ell R^{\ell i})(n)+R^{\ell i}([A_\ell,n]),\\
k_\varphi^0&=2K_{mn}u^{mn}(n)/\varphi+\frac{4\kappa}{\varphi\sqrt q}\mathscr D(u(n),V).
\end{align}
Thus $E=-\mathcal K^\dagger$, $E_1^i=(k^i)^T$, $E_0=(\partial_i k^i-k^0)^T$. In the metric block $B_{mn}^i=f_1(\partial n/\partial(\partial_i q_{mn}))-R^{ij}{}_{,q_{mn}}(n)g_j$, where the derivative of $n$ follows from $\partial\beta(X)/\partial(\partial_iq_{mn})=-B^i{}_{,q_{mn}}(X)$. The direct coefficient $j_0^i=C^{mn}B_{mn}^i$ has no other momentum contribution; the algebraic odd self-contraction of the even scalar vanishes. All these expressions use the unnormalized secondary. With $b^i=E_0^TQ_7E_1^i$ and $M^{ij}=(E_1^i)^TQ_7E_1^j=-M^{ji}$, expansion of $E^\dagger Q_7E$ cancels the second derivatives exactly. Consequently
\begin{equation}\label{eq:secondary_first_order}
J_*=2j_*^i\partial_i+\partial_i j_*^i,\qquad j_*^i=j_0^i+b^i-\tfrac12\partial_j M^{ji}.
\end{equation}
Integration by parts gives $J_*^\dagger=-J_*$: for $A^i\partial_i+B$, skew-adjointness requires $2B=\partial_i A^i$. The symbol is $2ij_*^ik_i$; it may vanish on special configurations. For the unit pair $(C_\nu,S_u)$, set $j_u^i=w^2j_*^i$, $z_\nu=w^2z$. The upper-left inverse block is
\begin{equation}
z_\nu^{-1}J_uz_\nu^{-1}=2(j_u^i/z_\nu^2)\partial_i+\partial_i(j_u^i/z_\nu^2).
\end{equation}
Appendix~\ref{app:regular_background_evolution} evaluates this prescription on the explicit constrained data. All differential adjoints use the stated boundary conditions.

\paragraph{Tangent Poisson operator.}\label{app:poisson-projector}
For eight independent even constraints $\chi$, write $L=L_\chi$, $J=\mathbb J_{\rm prod}$ and $C=LJL^\dagger$. On a domain admitting the inverse, define
\begin{equation}
\mathbb P=1-JL^\dagger C^{-1}L,\qquad
\mathbb J_{\rm st}=J-JL^\dagger C^{-1}LJ=\mathbb P J\mathbb P^\dagger.
\end{equation}
Using $LJL^\dagger=C$ and graded skew-adjointness gives $\mathbb P^2=\mathbb P$, $L\mathbb P=0$ and $\mathbb J_{\rm st}L^\dagger=0$. The projector $\mathbb P$ is determined by
the constraint differentials and the
product-reduced Poisson tensor.
It is a tangent projection associated with
the second-class reduction, not in general
an orthogonal projection with respect to
an auxiliary positive-definite metric. Block inversion of $C$ gives the six-primary Dirac correction followed by the residual Schur correction of Sec.~\ref{subsec:complete_regular_reduction}, in that order. For configuration torsion smearings the resulting bracket is $L_{T_f}\mathbb J_{\rm st}L_{T_h}^\dagger$. This is a second-class projection, not a quotient by hypersurface gauge transformations. If the inverse is specified only weakly, all identities are restricted to the constraint surface.

\section{Functional constraint stabilization and the deformation algebra}
\label{app:constraint-inventory}
\subsection{Legendre map and secondary stabilization}
\label{app:D-stabilized}\label{app:D-count-final}\label{app:functional_stabilization}
With the inverse domains of Sec.~\ref{sec:dirac_bergmann}, for the original-chart five-by-five operator let $A_{IJ}=\{L_I(x),L_J(y)\}$ and $Q_{\alpha I}=\delta_{\alpha I}-r_\alpha\delta_{6I}$. Then
\begin{align}
	\Delta_{\alpha\beta}(x,y)={}&Q_{\alpha I}(x)A_{IJ}(x,y)Q_{\beta J}(y)\notag\\
	&-Q_{\alpha I}(x)\{L_I(x),r_\beta(y)\}L_6(y)\notag\\
	&-Q_{\beta J}(y)\{r_\alpha(x),L_J(y)\}L_6(x).
\end{align}
The ratios commute, but $L_6$ is not constrained; the last two terms survive. Pointwise odd-matrix rank does not determine the differential kernel \cite{BlixtReview2021,GolovnevGuzman2021}.

Let $\chi$ be all independent remaining primaries and $\sigma$ the secondaries. With $H_T=H_c+\int u^m\chi_m$, the simultaneous multiplier equations are
\begin{equation}
	\mathcal M u\approx-f,\qquad
	\mathcal M=\begin{pmatrix}\{\chi,\chi\}\\\{\sigma,\chi\}\end{pmatrix},\qquad
	f=\begin{pmatrix}\{\chi,H_c\}\\\{\sigma,H_c\}\end{pmatrix}.
\end{equation}
Left null smearings give necessary compatibility conditions; range, boundary conditions and independence must also be checked. For $b=0$,
\begin{equation}
	\dot S_\phi\approx\{S_\phi,H_c\}+\int u^I\{S_\phi,C_I\}-u_\phi\mathscr D/F_{TT}+\text{other primary terms}.
\end{equation}
At $\mathscr D=0$ the scalar multiplier drops out, but simultaneous elimination and branch tangency are still required. The $F_T=0$ stratum needs a separate Legendre analysis \cite{Dirac1964,HenneauxTeitelboim1992,BlixtReview2021}.

\paragraph{Exceptional null systems and tangency.}
\label{app:D-exceptional}
The augmented primary null equations are $D_6N=\beta\ell$ and $\beta^TN=0$. For $a\ne0$ and $\beta\ne0$ with $r_K=r_J=0$, the three independent null vectors are $(k,0)$, $(j,0)$ and $(D_6^+\beta,1)$. The last one includes a scalar variation. For $a=0$, $\beta\ne0$, the null space is $\beta^\perp\times\{0\}$. Thus the secondary candidates are $S_A=N_A^If_I+\ell_A f_\varphi$, not simply the projected Lorentz sources in every branch. On a spatially open region with $\partial_i\varphi=0$, the Lorentz sources vanish, and preservation of homogeneity requires $\partial_i\dot\varphi=0$.

After transverse reduction solve $Y_A+Z_{AB}\mu^B\approx0$ together with branch tangency $\{g_s,H_T\}\approx0$ (e.g. $g_s=r_K,r_J,\beta_I$). These equations can fix multipliers; no exceptional physical count follows \cite{Dirac1964,HenneauxTeitelboim1992,BlixtReview2021}.

\paragraph{Reference metric Legendre map and scalar source}
\label{app:reference_metric_legendre}

Define
\begin{equation}
	g_i=\partial_i\varphi,\qquad
	A_0^c=\Lambda^{-1}\omega_0\Lambda,\qquad
	\Omega=\Lambda^{-1}\dot\Lambda,
\end{equation}
and
\begin{equation}
	A_i=\Lambda^{-1}(\partial_i+\omega_i)\Lambda.
\end{equation}
The relative Lorentz density is
\begin{equation}
	\mathcal L_\Lambda
	=
	a(\Omega)+a(A_0^c)
	-B^i(A_i)\dot\varphi+H^{ij}(A_i)g_j,
	\label{eq:appF_relative_density}
\end{equation}
where
\begin{equation}
	a(X)=B^i(X)g_i.
\end{equation}

Since this density contains no reference metric velocity,
the metric momentum satisfies
\begin{equation}
	p_m=W_{mn}\dot q^n+d_m,
\end{equation}
where $d_m$ includes the reference torsion and fermionic
momentum shifts. In symmetric tensor notation,
\begin{align}
	W^{ij,kl}
	&=
	\frac{\varphi\sqrt q}{4\kappa N}
	\left(q^{i(k}q^{l)j}-q^{ij}q^{kl}\right),\\
	(W^{-1})_{ij,kl}
	&=
	\frac{4\kappa N}{\varphi\sqrt q}
	\left(q_{i(k}q_{l)j}-\frac12q_{ij}q_{kl}\right).
	\label{eq:appF_DeWitt_inverse}
\end{align}
Direct contraction gives
\begin{equation}
	W^{ij,mn}(W^{-1})_{mn,kl}
	=\delta^i_{(k}\delta^j_{l)}.
\end{equation}
The trace eigenvalue $-2$ fixes the coefficient
$-1/2$ in its inverse.

The full reference metric velocity is
\begin{equation}
	V^m=(W^{-1})^{mn}(p_n-d_n).
\end{equation}
In particular,
\begin{equation}
	\dot B^i(X)=B^i{}_{,m}(X)V^m
\end{equation}
for a constant body generator $X$ in the zero-shift normal
calculation; for nonzero shift the Lie-transport contribution
must be restored.

The reference torsion decomposition gives
\begin{equation}
	\mathcal L_{\mathrm{ref}}
	=
	\frac{\varphi N\sqrt q}{2\kappa}
	\left(K_{ij}K^{ij}-K^2-{}^{(3)}T\right)
	+\frac{\sqrt q}{\kappa}\varphi t^i\partial_iN
	+\frac{N\sqrt q}{2\kappa}U
	+\mathcal L_F.
\end{equation}
Here $t^i$ is the reference spatial torsion trace.
The shift-subtracted metric inversion is
\begin{equation}
	K_{ij}
	=
	\frac{2\kappa}{\varphi\sqrt q}
	\left(P_{ij}-\frac12q_{ij}P\right).
	\label{eq:appF_metric_inversion}
\end{equation}

Integration by parts of the lapse-gradient term
produces the lapse-linear normal Hamiltonian of
Eq.~\eqref{eq:sec6_normal_density}.

For the scalar Euler equation, put
\begin{equation}
	\frac{J^\mu}{\kappa}=H^{\nu\mu}(A_\nu).
\end{equation}
Variation with respect to $\varphi$ yields
\begin{equation}
	E_\varphi
	=
	-\frac{h}{2\kappa}(\bar T-U')
	-\partial_\mu[H^{\nu\mu}(A_\nu)].
\end{equation}
Using flatness,
\begin{equation}
	\dot A_i=D_i^AA_0,\qquad
	D_i^AX=\partial_iX+[A_i,X],
\end{equation}
and $H^{i0}=-B^i$, one obtains
\begin{equation}
	E_\varphi=f_\varphi+\beta(\Omega),
\end{equation}
where
\begin{align}
	f_\varphi={}&
	-\frac{h}{2\kappa}(\bar T-U')
	+\dot B^i(A_i)+\beta(A_0^c)
	\nonumber\\
	&-\partial_j[H^{ij}(A_i)].
	\label{eq:appF_scalar_source}
\end{align}
The derivative terms $B^i(\partial_iA_0)$ cancel
between the temporal and spatial divergences.

These identities establish the regular reference
Legendre inversion for $\varphi\neq0$ and the
lapse-linear canonical representative
\cite{ADM1962,Oko2013,MalufRocha2001}.

\paragraph{Lorentz sources and projectable secondary constraint.}
\label{app:lorentz_secondary_source}

For a constant body generator $X$, the Lorentz source is
\begin{align}
	f(X)={}&
	-B^i{}_{,m}(X)V^mg_i
	-(\partial_iH^{ij})(X)g_j
	\nonumber\\
	&+H^{ij}([A_i,X])g_j
	+a([A_0^c,X]).
	\label{eq:appF_lorentz_source}
\end{align}
The second derivative of $\varphi$ cancels by the
antisymmetry of $H^{ij}$.

Let
\begin{equation}
	C[\eta]=\int_\Sigma d^3x\,
	\eta^I(L_I-a_I)
\end{equation}
for an external Lorentz smearing $\eta$.
Its Hamiltonian flow satisfies
\begin{align}
	\delta_\eta\Lambda&=\Lambda\eta,\\
	\delta_\eta A_i&=D_i^A\eta,\\
	\delta_\eta p_m&=B^i{}_{,m}(\eta)g_i,\\
	\delta_\eta p'_\varphi&=-\partial_i[B^i(\eta)],\\
	\delta_\eta V^m
	&=(W^{-1})^{ml}B^i{}_{,l}(\eta)g_i.
\end{align}

The scalar momentum variation follows by integrating the gradient of $\delta\varphi$ by parts. In particular,
\begin{equation}
 \delta_\eta P_\varphi=-\partial_i[B^i(\eta)]+B^i(D_i^A\eta)=\beta(\eta).
\end{equation}
Smearing derivatives cancel.

Define
\begin{equation}
 u_m(X)=B^i{}_{,m}(X)g_i.
\end{equation}
Direct functional differentiation gives
\begin{align}
	\delta_\eta\beta(X)
	&=B^i([D_i^A\eta,X]),\\
	\delta_\eta f(X)
	&=-u_m(X)(W^{-1})^{ml}u_l(\eta)
	\nonumber\\
	&\quad+H^{ij}([D_i^A\eta,X])g_j
	+a([[A_0^c,\eta],X]).
	\label{eq:appF_lorentz_source_variation}
\end{align}
For field-dependent smearings, the additional terms
are proportional to the primary constraints or their
spatial derivatives and vanish on the smooth primary
surface.

In the local BF-reduced chart, introduce
\begin{equation}
	R^{ij}(X)=\frac{\sqrt q}{\kappa}
	e_I{}^ie_J{}^jX^{IJ}.
\end{equation}
The unit-normal source becomes
\begin{align}
	f_1(X)={}&
	-2B^i{}_{,q_{kl}}(X)K_{kl}g_i
	-\partial_i[R^{ij}(X)g_j]
	\nonumber\\
	&+R^{ij}([A_i,X])g_j.
	\label{eq:appF_unit_normal_source}
\end{align}

For an arbitrary lapse,
\begin{equation}
	f_N(X)=Nf_1(X)-(\partial_iN)R^{ij}(X)g_j.
\end{equation}
Set
\begin{equation}
	k=(a,0),\qquad j=(0,a),\qquad
	n=r_Jk-r_Kj.
\end{equation}
The rotational component of $n$ is parallel to $a$.
Consequently,
\begin{equation}
	R^{ij}(n)g_j=0,
\end{equation}
because the corresponding rotational contraction is
proportional to $a\times a$.

The null compatibility condition is therefore
\begin{equation}
	n^If_N(J_I)=NS,
	\qquad
	\boxed{S=n^If_1(J_I)\approx0.}
	\label{eq:appF_projectable_secondary}
\end{equation}
Here $J_I$ in the contracted expression denotes a
generic element of the ordered six-dimensional Lorentz
basis.

The secondary is independent of the lapse and its
spatial derivatives. The shift contribution is
generated by the compensated spatial transport.

This projectability is essential for separating the
gravitational second-class reduction from the temporal
first-class chains
\cite{Dirac1964,HenneauxTeitelboim1992,
	GolovnevGuzman2021}.

\paragraph{Residual stabilization and functional regularity}
\label{app:residual_stabilization}

On the generic primary stratum,
\begin{equation}
	a^2>0,\qquad r_K^2+r_J^2>0,
\end{equation}
the null generator is
\begin{equation}
	n=r_Jk-r_Kj,
	\qquad [k,j]=0.
\end{equation}
Since $\delta_\eta a=0$ under the primary Lorentz flow,
\begin{equation}
	\delta_\eta n
	=
	kB^i([D_i^A\eta,j])
	-jB^i([D_i^A\eta,k]).
\end{equation}

For $S=n^If_{1I}$, the product rule gives
\begin{equation}
	\delta_\eta S=f_1(\delta_\eta n)+n^I\delta_\eta f_{1I}.
\end{equation}
Choose $\eta=\mu n$. Then
\begin{equation}
	D_i^A(\mu n)=n\partial_i\mu+\mu D_i^An.
\end{equation}
The coefficient of $\partial_i\mu$ is
\begin{align}
	\mathcal L^i={}&
	f_1(k)B^i([n,j])-f_1(j)B^i([n,k])
	\nonumber\\
	&+R^{ij}([n,n])g_j=0.
\end{align}
All three commutators vanish identically.

In the unreduced expression the temporal contribution also vanishes because
\begin{equation}
	a([Y,n])=0
\end{equation}
for every Lorentz generator $Y$, as follows from
$D_6n=0$.

For field-dependent $n$, the extra Hamiltonian-flow terms are proportional to the primaries and their spatial derivatives. On the smooth primary surface $\{\Phi,C[\mu n]\}_*\approx0$, so the paired-primary Dirac correction vanishes weakly. Consequently,
\begin{equation}
	\boxed{
		\{S,C[\mu n]\}_4\approx z\mu,
	}
	\label{eq:appF_ultralocal_residual}
\end{equation}
where
\begin{align}
	z={}&
	f_1(k)B^i([D_i^An,j])
	-f_1(j)B^i([D_i^An,k])
	\nonumber\\
	&-u_m(n)(W^{-1})^{ml}u_l(n)
	+R^{ij}([D_i^An,n])g_j.
	\label{eq:appF_residual_coefficient}
\end{align}

In the unit-normal metric representative the explicit DeWitt contraction in this expression vanishes. Indeed, with $c_q=\sqrt q/\kappa$ and $\widehat g^i=g^i/\sqrt{g_jg^j}$, differentiation at fixed Lorentz generator gives
\begin{align}
 u^{ij}(k)&=\tfrac12 a^2(q^{ij}-\widehat g^i\widehat g^j), & u^{ij}(j)&=0,\\
 u^{ij}(n)&=r_Ju^{ij}(k).
\end{align}
A reference rotation in the triad variation drops out of the contraction with $a_I$. The transverse projector has trace and squared norm both equal to two. Hence
\begin{equation}
 u^{ij}(n)u_{ij}(n)-\tfrac12(q_{ij}u^{ij}(n))^2=0.
\end{equation}
The inverse in Eq.~\eqref{eq:appF_DeWitt_inverse} therefore gives $u(n)W^{-1}u(n)=0$ for unit lapse. Equivalently,
\begin{align}
 z={}&f_1(k)B^i([D_i^An,j])-f_1(j)B^i([D_i^An,k])\notag\\
 &+R^{ij}([D_i^An,n])g_j.
\end{align}
On $S=0$, set $Y_i=r_JD_i^Ak-r_KD_i^Aj$ and
$\chi_f=(r_Kf_1(k)+r_Jf_1(j))/(r_K^2+r_J^2)$. Since $[k,j]=0$, derivatives of $r_K,r_J$ drop out of the commutators, and $f_1(k)=\chi_f r_K$, $f_1(j)=\chi_f r_J$. Consequently,
\begin{equation}
 z\approx R^{ij}([Y_i,n])g_j-\chi_f B^i([Y_i,n]).
\end{equation}
Here $f_1(n)$ denotes contraction of the fixed-basis sources, not differentiation of a field-dependent generator inside the divergence. Metric momenta may still enter through $f_1(k)$.

Thus the residual operator is weakly ultralocal
in this regular chart:
\begin{equation}
	Z(x,y)\approx z(x)\delta^3(x-y).
\end{equation}

Under a smooth nonzero rescaling $n\mapsto\rho n$,
one has, weakly on the secondary surface,
\begin{equation}
	S\mapsto\rho S,\qquad z\mapsto\rho^2z.
\end{equation}
The derivatives of $\rho$ multiply either constraints
or the vanishing commutators above.

For independence, see Sec.~\ref{sec:dirac_bergmann} and Appendix~\ref{app:regular_background_evolution}.

\subsection{Primary pseudoinverse and Schur reduction}
\label{app:stabilized_dirac_inverse}

With $C_7Q_7=Q_7C_7=I_7-\nu\nu^T$, the full Schur inverse follows from Sec.~\ref{subsec:complete_regular_reduction}. In the adapted basis the off-diagonal block is $E=(0,E_S)$ weakly; hence $E\mathcal B^{-1}E^\dagger=0$ by the zero secondary--secondary entry of the residual inverse. The upper-left block is therefore $D_P^{-1}$.

For the product-reduced Poisson tensor $\mathbb J$ (with graded
adjoints on the full super phase space),
the stabilized tensor is
\begin{equation}
	\mathbb J_{\mathrm{st}}
	=
	\mathbb J-\mathbb J L_\chi^\dagger
	C_8^{-1}L_\chi\mathbb J,
	\qquad L_\chi=d\chi.
\end{equation}
It inverts the restricted symplectic form and annihilates $d\chi$
\cite{Dirac1964,HenneauxTeitelboim1992}.

\subsection{Stabilized hypersurface-deformation algebra}
Here weak equality is on $\Gamma_{\rm SC}$; the current calculation must not use $H=D=0$.
\label{app:temporal_primary_currents}

Let
\begin{equation}
	H[N]=\int_\Sigma d^3x\,N\mathcal H_\perp,
	\qquad
	\Phi=(C_I,P_\varphi).
\end{equation}
On the primary surface, write
\begin{equation}
	b[N]=\{H[N],\Phi\}_*
	=Nu+\partial_iN\,v^i.
\end{equation}
The Lorentz components satisfy
\begin{equation}
	u_X=-f_1(X),\qquad
	v_X^i=R^{ij}(X)g_j.
\end{equation}

The scalar primary generates
\begin{align}
	\delta_\rho\varphi&=\rho,\\
	\delta_\rho q_{ij}&=0,\\
	\delta_\rho\Lambda&=0,\\
	\delta_\rho p^{ij}
	&=-\rho B^k{}_{,q_{ij}}(A_k).
\end{align}
Its Lorentz momentum variation must also be retained
when evaluating general bracket numerators.

The scalar flow changes the relative and scalar momenta, but $H[N]$ in the local chart is independent of $L,p'_\varphi$. Its metric momentum variation and all explicit undifferentiated scalar variations contribute only to $u_\varphi$. The term $-N(c_qt^i+R^{ji}(A_j))\partial_i\varphi$ yields, by integration by parts,
\begin{equation}
 v_\varphi^i=c_qt^i+R^{ji}(A_j).
\end{equation}
The identity $-e_I{}^i\partial_j(c_qe_I{}^j)=c_qt^i$ follows by differentiating the inverse triad and its determinant. In the rotational part of $A_j$, the Lorentz commutator gives $e_I{}^i B^j([A_j,K_I])=R^{ji}(A_j)$; its boost part gives zero. Thus the scalar numerator is exactly $e_I{}^i\beta_{K_I}$, including the reference torsion trace.

The triad identities give
\begin{equation}
	v_K^i=0,\qquad
	v_J^i=a\times e^i,\qquad
	v_\varphi^i=e_I{}^i\beta_{K_I}.
	\label{eq:appF_primary_numerator}
\end{equation}
Here $e^i$ denotes the internal vector with components
$e_I{}^i$.

A useful exact identity is
\begin{equation}
	C_7
	\begin{pmatrix}
		e^i\\0\\0
	\end{pmatrix}
	=v^i.
	\label{eq:appF_primary_current_identity}
\end{equation}
Indeed, its Lorentz part is
$(0,a\times e^i)^T$, while its scalar part is
$\beta_K^Te^i$.

Since
\begin{equation}
	Q_7C_7=I_7-\nu\nu^T,
\end{equation}
one obtains
\begin{equation}
	Q_7v^i
	=
	(I_7-\nu\nu^T)
	\begin{pmatrix}
		e^i\\0\\0
	\end{pmatrix}.
\end{equation}
The secondary compatibility condition implies
$u^T\nu\approx0$. Consequently,
\begin{equation}
	\boxed{
		u^TQ_7v^i\approx-e_I{}^if_1(K_I).
	}
	\label{eq:appF_exact_primary_current}
\end{equation}
Moreover,
\begin{equation}
	(v^i)^TQ_7v^j=0,
\end{equation}
because the boost--boost block of $D_6$ vanishes.

Define
\begin{equation}
	s_i=N\partial_iM-M\partial_iN.
\end{equation}
With the row convention $b[N]=\{H[N],\Phi\}_*$ the correction is $+\int b[N]Q_7b[M]^T$, since $\{\Phi,H[M]\}_*=-b[M]^T$. The $NM$ term vanishes by antisymmetry and the double-gradient term vanishes by the boost--boost identity. The two cross terms combine into $s_i u^TQ_7v^i$. The primary Dirac correction to the normal-normal bracket is therefore
\begin{equation}
	\Delta_P[N,M]
	\approx\int_\Sigma d^3x\,s_iJ_P^i,
	\qquad
	J_P^i=-e_I{}^if_1(K_I).
\end{equation}

The remaining residual correction vanishes weakly
for the normal-normal bracket because
\begin{equation}
	\{C_\nu,H[N]\}_4\approx0
\end{equation}
on the secondary surface and the secondary-secondary
entry of $\mathcal B^{-1}$ is zero.

This conclusion does not require
$\{S,H[N]\}_4=0$ before final stabilization.
It follows from the ordered residual inverse and
the null-primary compatibility equation.

\paragraph{Metric and relative Lorentz deformation currents.}
\label{app:metric_lorentz_currents}

Using the spatial curvature identity, write the
normal Hamiltonian as
\begin{align}
	H[N]=\int_\Sigma d^3x\,N
	\bigg[
	&\frac{2\kappa}{\varphi\sqrt q}
	\left(P^{ij}P_{ij}-\frac12P^2\right)
	\nonumber\\
	&-\frac{\sqrt q}{2\kappa}
	(\varphi\,{}^{(3)}R+U)
	\nonumber\\
	&-\left(c_qt^i+R^{ji}(A_j)\right)g_i
	+\mathcal H_F
	\bigg],
\end{align}
where $c_q=\sqrt q/\kappa$.

In a symmetric reference section,
\begin{equation}
	\delta e^I{}_i=\frac12e^{Ij}\delta q_{ij}.
\end{equation}
The inverse-triad variation is
\begin{equation}
	\delta e_I{}^i=-\frac12q^{ik}e_I{}^j\delta q_{kj}.
\end{equation}
It follows that
\begin{equation}
	\delta B^i(K_I)
	=
	\frac12B^i(K_I)q^{kl}\delta q_{kl}
	-\frac{\sqrt q}{2\kappa}
	q^{ik}e_I{}^l\delta q_{kl}.
\end{equation}

Only terms differentiating a lapse can survive antisymmetrization. For the curvature potential the derivative part of the covariant-metric functional derivative is
\begin{equation}
 \frac{\delta H_{\rm curv}[N]}{\delta q_{kl}}\bigg|_{D D(N\varphi)}
 =\frac{\sqrt q}{2\kappa}(q^{kl}D^2-D^kD^l)(N\varphi).
\end{equation}
Contracting with $2MK_{kl}$ and subtracting the exchanged term gives $2P^{ij}D_i s_j+4P^{ij}g_i s_j/\varphi$ before integration by parts. Here $D_i$ acts on tensor densities with their appropriate weight. The curvature variation produces the current
\begin{equation}
	J_{\mathrm{curv}}^i
	=
	-2D_jP^{ij}+\frac4\varphi P^{ij}g_j.
\end{equation}
The derivative part of the torsion-trace variation
is
\begin{equation}
	\delta t_i\big|_{\partial\delta q}
	=
	\frac12q^{jk}
	(\partial_j\delta q_{ki}-\partial_i\delta q_{jk}).
\end{equation}
The lapse-gradient part of the functional derivative of $-\int Nc_qt^ig_i$ is $c_q[g^{(k}D^{l)}N-q^{kl}g^iD_iN]/2$. Its contraction with the kinetic derivative $2MK_{kl}$, followed by antisymmetrization, gives $-c_q(K^{ij}-q^{ij}K)g_j s_i$. After integration by parts, its contribution is
\begin{equation}
	J_{\mathrm{tor}}^i=-\frac2\varphi P^{ij}g_j.
\end{equation}

The metric part of the primary current follows from
\begin{equation}
	2e_I{}^iB^j{}_{,q_{kl}}(K_I)K_{kl}g_j
	=
	\frac{\sqrt q}{\kappa}
	(Kq^{ij}-K^{ij})g_j.
\end{equation}
Using the metric Legendre inversion,
\begin{equation}
	J_{\mathrm{prim,met}}^i
	=-\frac2\varphi P^{ij}g_j.
\end{equation}

The explicit scalar-gradient terms cancel:
\begin{equation}
	\boxed{
		J_{\mathrm{met}}^i
		=
		J_{\mathrm{curv}}^i+J_{\mathrm{tor}}^i
		+J_{\mathrm{prim,met}}^i
		=-2D_jP^{ij}.
	}
\end{equation}

In the product chart the bare relative potential contains no $L$ or $p'_\varphi$; its metric dependence is algebraic and its bare kinetic cross term is symmetric in the lapses. Its surviving relative current is thus supplied by the primary correction. Writing $A_j=A_j^{K_I}K_I+A_j^{J_I}J_I$, only the boost part contributes to $R^{jk}([A_j,K_I])$. The identity $\epsilon_{LIQ}\epsilon_{ABQ}=\delta_{LA}\delta_{IB}-\delta_{LB}\delta_{IA}$ gives the relative Lorentz contraction
\begin{align}
	J_\Lambda^i
	&=-e_I{}^iR^{jk}([A_j,K_I])g_k\\
	&=q^{ik}
	\left[B^j(A_k)g_j-B^j(A_j)g_k\right].
	\label{eq:appF_relative_current}
\end{align}
On the primary surface,
\begin{equation}
	L_X=a(X),\qquad
	p'_\varphi=-B^j(A_j),
\end{equation}
and therefore
\begin{equation}
	\int_\Sigma d^3x\,s_iJ_\Lambda^i
	=
	\int_\Sigma d^3x\,
	\left[L_{V^iA_i}+p'_\varphi V^ig_i\right],
\end{equation}
where $V^i=q^{ij}s_j$.

The rotation compensator contributes no additional
vacuum term on the primary surface because
$B^i(J_I)=0$.

\paragraph{Stabilized metric brackets.}

Since $\{q_{ij},\Phi_A\}_*=0$, the paired-primary reduction leaves metric brackets unchanged. The unit null coefficient is configuration dependent, hence $\{q_{ij},C_\nu\}_4\approx_P0$. The residual inverse has zero secondary--secondary entry, so $\{q_{ij},q_{kl}\}_{\rm st}=0$ on $\Gamma_{\rm SC}$. Differentiating the kinetic density with respect to $p^{ij}$ gives $2NK_{ij}$. The only possible residual correction is $\{q_{ij},S_u\}_4(Z_\nu^\dagger)^{-1}\{C_\nu,H[N]\}_4$, which vanishes by secondary stabilization. Spatial preservation of the ideal gives the spatial bracket. Consequently
\begin{align}
 \{q^{ij},H[N]\}_{\rm st}&=-2NK^{ij}, &
 \{q^{ij},D[V]\}_{\rm st}&=\mathcal L_Vq^{ij}.
\end{align}
These identities enter the coefficient Jacobi test \cite{ADM1962,Teitelboim1973,Oko2014}.

\paragraph{Fermionic contributions to the deformation algebra.}
\label{app:fermionic_deformation}

At a fixed smooth metric $q_0$, choose the local
reference section
\begin{equation}
	e(q,x)=e_0(x)(q_0^{-1}q)^{1/2}.
\end{equation}
Its metric variation is symmetric at $q_0$.
The reference fermionic momentum shift therefore
vanishes there:
\begin{equation}
	a_F^{ij}\big|_{q_0}=0.
\end{equation}
Its metric derivatives need not vanish.

For clarity, use a condensed symmetric metric index $A$ and write
$a_{F,A}=\xi^\dagger m_A\xi$, $P_A=p_A-a_{F,A}$. The
section is symmetric at every spatial point of the chosen metric
$q_0$, so $m_A|_{q_0}=0$ for all spinors, but $\partial_Bm_A$
need not vanish. The product brackets of Sec.~3 then give
\begin{align}
 \{P_A,P_B\}_*&=\xi^\dagger\mathcal F_{AB}\xi, &
 \{\xi,P_A\}_*&=im_A\xi,
\end{align}
with the ultralocal delta distributions understood. The
kinetic--kinetic curvature contribution is
\begin{equation}
 \Delta_{\mathcal F}[N,M]=4\int_\Sigma d^3x\,NM
 K_{ij}K_{kl}\xi^\dagger\mathcal F_{(ij)(kl)}\xi=0,
\end{equation}
by antisymmetry in the metric index pairs. The mixed elementary
spinor--momentum brackets vanish at $q_0$ because $m_A=0$.
Thus neither the curvature nor the mixed elementary brackets
supply a deformation current. This argument does not set the
metric derivatives of the shift to zero.

\paragraph{Metric variation of the Dirac Hamiltonian.}
Put $h_{ij}=\delta q_{ij}$ in this paragraph only, and let $\gamma_{IJi}$
be the spatial Levi--Civita rotation connection. Differentiating
the spatial tetrad postulate in the symmetric section gives
\begin{equation}
 \delta\gamma_{IJi}=\tfrac12 e_I{}^j e_J{}^k
 (D_k h_{ij}-D_j h_{ik}).
 \label{eq:appC_spin_connection_variation}
\end{equation}
Indeed, the variation of the Christoffel symbol contributes
$e_I{}^j e_J{}^k(D_i h_{jk}+D_k h_{ij}-D_j h_{ik})/2$;
the variation of the triad cancels the first term.
The integrated normal Hamiltonian can be written without
lapse derivatives as
\begin{equation}
 H_F[N]=\int d^3x\,N\left[-\frac i2
 \big(\xi^\dagger\alpha^i\nabla_i\xi
 -(\nabla_i\xi)^\dagger\alpha^i\xi\big)
 +m\xi^\dagger\Gamma^{\hat0}\xi\right].
 \label{eq:appC_symmetric_matter}
\end{equation}
At fixed half-density spinor the density-connection variation
cancels between the two derivative terms. The spin-connection
variation is proportional to
$\xi^\dagger\Gamma^{\hat0}\Gamma^{KIJ}\xi\,e_K{}^i
\delta\gamma_{IJi}$, since the anticommutator of the spatial
Clifford vector with the rotation bivector is a Clifford triple.
Writing $D_J h_{KI}=e_J{}^j e_K{}^k e_I{}^l D_j h_{kl}$,
its derivative coefficient is proportional to
\begin{equation}
 \Gamma^{KIJ}(D_J h_{KI}-D_I h_{KJ})=0.
\end{equation}
Both terms vanish separately by symmetry of $h_{KI}$ and
$h_{KJ}$. The remaining variation of $\alpha^i$ is algebraic,
$\delta\alpha^i=-q^{ik}\alpha^j h_{kj}/2$. Consequently at $q_0$
the remaining algebraic metric derivative is the symmetric density
\begin{equation}
 \mathcal E_F^{ij}=\frac{i}{4}q^{k(i}\left[\xi^\dagger\alpha^{j)}\nabla_k\xi-(\nabla_k\xi)^\dagger\alpha^{j)}\xi\right].
 \label{eq:appC_explicit_matter_metric_derivative}
\end{equation}
It is independent of the lapse, and
\begin{equation}
 \left.\frac{\delta H_F[N]}{\delta q_{ij}}\right|_{\xi,q_0}
 =N\mathcal E_F^{ij}.
\end{equation}
The canonical mixed contribution is therefore
\begin{align}
 \Delta_{gF}[N,M]
 &=\{H_g[N],H_F[M]\}_*+\{H_F[N],H_g[M]\}_*\notag\\
 &=\int d^3x\,(-2NK_{ij}M\mathcal E_F^{ij}
 +2MK_{ij}N\mathcal E_F^{ij})=0.
 \label{eq:appC_mixed_cancellation}
\end{align}
Here the elementary mixed brackets have already been evaluated
at $q_0$. The pure matter bracket has no metric-momentum term.

\paragraph{Matter terms in the Dirac corrections.}
In the relative chart $H_F[N]$ depends only on $q,\xi,\xi^\dagger$,
whereas $\Phi=(C_I,\mathcal P_\varphi)$ depends on $q,\Lambda,\varphi$
and their relative or scalar momenta, but not on $p^{ij}$ or the
spinors. Hence $\{H_F[N],\Phi\}_*=0$ exactly in this chart.
The primary correction is therefore precisely the current
computed above, with the full kinetic momentum $P^{ij}$.
The secondary may depend on the spinors through $P^{ij}$;
no vanishing of $\{H_F[N],S_u\}_4$ is assumed. Rather,
$\{H[N],C_\nu\}_4\approx_{\rm SC}0$ and the zero
secondary--secondary entry of the residual inverse eliminate
all residual normal--normal corrections, including the mixed
ones. These statements use the regular projectable normal
representative, not the hypersurface constraint equations.
Since $q_0$ was arbitrary, the symmetric-section calculation
applies locally at any smooth metric. Reference covariance below
transports the result to other smooth reference sections on the
second-class surface.

Let $\nabla_i$ denote the spatial spin covariant
derivative on half-densities and define
\begin{equation}
	\alpha^i=\Gamma^{\hat0}\Gamma^Ie_I{}^i.
\end{equation}
The smeared Dirac operator is
\begin{equation}
	\mathscr D_N
	=
	-i\alpha^i\left(N\nabla_i+\frac12\partial_iN\right)
	+Nm\Gamma^{\hat0}.
	\label{eq:appF_dirac_operator}
\end{equation}
The integrated matter Hamiltonian is
\begin{equation}
	H_F[N]=\int_\Sigma d^3x\,\xi^\dagger\mathscr D_N\xi.
\end{equation}

At a point in normal coordinates and a parallel
spin frame, the second-order terms in
$[\mathscr D_N,\mathscr D_M]$ cancel because their
common coefficient is $NM$.
Write $\mathscr D_N=N\mathscr D_1-i\alpha^i\partial_iN/2$. At the chosen point $\alpha^i\alpha^j=q^{ij}-\Gamma^{IJ}e_I{}^ie_J{}^j$. The first-order coefficient in the commutator is $-V^i\nabla_i$: the antisymmetric Clifford part from $N\mathscr D_1$ cancels against the two half-gradient cross terms. The zero-order Hessian terms give $-(N D^2M-M D^2N)/2=-D_iV^i/2$, since the Hessians of scalar lapses are symmetric. The commutator of the two half-gradient terms gives $[\Gamma^{IJ}e_I{}^ie_J{}^j(\partial_iN)(\partial_jM)]/2$. Terms proportional to $NM$, including the common spin curvature, cancel. The mass cross term vanishes by
\begin{equation}
	\{\alpha^i,\Gamma^{\hat0}\}_+=0.
\end{equation}

The covariant result is
\begin{align}
	[\mathscr D_N,\mathscr D_M]
	={}&-V^i\nabla_i-\frac12D_iV^i
	\nonumber\\
	&+\frac12\Gamma^{IJ}e_I{}^ie_J{}^j
	(\partial_iN)(\partial_jM),
	\label{eq:appF_dirac_commutator}
\end{align}
where
\begin{equation}
	V^i=q^{ij}(N\partial_jM-M\partial_jN).
\end{equation}

The reference compensator is
\begin{equation}
	r_V=-V^i\gamma_i+\Omega_V,
\end{equation}
with
\begin{equation}
	(\Omega_V)_{IJ}
	=
	\frac12e_I{}^ie_J{}^j(D_jV_i-D_iV_j).
\end{equation}
For the vector field $V$ above,
\begin{equation}
	(\Omega_V)_{IJ}\Gamma^{IJ}
	=
	-2\Gamma^{IJ}e_I{}^ie_J{}^j
	(\partial_iN)(\partial_jM).
\end{equation}

With
\begin{equation}
	\Sigma(X)=-\frac14X_{IJ}\Gamma^{IJ},
\end{equation}
the negative commutator becomes
\begin{equation}
	-[\mathscr D_N,\mathscr D_M]
	=
	V^i\partial_i+\frac12\partial_iV^i-\Sigma(r_V).
\end{equation}

For even bilinears $F_A=\int\xi^\dagger A\xi$ the reduced graded bracket is $\{F_A,F_B\}_{D,F}=-i\int\xi^\dagger[A,B]\xi$ on the common test domain. The sign follows from $\{\xi,\xi^\dagger\}_{D,F}=-i\delta^3$. The graded bilinear bracket therefore yields
\begin{align}
	\{H_F[N],H_F[M]\}_{D,F}
	={}&
	\frac i2\int_\Sigma d^3x\,
	\left[
	\xi^\dagger V^i\partial_i\xi
	-(V^i\partial_i\xi^\dagger)\xi
	\right]
	\nonumber\\
	&-i\int_\Sigma d^3x\,
	\xi^\dagger\Sigma(r_V)\xi.
\end{align}
This is precisely the compensated fermionic spatial
generator.

Combining the metric, Lorentz, and fermionic contributions,
including Eq.~\eqref{eq:appC_mixed_cancellation} and both
Dirac corrections, gives
\begin{equation}
	\boxed{
		\{H[N],H[M]\}_{\mathrm{st}}
		\approx D[V].
	}
\end{equation}

\paragraph{Spatial preservation of the constraint ideal.}\label{app:spatial_ideal}
Write $r=r_V$, $\delta_V F=\{F,D[V]\}_*$ and $\mathscr L_V^{(w)}=V^i\partial_i+w\partial_iV^i$. For fixed internal $X$, the compensated cotangent lift gives
\begin{align}
 \delta_V A_i&=\mathcal L_V A_i+D_i^A r, &\delta_V g_i&=\mathcal L_V g_i,\\
 \delta_V B^i(X)&=\mathcal L_V^{\rm dens}B^i(X)+B^i([r,X]),\\
 \delta_V L_X&=\mathscr L_V^{(1)}L_X+L_{[r,X]}.
\end{align}
Here $\mathcal L_V^{\rm dens}$ is the vector-density Lie derivative. Contraction with $g_i$ and subtraction yield
\begin{equation}
 \delta_V C_X=\mathscr L_V^{(1)}C_X+C_{[r,X]},\qquad
 \delta_V\mathcal P_\varphi=\mathscr L_V^{(1)}\mathcal P_\varphi.
\end{equation}
The latter follows because the commutator terms in $\delta_V B^i(A_i)$ cancel and $B^i(\partial_i r)=0$ for a rotation. In $\delta_V\beta(X)$ the terms $-B^i([\partial_i r,X])$ and $+B^i([\partial_i r,X])$ cancel; Jacobi combines the remaining commutators into $B^i([A_i,[r,X]])$. Thus $a,\beta$ transform as internal densities of weight one, with the same orthogonal rotation. Consequently $r_K,r_J$ have weight two, $n$ weight three and $\nu=n/\|n\|$ weight zero. The contraction $C_\nu$ has weight one.

The spatial generator of Sec.~5 is the compensated cotangent lift, including the half-density spinor. The boundary translation $\mathcal G_\partial=-\int\varphi B^i(A_i)$ is invariant: the commutator terms cancel as above, giving $\delta_V\mathcal G_\partial=-\int\partial_i(V^i\varphi B^j(A_j))=0$. Hence its exact momentum translation commutes with the cotangent lift. For a time-independent spatial transformation $\Phi$, covariance of the translated Lagrangian (including the Levi--Civita Dirac term) implies $\mathbb FL'\circ T\Phi=T^*\Phi\circ\mathbb FL'$, where $T^*\Phi$ denotes the forward cotangent lift. This follows by differentiating the invariant Lagrangian in the velocity directions. The Legendre energy is invariant under the same lift; on the regular primary image its projectable normal representative is unique modulo primary constraints, provided the lapse and shift transform as scalar and vector and the boundary representative is retained. Returning the pulled-back triad to the reference section uses the rotation of Appendix~\ref{app:reference_section_covariance}. Its canonical lift shifts $P^{ij}$ by $C_{\alpha_{ij}}$, so the kinetic density changes only by primary terms. The curvature and scalar potential are scalar densities, and the combination $c_qt^i+R^{ji}(A_j)$ is a reference-independent vector density. The Dirac normal operator transforms by the spin lift and half-density pullback, with lapse transformed as a scalar. Thus, with the boundary representative and spinor shift consistently included,
\begin{equation}
\{H[N],D[V]\}_*+H[\mathcal L_VN]\approx_P0.
\end{equation}
For the secondary, use $\{H[N],C_\nu[\mu]\}_*\approx_P-\int N\mu S_u$, $S_u=S/\|n\|$, where $\approx_P$ restricts only to the primary surface. The null-primary and spatial flows are tangent to the primary surface, so Jacobi applied to the preceding weak identities gives
\begin{equation}
 -\int N\mu\,\delta_V S_u\approx_P\int S_u\mathcal L_V(N\mu),\qquad
 \delta_V S_u\approx_P\mathscr L_V^{(1)}S_u,\quad
 \delta_V S\approx_P\mathscr L_V^{(4)}S.
\end{equation}
Hence $\{\chi_A,D[V]\}_*$ belongs to the differential constraint ideal and $\{D[V],F\}_{\rm st}\approx_{\rm SC}\{D[V],F\}_*$. The spatial deformation brackets follow from the cotangent-lift algebra and the stated scalar-density property. No hypersurface constraint is imposed \cite{Teitelboim1973,HenneauxTeitelboim1992,Oko2014}.

\subsection{Reference covariance and constraint determinants}
\label{app:reference_section_covariance}

Consider two smooth local reference triads related by
\begin{equation}
	\widetilde e=O(q,x)e,\qquad O\in SO(3),
\end{equation}
with a smooth spin lift $U=S(O)$.

The physical coframe and spinor are unchanged under
\begin{equation}
	\widetilde\Lambda=\Lambda O^{-1},
	\qquad
	\widetilde\xi=U\xi.
\end{equation}
Define
\begin{equation}
	\alpha=O^{-1}\delta O,\qquad
	\eta_i=O^{-1}\partial_iO.
\end{equation}
Then
\begin{align}
	\widetilde A_i&=\operatorname{Ad}_O(A_i-\eta_i),\\
	\widetilde\Lambda^{-1}\delta\widetilde\Lambda
	&=
	\operatorname{Ad}_O
	(\Lambda^{-1}\delta\Lambda-\alpha).
\end{align}

The canonical lift is
\begin{align}
	\widetilde L_{\operatorname{Ad}_OX}&=L_X,\\
	\widetilde p^{ij}
	&=
	p^{ij}+L_{\alpha_{ij}}
	-i\xi^\dagger\Sigma(\alpha_{ij})\xi,
\end{align}
where
\begin{equation}
	\alpha_{ij}=O^{-1}\frac{\partial O}{\partial q_{ij}}.
\end{equation}
The Lorentz one-form changes by $-L_\alpha$, while
the fermionic one-form changes by
$i\xi^\dagger\Sigma(\alpha)\xi$.
Both terms are cancelled by the metric momentum
translation.

The reference fermionic shift transforms as
\begin{equation}
	\widetilde a_F^{ij}
	=
	a_F^{ij}-i\xi^\dagger\Sigma(\alpha_{ij})\xi.
\end{equation}
Consequently,
\begin{equation}
	\widetilde P^{ij}=P^{ij}+L_{\alpha_{ij}}
	=P^{ij}+C_{\alpha_{ij}},
\end{equation}
because $\alpha_{ij}$ is a rotation.

The primary relations satisfy
\begin{equation}
	\widetilde C_{\operatorname{Ad}_OX}=C_X,
	\qquad
	\widetilde P_\varphi=P_\varphi.
\end{equation}

For the spatial compensator,
\begin{equation}
	\widetilde r_V
	=
	\operatorname{Ad}_O(r_V+\eta_V-\alpha_V),
\end{equation}
where
\begin{equation}
	\eta_V=V^i\eta_i,\qquad
	\alpha_V=O^{-1}\delta_qO[\mathcal L_Vq].
\end{equation}
The induced Lorentz and spinor translations cancel
the metric momentum translation in the spatial
generator. Thus
\begin{equation}
	\widetilde D[V]=D[V].
\end{equation}

The spatial torsion trace transforms as
\begin{equation}
	\widetilde t^i=t^i+e_I{}^je_J{}^i(\eta_j)^{IJ},
\end{equation}
while
\begin{equation}
	\widetilde R^{ji}(\widetilde A_j)
	=
	R^{ji}(A_j)-R^{ji}(\eta_j).
\end{equation}
Hence
\begin{equation}
	c_q\widetilde t^i+
	\widetilde R^{ji}(\widetilde A_j)
	=
	c_qt^i+R^{ji}(A_j).
\end{equation}

Writing $\ell^{ij}=C_{\alpha_{ij}}$, the normal
Hamiltonian difference is
\begin{align}
	\widetilde H[N]-H[N]
	=\int_\Sigma d^3x\,N
	\bigg[
	2K_{ij}\ell^{ij}
	+\frac{2\kappa}{\varphi\sqrt q}
	\left(\ell_{ij}\ell^{ij}-\frac12\ell^2\right)
	\bigg].
\end{align}
It belongs to the primary constraint ideal.

Null projection removes this primary-ideal change of source; stabilized surfaces and brackets transform as in Sec.~\ref{subsec:reference_section_independence}.

Time-dependent reference changes additionally require the time-dependent canonical Hamiltonian term
\cite{Dirac1964,HenneauxTeitelboim1992,
	DonnellyFreidel2016,GolovnevGuzman2021}.

\paragraph{Constraint determinants and measure variations}
\label{app:constraint_measure_variations}

In the orthonormal basis of Sec.~\ref{subsec:regular_constraint_measure}, a unit-determinant congruence removes $\beta_R$ because $\beta_R^TD_R^{-1}\beta_R=0$. The remaining scalar pair has entries $\pm\sqrt\sigma$, giving $\det D_P=(a^2)^2\sigma$, independently of the residual self-bracket.

\paragraph{Intrinsic primary inverse variation.}

For a smooth constant-rank perturbation of $C_7$,
let
\begin{equation}
	P_\nu=\nu\nu^T,\qquad \nu^T\nu=1.
\end{equation}
Differentiating $C_7\nu=0$ gives
\begin{equation}
	\boxed{
		\delta\nu=-Q_7(\delta C_7)\nu.
	}
	\label{eq:appF_null_variation}
\end{equation}
The normalization condition implies
$\nu^T\delta\nu=0$.

The Moore--Penrose variation is
\begin{align}
	\delta Q_7={}&
	-Q_7(\delta C_7)Q_7
	+Q_7^2(\delta C_7)P_\nu
	+P_\nu(\delta C_7)Q_7^2.
	\label{eq:appF_inverse_variation}
\end{align}
Equation (C.137) follows by differentiating
$C_7Q_7=I-P_\nu$ and
$Q_7P_\nu=P_\nu Q_7=0$, using
\[
\delta P_\nu
=-Q_7(\delta C_7)P_\nu
-P_\nu(\delta C_7)Q_7.
\]
The formula applies to variations preserving
the rank-six stratum.

Furthermore,
\begin{equation}
	\delta P_a
	=
	\frac{\delta a\,a^T+a\,\delta a^T}{a^2}
	-2\frac{a^T\delta a}{a^2}P_a,
\end{equation}
and
\begin{align}
	\delta\beta_0&=(\delta P_0)\beta+P_0\delta\beta,\\
	\delta\sigma&=2\beta_0^T\delta\beta_0.
\end{align}
Thus
\begin{equation}
	\boxed{
		\delta\ln\operatorname{pdet}C_7
		=
		4\frac{a^T\delta a}{a^2}
		+\frac{\delta\sigma}{\sigma}.
	}
\end{equation}

For the determinant factor alone,
\begin{equation}
	\delta\ln\mu_{\mathrm{det}}
	=
	\frac12\operatorname{Tr}
	(C_8^{-1}\delta C_8).
\end{equation}
This trace identity is exact for the
finite-dimensional regulated constraint matrix.
Its continuum interpretation requires a
compatible determinant and trace prescription;
no unregularized functional trace is assumed
to exist. In the ultralocal regular chart,
\begin{equation}
	\delta\ln\mu_{\mathrm{det}}
	=
	\operatorname{Tr}_\Sigma
	\left[
	\frac{\delta(a^2)}{a^2}
	+\frac12\frac{\delta\sigma}{\sigma}
	+\frac{\delta z_\nu}{z_\nu}
	\right].
	\label{eq:appF_measure_variation}
\end{equation}
The delta distributions must be varied separately; their basis transformation cancels that of the determinant factor as in Sec.~\ref{subsec:regular_constraint_measure}.

A vanishing determinant signals chart failure, not zero physical weight
\cite{Dirac1964,HenneauxTeitelboim1992,HT1990}.

\section{Classical BV master equation and reducibility}
\label{app:BV-proof}
\subsection{Total-degree algebra and canonical odd symplectic structure}
For the superfields of Sec.~\ref{sec:internalBVrevised}, the product is $(\alpha_p a)(\beta_q b)=(-1)^{g_aq}(\alpha_p\wedge\beta_q)ab$, and the bracket uses total degree $p+g_a$. The field degrees $(p,g)$ are $(1,0),(2,0),(0,1),(1,1),(0,2)$ for $(\omega,\lambda,c,\eta,\rho)$; their antifields have degrees $(4-p,-1-g)$. Use the canonical component antibracket $F\overleftarrow\delta_\Phi\overrightarrow\delta_{\Phi^*}G-F\overleftarrow\delta_{\Phi^*}\overrightarrow\delta_\Phi G$, with ordinary antifield-first wedge dualization. For $s=(S,\cdot)$ the linear term is $(-1)^{\epsilon_\Phi}\Phi^*s\Phi$. Their ordinary component action is
\begin{align}
S_{\rm BF}^{\rm BV}=\int_M\big[&\langle\lambda,R\rangle-\langle\omega^*,D_\omega c\rangle-\tfrac12\langle c^*,[c,c]\rangle+\langle\lambda^*,[c,\lambda]+D_\omega\eta\rangle-\langle\eta^*,[c,\eta]+D_\omega\rho\rangle\nonumber\\
&+\langle\rho^*,[c,\rho]\rangle-\tfrac12\langle\rho,[\lambda^*,\lambda^*]\rangle\big].
\end{align}
For Lie-algebra-valued homogeneous forms $X=X^aT_a$ and $Y=Y^bT_b$, define the ordinary bracket by
\[
[X,Y]_o=X^a\wedge Y^b[T_a,T_b],
\]
where the form product retains the ordinary Grassmann ordering of the coefficients. Thus
\[
[X,Y]_o
=-(-1)^{p_Xp_Y+g_Xg_Y}[Y,X]_o.
\]
 For homogeneous forms the total and ordinary brackets obey
\begin{equation}
 [X,Y]_{\rm t}=(-1)^{g_Xp_Y}[X,Y]_{\rm o}.
 \label{eq:BV_total_ordinary_conversion}
\end{equation}
Expanding the signed superfields and integrating by parts gives the component action above. The simultaneous reversal of $\rho,\rho^*$ preserves their canonical pairing. For the signed embedding $\mathscr I:(\Phi,\Phi^*)\mapsto(\mathcal A,\mathcal B)$, define the transported odd symplectic potential using ordinary antifield-first pairings:
\begin{align}
\Theta_{\rm SF}=\int_M\big[\langle\mathcal B_3,\delta\mathcal A_1\rangle+\langle\mathcal A_2,\delta\mathcal B_2\rangle-\langle\mathcal B_4,\delta\mathcal A_0\rangle
+\langle\mathcal A_3,\delta\mathcal B_1\rangle+\langle\mathcal A_4,\delta\mathcal B_0\rangle\big]_{\rm can}.
\end{align}
Subscripts denote form degrees. Substitution gives $\mathscr I^*\Theta_{\rm SF}=\int_M\sum_\Phi\langle\Phi^*,\delta\Phi\rangle_{\rm can}$, hence $\mathscr I^*\delta\Theta_{\rm SF}=\Omega_{\rm can}$. This is canonical transport, not an independent variation-bigraded AKSZ identification. The component calculation below uses the canonical antibracket; boundary terms are discarded \cite{BV1981,BVReducible1983,HenneauxTeitelboim1992}.

\paragraph{Master equation.}
Set $\mathcal F=d\mathcal A+\tfrac12[\mathcal A,\mathcal A]$ and $\mathcal D=d+[\mathcal A,\cdot]$. The graded Jacobi identity implies
\begin{equation}
	\mathcal D\mathcal F=0,\qquad \mathcal D^2X=[\mathcal F,X].
\end{equation}
The total differential $Q=-(-1)^p s$ on form-degree $p$ coefficients obeys $Qd+dQ=0$. The superfield differential is $Q\mathcal A=-\mathcal F$, $Q\mathcal B=-\mathcal D\mathcal B$, hence $Q^2\mathcal A=-\mathcal D\mathcal F=0$ and $Q^2\mathcal B=[\mathcal F,\mathcal B]-\mathcal D^2\mathcal B=0$. 

\paragraph{Component projections and reducibility.}
Write $\mathcal F_p$ and $\mathcal G_p$ for the form-degree $p$ projections of $\mathcal F$ and $\mathcal D\mathcal B$. With total products throughout, the signed embedding gives
\begin{align}
\mathcal F_0&=[c,c]/2,&\mathcal F_1&=dc+[c,\omega],\\
\mathcal F_2&=R+[c,\lambda^*],&\mathcal F_3&=D_\omega\lambda^*+[c,\eta^*],\\
\mathcal F_4&=D_\omega\eta^*-[c,\rho^*]+[\lambda^*,\lambda^*]/2,\\
\mathcal G_0&=-[c,\rho],&\mathcal G_1&=-D_\omega\rho+[c,\eta],\\
\mathcal G_2&=D_\omega\eta+[c,\lambda]-[\lambda^*,\rho],\\
\mathcal G_3&=D_\omega\lambda+[c,\omega^*]+[\lambda^*,\eta]-[\eta^*,\rho],\\
\mathcal G_4&=D_\omega\omega^*-[c,c^*]+[\lambda^*,\lambda]+[\eta^*,\eta]+[\rho^*,\rho].
\end{align}
The ten projected transformations are compactly encoded by
\begin{align}
 s(c,\omega,\lambda^*,\eta^*,\rho^*)&=(\mathcal F_0,-\mathcal F_1,\mathcal F_2,-\mathcal F_3,-\mathcal F_4),\\
 s(\rho,\eta,\lambda,\omega^*,c^*)&=(-\mathcal G_0,-\mathcal G_1,\mathcal G_2,-\mathcal G_3,-\mathcal G_4).
\end{align}

\paragraph{Canonical component verification.}
With the ordinary antifield-first dualization,
the convention $s=(S,\cdot)$ gives
\[
s\Phi=-\frac{\delta_R S}{\delta\Phi^*},
\qquad
s\Phi^*=\frac{\delta_R S}{\delta\Phi}.
\]
For the even master action, right and left
derivatives differ by $(-1)^{\epsilon_Z}$
for a variable $Z$.
Equivalently,
\[
\delta S
=
\int_M\sum_\Phi
\left[
\langle s\Phi^*,\delta\Phi\rangle
-\langle s\Phi,\delta\Phi^*\rangle
\right]_{\rm can},
\]
with variations ordered on the right in
canonical component coordinates.
Direct differentiation then yields the
following ordinary component transformations.
\begin{align}
 sc&=\tfrac12[c,c],&s\omega&=-D_\omega c,\\
 s\lambda&=[c,\lambda]+D_\omega\eta-[\lambda^*,\rho],\\
 s\eta&=[c,\eta]+D_\omega\rho,&s\rho&=[c,\rho],\\
 s\lambda^*&=R+[c,\lambda^*],\\
 s\eta^*&=-D_\omega\lambda^*+[c,\eta^*],\\
 s\rho^*&=-D_\omega\eta^*+[c,\rho^*]-\tfrac12[\lambda^*,\lambda^*],\\
 s\omega^*&=-D_\omega\lambda+[c,\omega^*]+[\lambda^*,\eta]+[\eta^*,\rho],\\
 sc^*&=-D_\omega\omega^*+[c,c^*]-[\lambda^*,\lambda]-[\eta^*,\eta]-[\rho^*,\rho].
\end{align}
For the connection derivative, use $\delta R=D_\omega\delta\omega$ and $\delta(D_\omega X)=[\delta\omega,X]$. Covariant integration by parts and invariance of the pairing give
\begin{equation}
 \delta_\omega S=\int_M\langle-D_\omega\lambda+[c,\omega^*]+[\lambda^*,\eta]+[\eta^*,\rho],\delta\omega\rangle.
\end{equation}
For the odd ghost derivative, move $\delta c$ to the right of the Grassmann coefficients before extracting the coefficient. The quadratic vertex varies as $-\int\langle\delta\lambda^*,[\lambda^*,\rho]\rangle$ with respect to $\lambda^*$. Equation~\eqref{eq:BV_total_ordinary_conversion} converts all ten expressions into the signed projections above; in particular the brackets $[c,\eta]$, $[c,\omega^*]$ and $[\lambda^*,\eta]$ reverse sign.

Setting antifields to zero gives $s_0^2\lambda=[R,\rho]$ and $s_0^2=0$ on the other fields. The full multiplier transformation satisfies
\begin{align}
 s^2\lambda={}&\tfrac12[[c,c],\lambda]-[c,[c,\lambda]]
 +D_\omega[c,\eta]-[D_\omega c,\eta]-[c,D_\omega\eta]\nonumber\\
 &+D_\omega^2\rho-[R,\rho]
 +[c,[\lambda^*,\rho]]-[[c,\lambda^*],\rho]+[\lambda^*,[c,\rho]]=0.
\end{align}
The four groups vanish by Jacobi, the covariant Leibniz rule, the curvature identity and graded Jacobi, respectively. Equivalently, writing $R^i{}_{\alpha}Z^\alpha{}_a=M_a{}^{ij}E_j$ for the reducibility identity, the contribution $-\lambda_i^*M_a{}^{ij}E_j\rho^a$ to $(S_1,S_1)/2$ cancels $(S_0,S_2)$ for $S_2=-\lambda_i^*\lambda_j^*M_a{}^{ij}\rho^a/2$. The remaining terms cancel by Lorentz covariance and $(S_2,S_2)=0$.
The super-Bianchi identities now imply nilpotency on all canonical components. By $s^2X=((S,S),X)/2$, nondegeneracy and the ghost number of the defect imply $(S,S)=0$ locally, modulo boundary terms. On a flat background the shift complex is the twisted de Rham sequence $\Omega^0(\mathrm{ad})\xrightarrow{D_\omega}\Omega^1(\mathrm{ad})\xrightarrow{D_\omega}\Omega^2(\mathrm{ad})$; its cohomology and boundary zero modes require separate treatment \cite{BVReducible1983,BarnichBrandtHenneaux2000}.

\subsection{Coframe and Dirac representation extension}
For $S_m=S_{\rm grav}+S_D$, Sec.~2 establishes Lorentz invariance and spin representation closure; the multiplier shift acts trivially. With the ordered term \eqref{eq:BVrevisedrep}, $sc=c^2$ implies $s^2h=0$ and $\Sigma(sc)=\Sigma(c)^2$ implies $s^2\Psi=s^2\bar\Psi=0$. In particular $s(\bar\Psi\Sigma(c))=(s\bar\Psi)\Sigma(c)-\bar\Psi\Sigma(sc)=0$. Noether invariance cancels ghost-linear terms and representation closure cancels ghost-quadratic terms. Thus
\begin{equation}
	S_{\rm int}^{\rm BV}=S_m+S_{\rm BF}^{\rm BV}+S_{\rm rep},\qquad (S_{\rm int}^{\rm BV},S_{\rm int}^{\rm BV})=0.
\end{equation}
For each internal antifield $Z^*$ the coupled equation adds $\delta_R(S_m+S_{\rm rep})/\delta Z$ to the pure BF contribution (zero for variables outside that sector). The auxiliary scalar has trivial internal transformation. The proof is independent of Hessian rank for differentiable $F$ \cite{Krssak2019,KrssakSaridakis2016,Maluf2003}.

\paragraph{Diffeomorphisms.}
Let $v$ be the odd diffeomorphism ghost and $Z$ the internal fields and ghosts (including $\phi$). Use the Lie derivative on forms and coordinate-scalar spinor components in a local trivialization. In a canonical convention with $s_v Z=\mathcal L_v Z$ and $s_v v=\tfrac12[v,v]$, define the cotangent-lift Hamiltonian
\begin{equation}
	S_{\rm diff}=\int_M\left(\sum_Z(-1)^{\epsilon_Z}\langle Z^*,\mathcal L_v Z\rangle_{\rm can}-\tfrac12 v^*_\mu[v,v]^\mu\right),
\end{equation}
with ghost-left Lie derivatives. Here $[v,v]^\mu=2v^\nu\partial_\nu v^\mu$. Indeed, $s_v^2Z=\mathcal L_{s_v v}Z-\mathcal L_v^2 Z=0$ and $s_v^2v=0$ by Jacobi. The cotangent lift acts on antifields as dual tensor densities. Its action on $S_{\rm int}^{\rm BV}$ is the integral of a Lie derivative of a top form, hence vanishes modulo the specified boundary term. Naturality gives $s_{\rm int}\mathcal L_vZ=-\mathcal L_v s_{\rm int}Z$ and $s_v s_{\rm int}Z=\mathcal L_v s_{\rm int}Z$. Thus the formal local functional $S_{\rm int}^{\rm BV}+S_{\rm diff}$ satisfies the classical master equation in this convention. Properness, global patching and boundary symmetries remain separate questions \cite{BV1981,BarnichBrandtHenneaux2000,DonnellyFreidel2016}.

\section{Regular constrained initial data and evolution}
\subsection{Explicit regular constrained data and initial Hamiltonian evolution}
\label{app:regular_background_evolution}

For $F=T+\alpha T^2$, $U=(\varphi-1)^2/(4\alpha)$, choose Cartesian initial data
\begin{align}
 q_{ij}&=\delta_{ij}, & P^{ij}&=0, & \xi&=0,\\
 \varphi&=\varphi_0+gs, & s&=x^3, & g&\ne0,\\
 \Lambda_0&=\exp[x^1t(s)J_2], & A_1&=tJ_2, & A_2&=0,\quad A_3=rJ_2,
\end{align}
where $r=x^1t'$.

The primary coefficients are
\begin{equation}
	a=\frac g\kappa(0,0,1)^T,
	\qquad
	\beta=\frac1\kappa(-r,0,t,0,0,0)^T.
\end{equation}
The explicit primary matrix is
\begin{equation}
	C_7=\frac1\kappa
	\begin{pmatrix}
		0&0&0&0&-g&0&r\\
		0&0&0&g&0&0&0\\
		0&0&0&0&0&0&-t\\
		0&-g&0&0&0&0&0\\
		g&0&0&0&0&0&0\\
		0&0&0&0&0&0&0\\
		-r&0&t&0&0&0&0
	\end{pmatrix}.
	\label{eq:appF_explicit_C7}
\end{equation}
Its unit null direction may be oriented as $J_3$. Relative to $\nu=n/\|n\|$, this may reverse both $C_\nu$ and $S_u$; their simultaneous sign reversal leaves $Z_\nu$, the stabilized bracket and the constraint determinant unchanged.

The normal constraint is
\begin{equation}
	\mathcal H_\perp=-\frac{U+2tg}{2\kappa}.
\end{equation}
It vanishes for
\begin{equation}
	\boxed{
		t=-\frac{U}{2g}
		=-\frac{(\varphi-1)^2}{8\alpha g}.
	}
\end{equation}
The spatial constraints, scalar primary momentum and Lorentz compatibility sources vanish for these data at unit lapse and zero shift.

The invariants of Sec.~5 become
\begin{align}
	a^2&=\frac{g^2}{\kappa^2},\\
	\sigma&=\frac{t^2}{\kappa^2},\\
	\operatorname{pdet}C_7&=\frac{g^4t^2}{\kappa^6},\\
	z_0&=-\frac{gt}{\kappa}
	=\frac{(\varphi-1)^2}{8\alpha\kappa}.
\end{align}
Both residual constraints are unit-normalized
along $J_3$, with a consistent orientation.
Thus $z_0$ denotes the initial-data value
of the normalized coefficient $z_\nu$ of
Eq.~(5.66), rather than the polynomial
coefficient $z$.

The unit-normal determinant factor is $a^2\sqrt\sigma|z_0|=|g|^3t^2/\kappa^4=|g|(\varphi-1)^4/(64\alpha^2\kappa^4)$.

\paragraph{Initial multipliers.}

For $N=1$ and $N^i=0$, the primary consistency
equations give
\begin{equation}
	\dot\varphi=0,\qquad
	\Omega=bK_3,
\end{equation}
where
\begin{equation}
	\boxed{
		b=-\frac{U'}{2t}
		=\frac{gU'}{U}
		=\frac{2g}{\varphi-1}.
	}
\end{equation}
The final rotational multiplier vanishes for the
initial data under consideration.

From $\dot A_i=D_i^A\Omega$,
\begin{equation}
 \dot A_1=tbK_1,\qquad \dot A_2=0,\qquad \dot A_3=b'K_3+rbK_1.
\end{equation}

Useful identities are
\begin{equation}
	tb=t'=-\frac{\varphi-1}{4\alpha},
	\qquad
	b'=-\frac{2g^2}{(\varphi-1)^2}.
\end{equation}

\paragraph{Metric evolution.}

Initially $\dot q_{ij}=0$.
The Hamiltonian momentum equation, including the
primary Lorentz multiplier contribution, gives
\begin{equation}
	\boxed{
		\dot P^{ij}
		=
		\frac1{4\kappa}
		\operatorname{diag}(U+2bg,2bg,U).
	}
	\label{eq:appF_background_momentum_evolution}
\end{equation}
Differentiating the metric Legendre inversion yields
\begin{equation}
	\ddot q_{ij}
	=
	\frac{4\kappa}{\varphi}
	\left(\dot P_{ij}-\frac12\delta_{ij}\dot P\right).
\end{equation}
Therefore
\begin{equation}
	\boxed{
		\ddot q_{ij}
		=
		\frac1\varphi\operatorname{diag}(0,-U,-2bg).
	}
	\label{eq:appF_background_metric_acceleration}
\end{equation}

The scalar equation is satisfied because
\begin{equation}
	T=-2tb=U'(\varphi).
\end{equation}
These are evolving initial data, not a stationary solution.

Initially,
\begin{equation}
	\partial_\tau(a^2)=0,\qquad
	\partial_\tau\sigma=0.
\end{equation}
For any sufficiently smooth solution evolving from
these data, nonzero regularity coefficients remain
nonzero locally in time by continuity.

This conditional persistence statement is not a local existence or hyperbolicity theorem. 

Since $\dot q=0$ initially, $\partial_\tau^2\sqrt q=-(U+2bg)/(2\varphi)$ there. This is negative for $\alpha>0$ and $\varphi>1$, but does not establish dynamical stability or collapse.

\paragraph{Secondary self-bracket on the initial slice.}
Orient the normalized null vector along $J_3$ and write $S_u=f_1(\nu)$. On these data $f_1(X)=0$ for every fixed generator, $u(J_3)=0$, and $\partial S_u/\partial P=0$. Hence $J_{0,u}=0$. The only remaining variation is $g_j R^{ij}([D_i^A\eta,J_3])$ for Lorentz smearings, and $R^{ij}([A_i,J_3])\partial_j\rho$ for a scalar smearing. With $R^{13}(J_2)=1/\kappa$, $R^{23}(J_1)=-1/\kappa$, the row in the ordered primary basis is
\begin{equation}
\mathcal K_u=\frac1\kappa(0,0,0,-g\partial_1,-g\partial_2,-gt,r\partial_2).
\end{equation}
In particular the scalar entry need not vanish when $r\ne0$. The $JJ$, $J\varphi$ and $\varphi\varphi$ blocks of $Q_7$ vanish identically along this slice (since $\beta$ is purely boost). Therefore $E_u=-\mathcal K_u^\dagger$ has support in an isotropic block of $Q_7$, and
\begin{equation}
J_{*,u}=J_{0,u}+E_u^\dagger Q_7E_u=0,\qquad
\mathcal B_u^{-1}=\begin{pmatrix}0&-\kappa/(gt)\\\kappa/(gt)&0\end{pmatrix}.
\end{equation}
This holds with variable coefficients on the slice. For spatial torsion the direct secondary numerator vanishes here, so $s_f=-U_fQ_7E_u$ in the unit--unit convention; substitution in Eq.~\eqref{eq:fully-stabilized-torsion} gives the remaining cross corrections.

\paragraph{Explicit nonvanishing torsion bracket.}
Set $\gamma=g/\kappa$ and use the ordered basis $(K_1,K_2,K_3,J_1,J_2,J_3,\varphi)$. Here $a=\gamma e_3$, $\beta_K=(-r,0,t)/\kappa$, $\beta_J=0$. Substitution in the primary pseudoinverse gives the only independent nonzero entries
\begin{equation}
(Q_7)_{K_1J_2}=1/\gamma,\quad (Q_7)_{K_2J_1}=-1/\gamma,\quad
(Q_7)_{K_3J_2}=r/(\gamma t),\quad (Q_7)_{K_3\varphi}=\kappa/t.
\end{equation}
The other entries follow by antisymmetry or vanish. For $T_f=\int f\mathscr T^0{}_{12}$ and $R_h=\int h\mathscr T^1{}_{13}$, with real compactly supported tests, the body variation $X=k^IK_I+j^IJ_I$ gives $\delta_X\mathscr T^0{}_{12}=\partial_1 k^2-\partial_2 k^1$ and $\delta_X\mathscr T^1{}_{13}=\partial_1j^2$ (the connection commutators have zero projection here). Integration by parts yields
\begin{equation}
U_f=(\partial_2f,-\partial_1f,0,0,0,0,0),\qquad
U_h=(0,0,0,0,-\partial_1h,0,0).
\end{equation}
Both annihilate the null line $J_3$, independently of its orientation. Since the lower-right entry of the residual inverse is zero, all residual corrections vanish for this pair, giving Eq.~\eqref{eq:explicit-torsion-example}. Taking $h=\partial_1\partial_2 f$ yields $(\kappa/g)\int(\partial_1\partial_2 f)^2$, nonzero for suitable $f$. Equivalently the evaluated distribution is $(\kappa/g)\partial_{x^1}\partial_{x^2}\delta^3(x-y)$. These are reference-relative, not hypersurface-invariant observables; $g\to0$ leaves the regular chart and does not by itself establish strong coupling.

\paragraph{Tangent independence test.}
For arbitrary metric and momentum variations at fixed $\varphi,\Lambda,\xi$, solve the primary linearizations by $\delta L=\delta a$, $\delta p'_\varphi=-\delta[B^i(A_i)]$. Add the null-primary Hamiltonian flow with smearing $\mu=-z_0^{-1}\delta S_u$. This makes the variation tangent to the secondary as well, since the null flow preserves the primaries and changes $S_u$ by $z_0\mu$. The projectability identities of Sec.~\ref{sec:dirac_bergmann} imply that this correction leaves the variations of the temporal constraints unchanged on the full constraint surface.
For $k\ne0$, the spatial momentum symbol $p^{ij}\mapsto-2ik_jp^{ij}$ is surjective: a right inverse of the full momentum symbol is $p^{ij}=\tfrac i2[(w^ik^j+k^iw^j)/k^2-(w\cdot k)k^ik^j/k^4]$. The normal metric symbol is $\varphi(k_ik_jh_{ij}-k^2h)/(2\kappa)$ with Fourier convention $\partial_i\mapsto ik_i$, with conformal right inverse $h_{ij}=-\kappa w_0\delta_{ij}/(\varphi k^2)$. Thus the mixed-order reduced symbol has rank four. More strongly, for compactly supported smooth smearings on a connected Cartesian patch, the momentum variation of $H[N]+D[V]$ gives $\partial_{(i}V_{j)}=0$: the normal kinetic derivative vanishes at $P=0$ and the relative spatial generator has no metric momentum correction. A compactly supported Euclidean Killing field is zero. For the remaining normal smearing take $\delta q_{ij}=f\delta_{ij}$. The curvature, scalar potential, torsion trace and relative terms respectively vary by
\begin{align}
 \delta H_R[N]&=\kappa^{-1}\int N\varphi\Delta f, &
 \delta H_U[N]&=-3(4\kappa)^{-1}\int NUf,\\
 \delta H_t[N]&=\kappa^{-1}\int Ng\partial_3 f, &
 \delta H_\Lambda[N]&=-(2\kappa)^{-1}\int Ntgf.
\end{align}
Here $\delta\sqrt q=3f/2$, $\delta{}^{(3)}R=-2\Delta f$, $\delta t_i=-\partial_i f$, and $\delta R^{ij}(X)=fR^{ij}(X)/2$. The kinetic term has zero first variation at $P=0$. Using $tg=-U/2$, the sum is
\begin{equation}
 \delta H[N]=\kappa^{-1}\int N(\varphi\Delta f+g\partial_3 f-Uf/2).
\end{equation}
The null-flow correction to the tangent variation leaves this result unchanged by projectability. Integrating by parts and using $\Delta\varphi=0$ gives $\delta H[N]=-\int f\mathcal M[N]$, where
\begin{equation}
 \mathcal M[N]=\kappa^{-1}[-\partial_i(\varphi\partial_iN)+UN/2].
\end{equation}
For $\alpha>0$, $\varphi>1$ and compact support (or homogeneous Dirichlet conditions),
\begin{equation}
 \int N\mathcal M[N]=\kappa^{-1}\int(\varphi|\nabla N|^2+UN^2/2)>0
\end{equation}
for nonzero real $N$. More explicitly, if $\phi\geq\phi_{\min}>0$
and $U\geq0$ on a bounded admissible patch,
then
\[
\langle N,\mathcal M[N]\rangle
\geq
\frac{\phi_{\min}}{\kappa}
\|\nabla N\|_{L^2}^2.
\]
With homogeneous Dirichlet conditions,
Poincar\'e's inequality gives coercivity
on $H_0^1$. Together with the Killing test this establishes slice independence conditional on projectability, not generic independence or evolution. The operator is not a temporal gauge bracket.

\end{document}